\documentclass{aa}  
\usepackage{siunitx}
\usepackage{graphicx}
\usepackage{txfonts}
\usepackage{amssymb}
\usepackage{booktabs} 
\usepackage{xcolor}
\usepackage{orcidlink}
\usepackage{subcaption}
\usepackage{xcolor} 
\usepackage{float}
\usepackage{siunitx}
\usepackage{hyperref}
\usepackage{url}
\usepackage{rotating}
\usepackage{makecell}
\usepackage[utf8]{inputenc}
\usepackage[english]{babel} %
\newcommand{\teff}{$T_{\rm eff}$}
\newcommand{\logg}{$\log g$}
\newcommand{\vsini}{$v\sin i$}
\newcommand{\rotfit}{{\tt ROTFIT}}

\usepackage[version=4]{mhchem}
\begin{document}

   \title{Exploring the radial velocity variations of RY Lup with VLT/ESPRESSO: Binary versus spot hypotheses}

   \subtitle{}

   \author{
    Hala Alqubelat\orcidlink{0009-0002-4535-1704}\inst{1}\thanks{Corresponding author: \email{hala.alqubelat@eso.org}} ,
    Claudia Di Maio\orcidlink{0000-0002-8669-1150}\inst{2} ,
    Antonio Frasca\orcidlink{0000-0002-0474-0896}\inst{3},
    Carlo F. Manara\orcidlink{0000-0003-3562-262X}\inst{1},
    Monika G. Petr-Gotzens\inst{1},
    Enrico~Ragusa\orcidlink{0000-0001-5378-7749}\inst{4},
    Evelyne Alecian \orcidlink{0000-0001-5260-7179}\inst{5},
    Louise D. Nielsen\inst{6}\orcidlink{0000-0002-5254-2499}{6}, 
    Lisa Drouglazet \orcidlink{0009-0002-5871-1529} \inst{5},
    Justyn Campbell-White\orcidlink{0000-0002-3913-3746}\inst{1}
    }

\institute{
    European Southern Observatory, Karl-Schwarzschild-Strasse 2, 85748 Garching bei München, Germany\and 
    INAF – Osservatorio Astronomico di Palermo, Piazza del Parlamento, 1, 90134 Palermo, Italy\and
    INAF – Osservatorio Astrofisico di Catania, Via S. Sofia 78, 95123 Catania, Italy\and
    Dipartimento di Fisica, Université degli Studi di Milano, Via Celoria 16, 20133 Milano, Italy \and 
    Univ. Grenoble Alpes, CNRS, IPAG, 38000 Grenoble, France\and
    University Observatory, Faculty of Physics, Ludwig-Maximilians- Universität München, Scheinerstr. 1, 81679 Munich, Germany}
    
   \date{Received 29 May 2026 ; Accepted  July 2026}

 \titlerunning{Testing the binarity hypothesis of RY Lup}
\authorrunning{H.Alqubelat et al}

\abstract{Stellar multiplicity is a possible cause for creating protoplanetary disc substructures, as tidal forces from a close-in spectroscopic stellar companion can carve out gaps and shape disc architecture. However, in young, active systems, the radial velocity (RV) signatures are often complicated by stellar activity. We investigate RY Lup, a classical T Tauri star hosting a disc with a $\sim 60$\,au cavity, where studies with \textit{Gaia} astrometry and VLT/SPHERE imaging hinted at an unseen companion. Using high-resolution VLT/ESPRESSO spectra and the least-squares deconvolution (LSD) technique, we analyse RV variations over 327 days. We detect significant line profile variations with a periodic signal of $\sim 3.75$ days, aligning with prior photometric estimates. The variations are compatible with a close-in binary system at $\sim 0.04$\,au and a mass ratio of $q \sim 0.6$. Combined analysis of RV data and ALMA dynamical mass estimates, using $\ce{^{13}CO}$ and $\ce{^{18}CO}$, reveals a highly misaligned system. The nearly face-on binary $i \sim 13^{\circ}$ is misaligned to both the inner and outer discs $i \sim 50^{\circ}$ and $\sim 70^{\circ}$, respectively. The derived orbital separation is compatible with the inner disc size, with the inner rim at $a=0.12$\,au, measured from VLTI/GRAVITY, which suggests a highly warped disc structure. Nonetheless, the short orbital period conflicts with the derived eccentricity ($e\sim0.23$).
To explore alternative explanations, we assess the impact of stellar spots on RV signals. While the LSD deformations can be modelled by different cool spot configurations, a $15\%$ dispersion in retrieved $v\sin i$ values— coupled with the lack of a significant periodic signal— suggests that spots alone cannot explain the observed variability. As neither hypothesis is ruled out, we recommend future combined RV and interferometric monitoring to clarify the nature of the spectroscopic variability.}

\keywords{stars: individual: RY Lup -- stars: pre-main sequence -- stars: binaries: spectroscopic -- planetary systems: protoplanetary disks -- stars: activity -- techniques: radial velocities}

   \maketitle
%

\section{Introduction}

The origin and early evolution of planetary systems are fundamentally tied to the physical processes occurring within protoplanetary discs. Observational studies, particularly in the millimeter regime, have revealed an ubiquitous presence of substructures (e.g., rings, gaps, and spirals) in these discs \citep[e.g.,][]{2020ARA&A..58..483A,2023ASPC..534..605B}.
While the presence of forming planets is the most common explanation for the carved discs (e.g., \citealt{2018A&A...617A..44K, 2024A&A...685L...1C, 2025A&A...704A.221V}), other mechanisms like stellar binary companions can carve large central cavities with size scales predictably with the binary separation as shown by simulations (e.g.,\citealt{1994ApJ...421..651A}; \citealt{2015MNRAS.452.2396M}; \citealt{2025A&A...698A.102R}). Observationally, systems like CS Cha and CoKuTau/4 host cavities that are carved by a 5 and 8 au binary, respectively (e.g.,\citealt{2018A&A...616A..79G,2024A&A...685A..52G,2008ApJ...678L..59I}). A particularly interesting subset of discs are those with cavities, often referred to as transition discs (TDs). These are characterised by a spectral energy distribution (SED) showing low near-infrared (NIR) excess, suggesting a significant reduction in dust mass within the inner disc, while displaying strong mid- and far-infrared (IR) excess from the outer disc. This indicates that the inner regions are cleared or substantially depleted of dust \citep{Espaillat2014, 2023EPJP..138..225V}. The origin of these cavities is still debated between photoevaporation \citep[e.g.,][]{2023MNRAS.523.3318P} or the presence of companions \citep[e.g.,][]{2025A&A...698A.102R}.

RY Lup (RA~15h\,59m\,28.39s, Dec~$-40^{\circ} 21' 51.3''$) is a K2-type T Tauri star located in the Lupus association at $158$~pc \citep{2018A&A...616A...1G}. The literature reports varying projected rotational velocity $v\sin i$ values from $v\sin i\sim16.3\pm5.3$\,\unit{km.s^{-1}}
\citep{2017A&A...600A..20A, 2017A&A...602A..33F} to $v\sin i$$\sim25.0 \pm 4.6$\,\unit{km.s^{-1}} 
\citep{1986A&A...165..110B}. The lower estimates by \citet{2017A&A...600A..20A} and \citet{2017A&A...602A..33F} are possibly due to the correlation between the $v\sin i$ and veiling, which is particularly pronounced in the 605-625\,nm spectral range used in their analysis.

The total mass of RY Lup is estimated between $1.3\text{--}1.5~M_{\odot}$, as derived from VLT/X-Shooter spectroscopy and $^{13}$CO and C$^{18}$O from ALMA \citep{2017A&A...600A..20A, 2018A&A...616A.100Y, 2021ApJ...908...46B}. The system is a well-known transition disc, characterised by a mass accretion rate of $\log \dot{M}_{\text{acc}} \approx -8.19\,M_{\odot}\,\text{yr}^{-1}$ \citep{2017A&A...600A..20A} and a large dust cavity with a radius of $\sim 60$\,au resolved with ALMA \citep{2016ApJ...828...46A}. UV-to-NIR studies confirm the presence of a gas gap within this millimeter-dust cavity, consistent with its evolutionary stage \citep{2018ApJ...855...98A}. High-resolution imaging and interferometry have revealed a complex architecture; while the outer disc is inclined at $\sim 67^{\circ}\text{--}70^{\circ}$ \citep{2018A&A...614A..88L, 2020ApJ...892..111F}, VLTI/GRAVITY and PIONIER data suggest an inner dust rim at $0.12$\,au \citep{2020A&A...642A.162G} inclined at $\sim 50^{\circ}$ \citep{2018A&A...616A.100Y,2022A&A...658A.183B}. This misalignment suggests a warped structure potentially carved by a massive planetary companion ($M \gtrsim 2~M_{\mathrm{Jup}}$) \citep{2018A&A...614A..88L}.

The system exhibits significant photometric and polarimetric variability with a 3.75-day period \citep{1989A&A...211..115G, 2009A&A...499..137M}. This behaviour is attributed to corotating circumstellar material causing partial occultations, consistent with the quasi-periodic, dipper light curve observed by \citet{2020MNRAS.496.3257B}.  Stellar spots have been excluded as the dominant driver, as the observed colour variations and stable spectral type are inconsistent with spot-induced modulation \citep{2009A&A...499..137M}. Furthermore, their analysis 
of low resolution spectra found no evidence of a companion. While the 3.75\,day periodicity is dominant, the phase is not consistently conserved, and additional random fluctuations suggest an evolving circumstellar environment.

Recent \textit{Gaia} proper motion anomaly modelling by \citet{2026A&A...705A.238V} suggests a companion of $M_{\star} \sim 100\,M_{\text{Jup}}$ with semi-major axis at $3\text{--}10$\,au, whereas \citet{2026arXiv260207731B} report no evidence for stellar-mass companions between $3$ and $25$\,au using \textit{Hipparcos}-\textit{Gaia} astrometry. Detecting close-in companions remains challenging for imaging techniques, motivating the use of high-resolution spectroscopy.

High-resolution spectroscopy, combined with robust techniques such as least-squares deconvolution (LSD; \citealt{1997MNRAS.291..658D}) is particularly well suited for rapidly rotating stars. The S/N increase provided by this method allows us to take into account the observations with low S/N across multiple spectral regions.

Both spectroscopic and photometric variations observed in T Tauri stars can arise from stellar activity, hidden companions, or a complex interplay of both. Studies of T Tauri stars have shown that often spots can cover up to 15--45$\%$ of the photosphere \citep{2008ApJ...678..472H, 2017ApJ...836..200G, 2022A&A...667A.124G, 2025ApJ...992L..33P}. In rapidly rotating stars, the distribution of cool magnetic or bright accretion spots distorts spectral line profiles, often mimicking the radial velocity (RV) signals of orbiting companions. This modulation is particularly problematic when the spectral profile is not fully resolved; in such cases, the shifting centroid of the distorted profile manifests as spurious RV variations. The resulting signals are modulated by the stellar magnetic cycle and often correlate with chromospheric activity indicators \citep{1997MNRAS.291..658D}. Techniques such as Doppler Imaging (DI) and Zeeman Doppler Imaging (ZDI) provide key tools for characterising these effects \citep{Strassmeier2009A&ARv..17..251S,2016A&A...587A..28H,2019A&A...625A..79H, 2026A&A...708A.230D}, alongside starspot characterisation modelling \citep{2024A&A...683A.239D}.

Disentangling intrinsic stellar noise from companion-induced signals is therefore critical for systems such as RY Lup. Utilising high-resolution VLT/ESPRESSO spectroscopy, this study investigates the RV variations of RY Lup to provide a comprehensive exploration of their origin—specifically exploring whether these shifts are products of stellar spots, a binary companion, or inherent single-star variability. This paper is structured as follows: In Sect.\ref{data_reduction}, we describe the observations and data reduction process. In Sect.\ref{rv_analysis}, we present the RV analysis, the orbital solution for a potential companion, stellar spots modelling and emission-line diagnostics. Finally, in Sect.\ref{discussion}, we discuss the implications for the evolution of the RY Lup system.


\section{Observations and data reduction}
\label{data_reduction}

Observations of RY Lup were obtained between 2022 May and 2023 April using the high-resolution, fibre-fed, cross-dispersed, échelle spectrograph ESPRESSO \citep[Echelle SPectrograph for Rocky Exoplanet and Stable Spectroscopic Observations,][]{2021A&A...645A..96P} mounted on the ESO Very Large Telescope
(VLT). The spectra were collected during the programs 106.20Z8.007 \citep[PENELLOPE Large Program][]{2021A&A...650A.196M}, 110.2483.001, and 111.250K.001 (PI: Manara). All observations were executed in Service mode with very loose constraints on moon phase and seeing to ensure data collection. In total, 21 spectra were obtained at unevenly spaced time separation across the three periods to cover a wide range of possible orbital periods. 

The data were collected in single-UT high-resolution mode, with a $2\times1$ pixel binning and a slow readout (\texttt{2x1\_SLOW}), providing a resolving power of $R \sim 140\,000$ over a spectral range of $370$ to $788\,\mathrm{nm}$. 
The observing blocks for RY Lup were set such that two consecutive exposures were taken per epoch such that they can be combined to help mitigating the impact of cosmic rays artifacts. For our subsequent analysis, we treated the exposures individually and ultimately selected one single exposure per epoch. The observations were performed in simultaneous-sky mode, where fibre A was centred on RY Lup and fibre B was pointed at the sky to allow for background subtraction. The individual exposure time was 340\,s throughout most of the campaign with the exception of the last four epochs in April 2023, where the exposure time was increased to 700\,s seconds per exposure. Seeing conditions during the observations remained stable with values falling between 0.3'' and 1.0'', yielding a S/N ratio range between $\sim$ $13$ and $97$ (Table~\ref{table_all}).

We used the dedicated ESPRESSO reduction pipeline v.3.3.12 \citep{2020ASPC..527..667M} within the ESO Reflex environment \citep{Freudling13} to reduce the RY Lup data. The pipeline carries out the bias, dark, flat field corrections, the wavelength calibration and extracts 2D spectra and merged ID spectra corrected to the barycentric velocity frame. For our analysis, we use the flux-calibrated, 1D merged spectra (S1D). No telluric correction was applied to the data, instead, we restricted our analysis to regions that are free of telluric lines.

\begin{table*}[h!] 
\hspace*{\fill}
\raggedleft
    \centering
    \caption{Log of the observations, RV measurements, errors ($\sigma$), the median S/N ratio across all orders, and seeing conditions of the observed spectra.}
    
        \begin{tabular}{lccccccc}
\hline
\hline
Observation date & MJD &$\phi$ &$RV_{1}$ (km s$^{-1}$)  & $RV_{2}$ (km s$^{-1}$) & S/N & Seeing (\arcsec)  \\
\hline

2022 May 27 & 59726.04793 &0.2061 & $8.0\pm0.35$ & $-15.43 \pm 0.48$  & 47 & 0.32 \\
2022 May 28 & 59727.04719 &0.4725 &$8.43 \pm 0.55$ & $-15.05 \pm 0.50$ & 65 & 0.59 \\
2022 May 30 & 59729.05996 &0.0091 &$-5.0 \pm 0.50$ & $16.34 \pm 0.77$  & 50 & 0.75 \\
2022 May 31 & 59730.98834 &0.5232 &$8.13 \pm 0.55$ & $-15.42 \pm 0.53$ & 30 & 0.62 \\
2022 June 4 & 59734.0292 &0.3339 &$11.20 \pm 0.38$ & $-12.0 \pm 0.32$ & 34 & 2.33 \\
2022 October 6 & 59858.00158 &0.3844 &$6.70 \pm 0.66$ & $-16.0 \pm 0.05$  & 25& 2.12 \\
2023 January 15 & 59959.33324 &0.3989 &$-11.44 \pm 0.30$ & $15.70 \pm 0.43$ & 13 & 1.56 \\
2023 January 30 & 59974.35934 &0.4048 &1$0.0 \pm 0.51$ & $-13.21 \pm 0.44$  & 39 & 0.43  \\
2023 February 3 & 59978.35017 &0.4688 &$7.42 \pm 0.63$ & $-14.0 \pm 0.47$ & 35 & 0.6 \\
2023 February 21 & 59996.25115 &0.2411 &$10.0 \pm 0.10$ & $-16.0 \pm 0.1$& 35 & 0.66  \\
2023 March 2 & 60005.22714 &0.6340 &$5.0 \pm 2.0$ & $-10.0 \pm 2.0$& 40 & 0.92 \\
2023 March 5 & 60008.30703 &0.4551 &$9.14 \pm 0.46$ & $-14.06 \pm 0.34$  & 30 & 0.83\\
2023 March 6 & 60009.22344 &0.6994 &$-10.0 \pm 2.0$ & $5.0 \pm 0.040 $& 23 & 1.85 \\
2023 March 10 & 60013.30837 &0.7885 &$-10.02 \pm 0.10$& $14.13 \pm 0.10$& 40 & 1.21 \\
2023 March 16 & 60019.38418 &0.4082 &$7.71 \pm 0.10$ & $-15.47 \pm 0.10$ & 70 & 0.57 \\
2023 March 24 & 60027.35871 &0.5342 &$-15.0 \pm 0.36$ & $13.13 \pm 0.33$ & 22 & 1.27 \\
2023 March 27 & 60030.1659 &0.2826 &$-13.61 \pm 0.10$ & $12.30 \pm 0.10$ & 33 & 0.52  \\
2023 April 13 & 60047.30848 & 0.8527 &$-10.0 \pm 0.49 $& $15.35 \pm 0.40$ & 84 & 0.53 \\
2023 April 14 & 60048.17495 &0.0837 &$-8.0 \pm 0.56$ & $15.03 \pm 0.56$ & 81 & 0.94 \\
2023 April 15 & 60049.39155 &0.4081 &$5.0 \pm 0.60$ & $-16.82 \pm 0.63$ & 97 & 0.5 \\
2023 April 19 & 60053.15715 &0.4120 &$9.73 \pm 0.42$ & $-13.83 \pm 0.57$ & 62 & 0.91 \\

\hline
\end{tabular}

\label{table_all}
\setlength{\tabcolsep}{0.5pt}

\footnotesize
    \tablefoot{The RV values of the spectra taken on March 2 and April 14 are excluded from the final orbital fitting shown in Fig.\ref{fig:rv_orbit_best.png} to constrain the argument of periastron. The 2023 January 15, March 24, and 27 are excluded from the orbital fitting analysis. }
\end{table*}

\section{Analysis}
\label{rv_analysis}

To explore the RV variations traced by the ESPRESSO spectra of RY Lup, we perform three analyses. First, we study the absorption lines assuming that RY Lup is a single star, then we test the hypothesis that the observed variations are driven by a binary system, and finally we consider that the observed variations in the absorption lines are caused by multiple spots on the stellar surface of a single star.

\subsection{Stellar properties of RY Lup as a single star}
\label{stellar_properties}

We used the \rotfit\ code \citep{2003A&A...405..149F} to estimate the photospheric properties as the effective temperature ($T_{\mathrm{eff}}$), surface gravity ($\log\, g$), the projected rotational velocity ($v \sin i$), the iron abundance ([Fe/H]), and veiling ($r$) at seven wavelengths ($\lambda\sim$ 400\,nm, $\lambda\sim$ 420\,nm, $\lambda\sim$ 450\,nm,  $\lambda\sim $500\,nm, $\lambda\sim$ 550, $\lambda\sim$ 600, and $\lambda\sim$ 650\,nm) of the RY Lup ESPRESSO spectra, following the same procedure as described by, e.g., \citet{2015A&A...575A...4F} and \citet{2021A&A...650A.196M}.

\rotfit\ searches within a grid of templates to find the one that minimises the $\chi^{2}$ relative to the target spectrum across several spectral regions. 
The templates grid consists of 44 high-signal-to-noise ratio HARPS spectra (R$\simeq$\,115\,000) retrieved from
the HARPS archive.
This grid expands upon the one used by \citet{2021A&A...650A.196M} by extending the coverage to late G-type stars (\teff$\simeq$5440\,K), and incorporating lower-gravity subgiant stars. 

To perform the fit, \rotfit\ first determines the RV of the target star via cross-correlation with the template spectra, shifting the observed spectrum to the rest frame. The code then convolves the observed spectrum with a Gaussian kernel of the proper width to match the HARPS resolution. Next, the templates are iteratively broadened by convolution with a rotational profile with increasing \vsini\ until the minimum $\chi^{2}$ is achieved. The procedure is repeated by introducing an artificial veiling, $r$, to the template spectra according to: 
\begin{equation}
     F_\lambda^r/F_{\rm c}=\frac{F_\lambda/F_{\rm c}+r}{1+r}\,,
\end{equation}

\begin{figure}
    \centering
    \includegraphics[trim=0 0 0 15  , clip,width=0.55\textwidth]{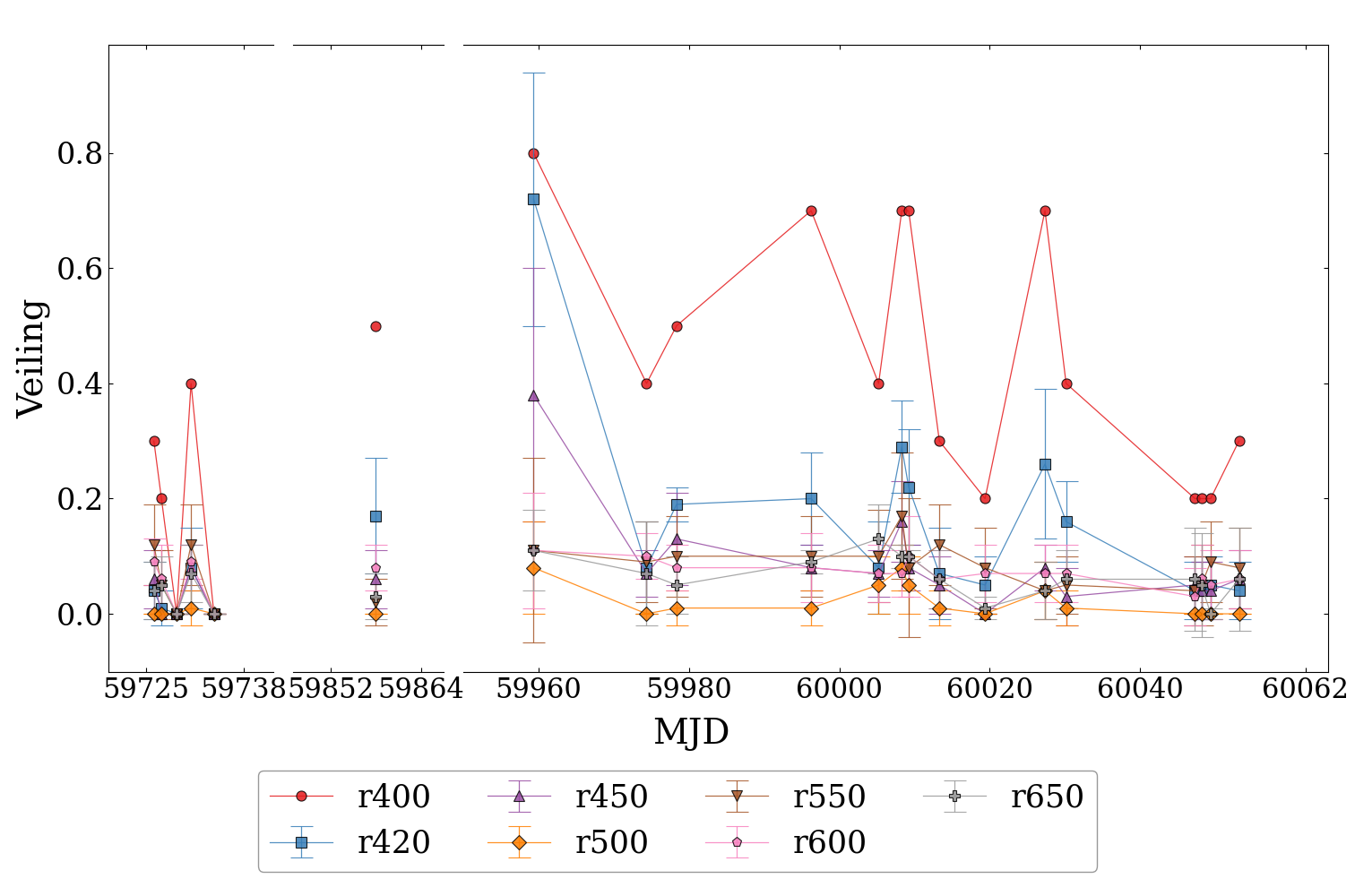}
    \vspace{-10pt}
    \caption{Measured veiling values at different wavelength ranges for the spectra of RY Lup.} 
    \label{fig:veiling}
\end{figure}

\noindent{where $F_{\rm c}$ represents the continuum flux. }
The optimal veiling for each analysed 100\,\AA-wide spectral segment is determined by minimising the $\chi^{2}$ residual.
We found zero or negligible $r$ across most spectral regions, except for $\lambda\sim$ 400 nm and $\lambda\sim$ 420 nm, where we measured a maximum value of $\sim 0.9$. 
The temporal evolution of $r$ at these specific wavelengths is shown in Fig.~\ref{fig:veiling}.

Finally, the fundamental stellar parameters were derived by taking a weighted average of the results from the individual spectral segments. For RY Lup, we derived a spectral type of K1 V-IV, \teff\,(K)= $5167\pm 24$, \logg\ (dex)$=4.21\pm 0.16$, \vsini\,(km s$^{-1}$) = $32.3 \pm 1.1$, and mean $\text{RV}$ (km s$^{-1}$)$=0.77\pm 1.40$.

\subsection{Analysis of LSD profiles with the hypothesis of a binary star}

\label{LSD binary}

        \begin{table}[h]
    \centering
    \caption{Spectroscopy orbital elements of RY Lup.
    }
    \label{table:orbital_param}
    \renewcommand{\arraystretch}{1.3} 
    \begin{tabular}{lc}
        \toprule
        \toprule
        Element & Value \\
        \midrule
        $P$ (d) & $3.75  \pm 0.05  $ \\
        $V_{0}$ (km s$^{-1}$) & $ -0.65  \pm 0.06 $ \\
        \textit{K$_1$} (km s$^{-1}$) & $ 12.45 \pm  0.26  $ \\
        \textit{K$_2$} (km s$^{-1}$) & $19.53 \pm  0.20 $ \\
        $e$ & $0.23  \pm 0.01 $  \\
        $\omega$ (deg)  & $ 47.17 \pm 0.12$ \\
        $T$$_0$ (MJD) & $59717.77\pm 0.15  $ \\
        \midrule
        $q = M_2/M_1$ & $0.65  \pm 0.02 $ \\
        $M_1 \sin^3 i$ (M$_\odot$) & $0.0100 \pm 0.00035 $ \\
        $M_2 \sin^3 i$ (M$_\odot$) & $0.0700 \pm 0.00028$ \\
        $a_1 \sin i$ (au) & $0.0040 \pm0.0001 $ \\
        $a_2 \sin i$ (au) & $ 0.0060\pm0.0001 $ \\

        $a \sin i$ (au) & $0.01010  \pm0.0001  $ \\

        $a \sin i$ ($R_{\odot}$) & $2.1700 \pm 0.0001 $ \\
        \bottomrule
    \end{tabular}
\end{table}

To determine the RV of RY Lup, we computed LSD profile for each ESPRESSO epoch following the methodology of \citet{1997MNRAS.291..658D}. This technique extracts a single high S/N average line profile from thousands of available spectral lines by deconvolving the observed spectrum with a theoretical line mask. We used K2 spectral type mask derived from a line list provided by the VALD database \citep{2015PhyS...90e4005R}, and covering the wavelength range of ESPRESSO. In the analysis, we excluded the telluric lines and all prominent emission lines. For this process, we used \textsc{specpolFlow}\footnote{\url{https://folsomcp.github.io/specpolFlow/}} \citep{2024AAS...24322503E} to first normalise the spectra and calculate the LSD profiles over a velocity range of $\pm 100$\,km\,s$^{-1}$ with a velocity step of $0.5$\,km\,s$^{-1}$.
 
\begin{figure}
    \centering
    \includegraphics[width=0.5\textwidth]{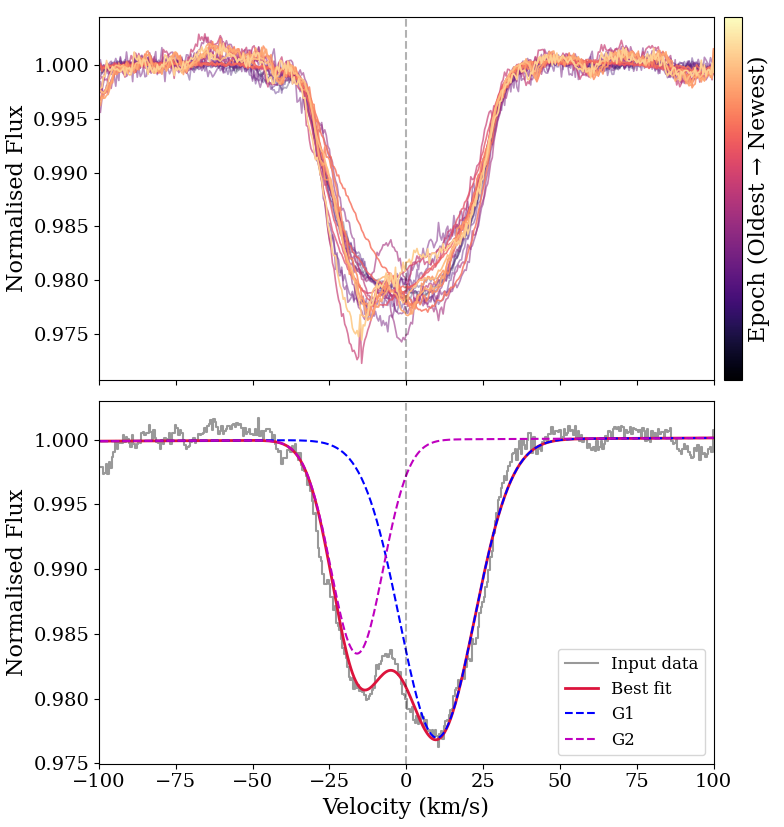}
        \vspace{-8pt}

    \caption{Top panel, RY Lup ESPRESSO LSD profiles over 327 days. Bottom panel, 2023 February 21 RY Lup LSD profile fitted with two Gaussian profiles in magenta and blue.
    }
    \label{fig:LSDs_SB2}
\end{figure}

 The LSD profiles of RY Lup are variable across the epochs as shown in Fig.\ref{fig:LSDs_SB2}. While several epochs clearly display bimodal signatures indicative of two distinct RV components (Fig.\ref{fig:LSDs_SB2}), others appear broader with complex peak morphologies, complicating the fitting process. In those cases, we observed additional structures in the LSD profiles. These features were particularly prevalent in the data from 2023 January and March (Fig.\ref{fig:total_LSD_grid}).
 The LSD profiles taken between 2022 May 27 and June 4 (Fig.\ref{fig:total_LSD_grid}), show variable peak morphologies. However, in the epoch taken early June, the line appears more distorted with two RV peaks. The LSDs of 2023 April show variable shapes and centroid shifts, with the last epoch of April 19 showing a whole profile shift (Fig.\ref{fig:total_LSD_grid}). The line profiles of the epochs taken during March (Fig.\ref{fig:total_LSD_grid}), especially the epochs of 24 and 27 appear deeper and more structured compared to the other RY Lup spectra in March as the 6th and 10th, which show two RV signatures. For the data taken in 2023 February and March, the profiles vary in shape, depth, and centroids (Fig.\ref{fig:total_LSD_grid}).

We modelled the LSD profiles of RY Lup using a dual-Gaussian fit to characterise its complex line morphology. This approach allowed us to resolve the profile into two possible distinct stellar components: A primary component characterised by broad, deep lines and a secondary component exhibiting narrower, shallower lines (Fig. \ref{fig:LSDs_SB2}).
We started the fitting procedure from the spectrum in which we identified two components in the LSD profiles, such as the one taken on 2023 February 21 (Fig.\ref{fig:total_LSD_grid}). We used the widths and amplitudes obtained in this initial fit as initial guess for the fit of the other epochs, while allowing the centroids to vary across all epochs. The observing seasons were defined as practical observational windows grouping epochs obtained within similar monthly periods: Season 1 (2022 May–June), Season 2 (2023 January–February), Season 3 (2023 March), and Season 4 (2023 April), with a single isolated epoch from 2022 October. The definition of our observing seasons is motivated by the uneven sampling of our data, aimed at ensuring each binned period contains sufficient epochs for a robust analysis. Individual monthly analysis is unfeasible for periods with sparse coverage, such as June 2022 (1 epoch) and October 2022 (1 isolated epoch). We group the data into distinct blocks based on temporal proximity and sampling density: Season 1 aggregates late Spring 2022 data (5 epochs in May, 1 in June); Season 2 forms a Winter 2023 baseline (2 epochs in January, 2 in February); while 2023 March (7 epochs) and 2023 April (4 epochs) contain enough coverage to be analysed as independent monthly windows.

We constrained the primary and secondary line widths to $\sim 11-13$\,km s$^{-1}$ and $\sim 6-9$\,km s$^{-1}$, and their depths to $45-65\%$ and $15-35\%$ relative to the continuum level. We then proceeded by sequentially fitting consecutive epochs to trace the change of RV of these components across different spectra and observing seasons.

Our fitting strategy prioritised capturing the RV peaks to constrain the centroids of the components.  As a result, the model did not always account for the shoulder of the LSD profiles (Fig.\ref{fig:total_LSD_grid}). In some cases, we note that we observe an additional component in the LSD profiles. For such complex profiles as shown in Season 2 and 3 (Fig.\ref{fig:total_LSD_grid}), we excluded the derived RVs from the orbital fitting analysis. To investigate the nature of the additional features present in the LSD profiles, in Sect.\ref{Spot modelling}, we explored a different explanation of the LSD behaviour based on star spots. In Fig.\ref{fig:total_comparison}, we show the LSD profiles across 1000 \text{\AA} of three different wavelength regions. While the shape of the LSD profiles is consistent across these ranges, the profiles in the blue part of the spectrum appear shallower compared to those in the red. This variation is unlikely to be caused by veiling, as increased veiling at shorter wavelengths would uniformly affect all parts of the LSD profile. Additionally, the \rotfit\ analysis (Sect.\ref{stellar_properties}) does not indicate significant veiling across the different wavelength regions, except for the the ranges at $\lambda\sim$ 400, and 420\,nm. 
Instead, a more plausible explanation is the presence of spots. The shallower blue wing at shorter wavelengths could be attributed to a spot-related bump or shoulder, which would be more prominent at blue wavelengths where the flux contrast between spots and the photosphere is greater.

\label{RV calculation and fitting}

\begin{figure}
    \centering
    \includegraphics[width=0.5\textwidth]{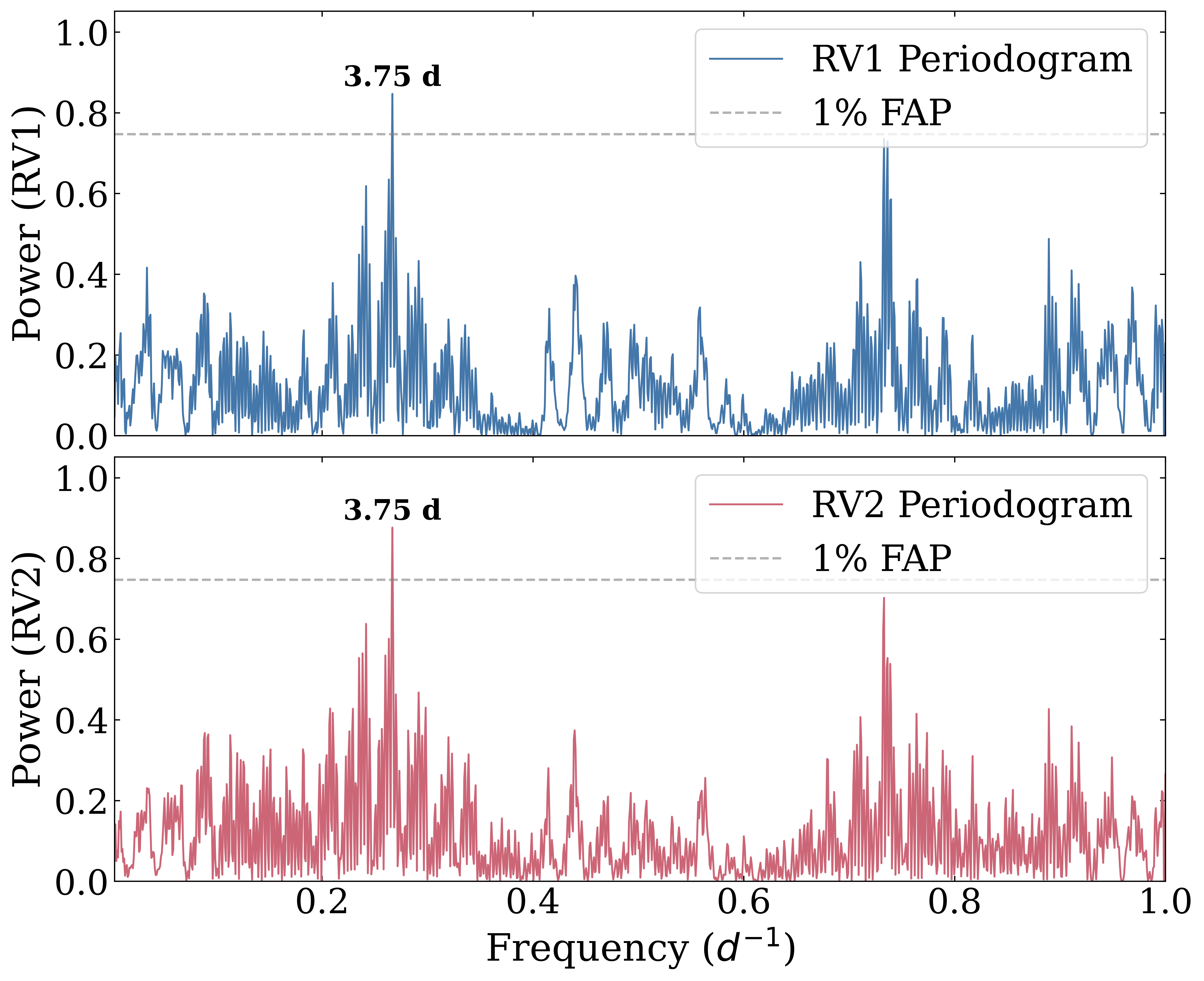}
    \vspace{-8pt}
    \caption{Lomb-Scargle periodogram of the $RV_{1}$ and $RV_{2}$ data of the ESPRESSO time-series of RY Lup spanning $\sim 327$ days. The 1\% FAP is shown as a dotted grey horizontal line.
    }
    \label{fig:periodogram}
\end{figure}

Next, we used Generalised Lomb-Scargle (GLS) periodogram \citep{Zechmeister2009A&A...496..577Z} to search for periodicity in the $RV_{1}$ and $RV_{2}$ values. We explored two frequency grids, a shorter one of 35 days and another one extending over the time span of the observations, i.e., 327 days.  We detect a period at $P=3.75$ days above the $1\%$ false alarm probability (FAP). In Season 4 (Fig.\ref{fig:total_LSD_grid}), we show an example of consecutive epochs taken over six days. The profiles shift happens across the period of $\sim 3.75$\,days, resulting in RV variations $> 15$\,\unit{km.s^{-1}}. To assess the period stability, we perform a bootstrapping approach that excludes a random subset of data points in each iteration. We run 50 iterations where a random subset of $80\%$ of the original observations were selected for each run. The 3.75\,days period remains the dominant signal across the different subsets. This value is consistent with the reported  period from the photometric and  polarimetric  observations \citep{2009A&A...499..137M}.

 We fitted the RV data using a a Markov Chain Monte Carlo (MCMC) approach to determine the orbital parameters of the binary system. The model fits the RV curves of a double spectroscopic binary system using a Keplerian orbital model to the data. The model takes as input the orbital period (P), the periastron passage ($T_{0}$), the systemic velocity ($V_{0}$), the semi-amplitudes of both components ($K_{1}$, $K_{2}$), the eccentricity ($e$), and the argument of periastron ($\omega$). Then, the model solves Kepler's equations to produce predicted curves of the RVs (\( v_1(t) \), \( v_2(t) \)) as functions of time (\( t \)). Detailed description of the used fitting procedure is described by \citet{2026A&A...706A.228A}.\\ 
 
\begin{figure}
    \centering
    \includegraphics[trim=0 20 0 70  , clip,width=\linewidth]{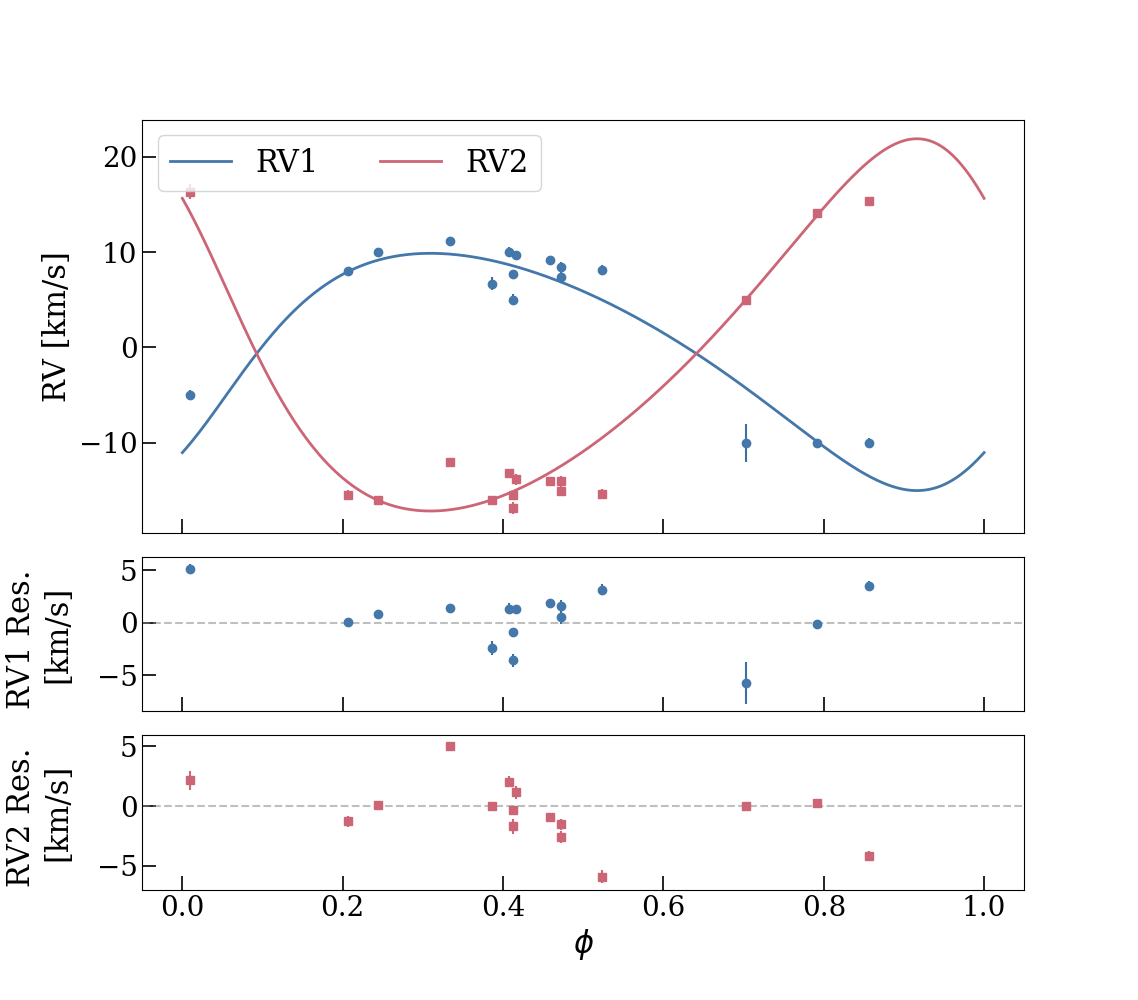}
    \vspace{-10pt}
    \caption{Orbital solution for RY Lup in phase. The blue and red lines fit the primary and secondary data taken with ESPRESSO, respectively. Below, the residuals of the fitted model in phase.
    }
    \label{fig:rv_orbit_best.png}
\end{figure}

  For RY Lup, we let the \texttt{emcee}\footnote{\url{https://emcee.readthedocs.io/en/stable/} } sampler runs for 300\,000 steps with an initial burn-in phase of 10\,000 steps, which are discarded.  The resulting corner plots revealed a bimodal distribution in the posterior probability (Fig.\ref{fig:corner_plot_orig}), showing two distinct orbital solutions. One is characterised by low eccentricity ($e<0.2$) (Fig.\ref{fig:cleaned_corner}), and another hints at a higher eccentricity ($e\sim0.5$). The low-eccentricity solution (Fig. \ref{fig:rv_orbit_best.png}) is statistically preferred, providing a better $\chi^{2}$ value, and more physically motivated given the short inferred period.

To constrain the argument of periastron ($\omega$), we excluded two RV data points at $\phi=0.2\, \text{and } 0.7$, corresponding to the April 14 and March 2. The resulting fit indicates an eccentric solution $e\sim0.23$, $\omega\sim 47.24 ^{\circ}$, period of $P\sim 3.75$\,days, and mass ratio of ($q\sim 0.64$). We adopt the parameters, which corresponds to the maximum likelihood values of this solution as our best estimates (Fig.\ref{fig:rv_orbit_best.png},\ref{fig:cleaned_corner}). These results are reported in the upper part of Table~\ref{table:orbital_param}.

\subsection{Spot modelling and emission lines analysis}
\label{Spot modelling}

\begin{figure}
     \centering
 \includegraphics[width=0.5\textwidth]{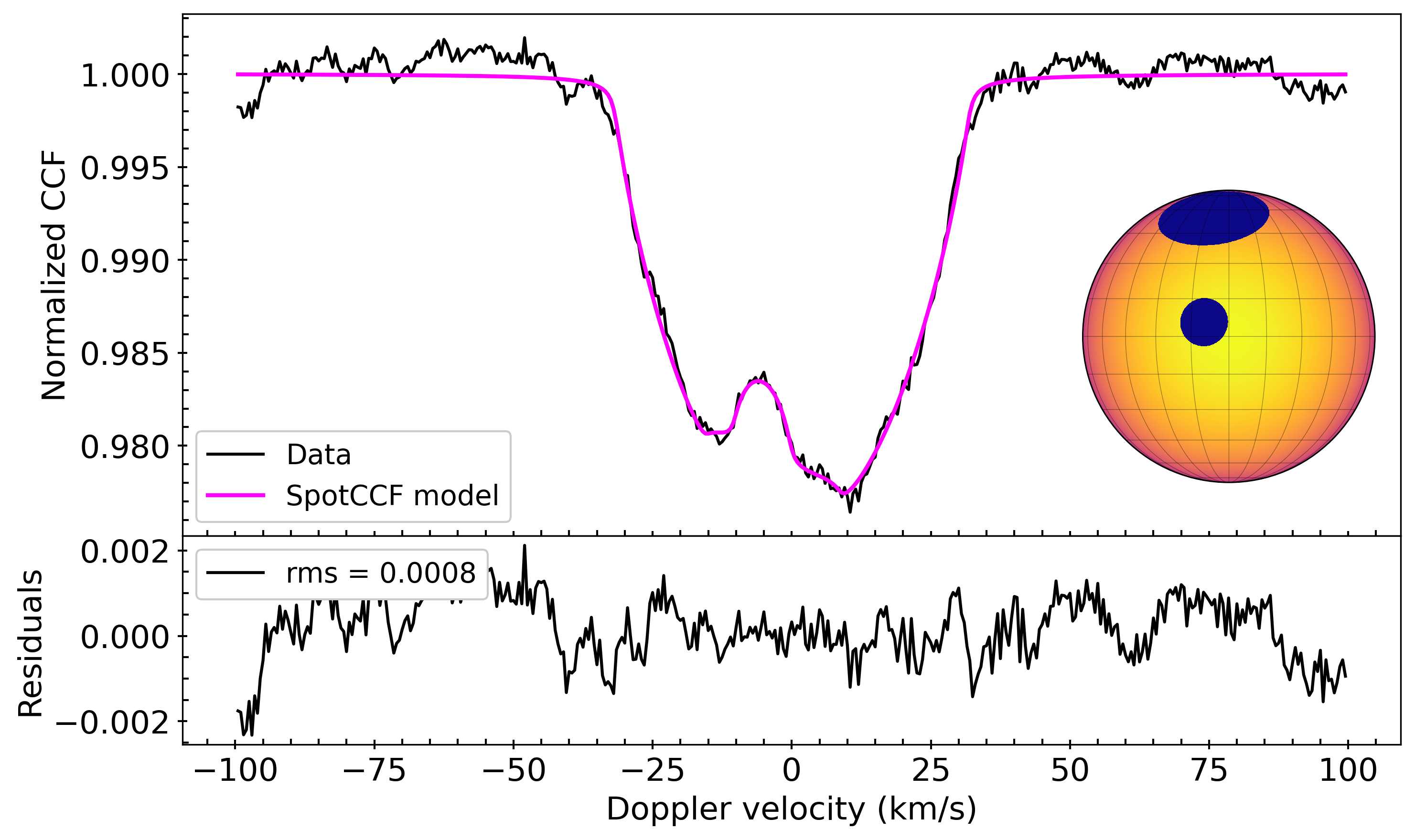} 
       \caption{Examples of LSD profiles of RY Lup fitted by \texttt{SpotCCF} assuming a two-spots model (top panel) and the corresponding residuals (bottom panel). The inset shows the location of the spots on the stellar disc, as defined by the fit; the grid indicates longitudes and latitudes from -90 to 90 degrees with 15-degree intervals.}
 \label{fig:Example_SpotCCF}
 \end{figure}

To assess the potential effect of stellar surface inhomogeneities on the LSD profiles, such as star spots, we employed the \texttt{SpotCCF} tool \citep{2024A&A...683A.239D}. This tool is specifically designed to model the deformation of mean stellar profiles in rapidly rotating stars by convolving a stellar rotation kernel, incorporating a linear limb darkening law, with a Lorentzian profile. 
The model includes several key parameters such as the centroid of the convolved function, representing the systemic radial velocity ($V_{0}$); the projected rotational velocity ($v \sin i$); the scale parameter of the Lorentzian profile ($\gamma$); and the total projected filling factor ($ff_{\mathrm{p, tot}}$), which represents the fractional area of the visible stellar disc covered by starspots, calculated by summing the individual projected contributions of each spot in the model configuration.
For configurations including surface features, each spot is further characterised by its latitude, longitude, and radius.
We adopted an iterative approach, testing models ranging from a spotless (zero-spot) configuration to a three-spot scenario, incrementing the complexity by one spot at each step. The tool utilises the \texttt{MultiNest} nested sampling algorithm via \texttt{PyMultiNest} \citep{Feroz2019OJAp....2E..10F, Buchner2014A&A...564A.125B} to explore the parameter space, using the logarithmic Bayesian evidence ($\ln Z$) to evaluate model performance. Optimal models were selected by comparing the Bayesian evidence of each configuration according to the Kass \& Raftery scale \citep{Kass10.2307/2291091}. To ensure an unbiased exploration of the parameter space, all parameters were allowed to vary within wide, non-informative priors.  A representative example of the fit performed by \texttt{SpotCCF}, assuming a two-spot configuration, is presented in Fig.\ref{fig:Example_SpotCCF}. 

In this analysis, we assume that RY Lup is a single active star. 
Under this assumption, the $v \sin i$ is a fundamental physical property that should remain intrinsically constant over the observed timescales. Consequently, one could justify fixing this parameter to a nominal literature value or imposing a narrow prior to stabilise the inversion process. However, we intentionally treated the $v \sin i$ as a free parameter for each individual observation. This approach serves as a critical diagnostic tool. By avoiding fixed constraints on the rotational broadening, we can evaluate the internal consistency of the single-star model. If the spot-induced deformations are correctly modelled, the retrieved $v \sin i$ should remain stable across all epochs. Conversely, significant temporal variations in this parameter would signal that the model is attempting to compensate for missing physics, potentially pointing toward unresolved binarity or other complex dynamical effects that distort the overall profile morphology. 

 \begin{figure}
     \centering
 \includegraphics[trim=0 0 0 20  , clip,width=\linewidth]{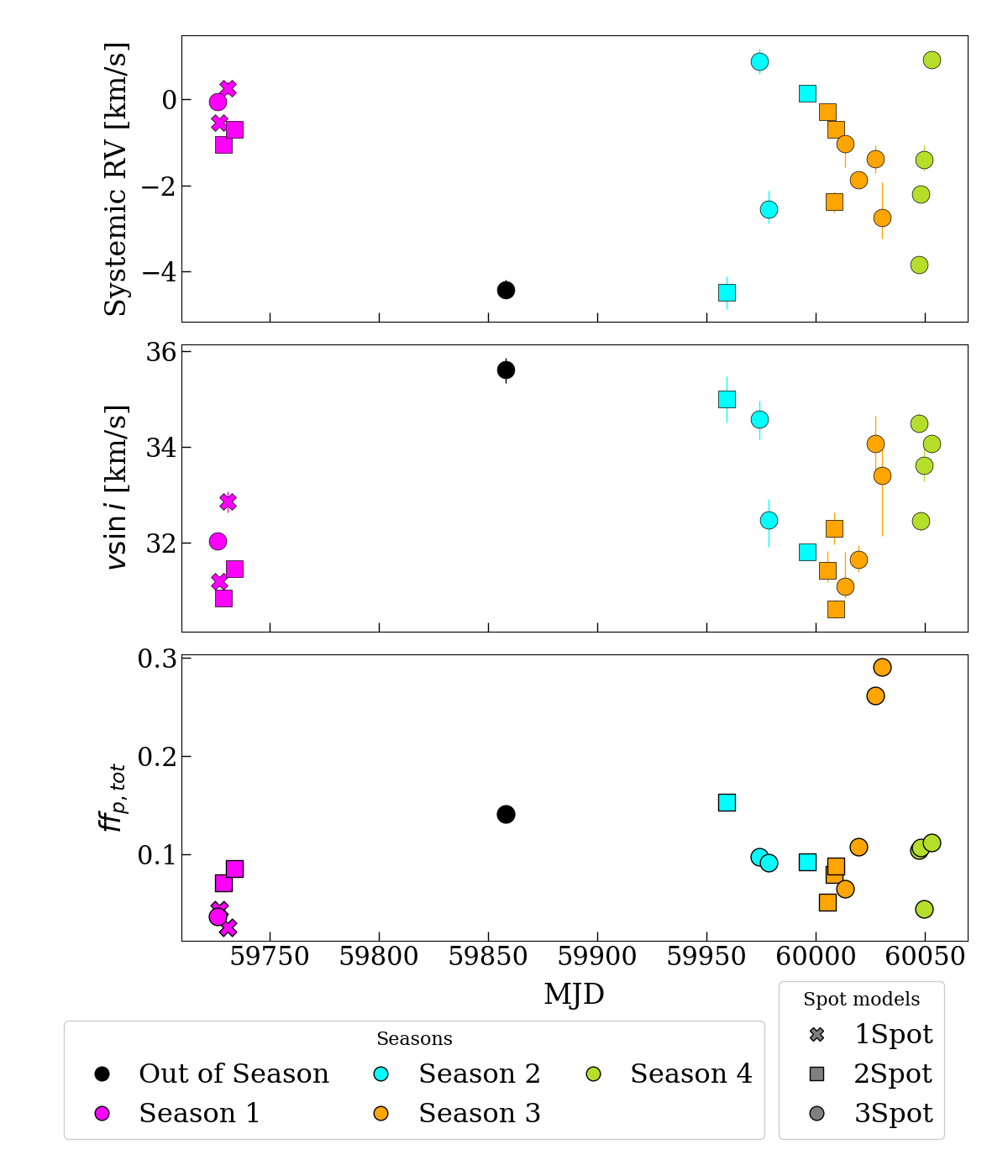}
 \vspace{-10pt}
       \caption{Temporal evolution of the systemic RV (top panel), the retrieved $v \sin i$ (middle panel), and $ff_{\mathrm{p, tot}}$ (bottom panel) derived from the LSD profiles of RY Lup. Marker colours denote the different observing seasons, consistent with Fig. \ref{fig:total_LSD_grid}). The 'Out of Season' data point (black circle) represents the isolated observation taken on 2022 October 6. Marker shapes indicate the optimal model configuration selected by Bayesian evidence: crosses, squares, and circles represent 1-, 2-, and 3-spot configurations, respectively.
       }
 \label{fig:timeseries_SpotCCF}
 \end{figure}

 Our results show that the Bayesian evidence consistently favours 2- or 3-spot configurations in the vast majority of cases (Fig. \ref{fig:Example_SpotCCF_extra}). A single-spot model is preferred only for two observations in Season 1, corresponding to the epochs with less pronounced profile deformations i.e., 2022 May 28 and 31 (Fig.\ref{fig:total_LSD_grid}). Despite this flexibility, as illustrated in Fig. \ref{fig:timeseries_SpotCCF}, the retrieved $v \sin i$ is not constant over time, ranging from a minimum of 30.6\,km s$^{-1}$ to a maximum of 35.6\,km s$^{-1}$, in line with the values derived in Sect.\ref{stellar_properties}. In the figure, the seasonal behaviour observed in RV and $v \sin i$ is further corroborated by $ff_{\mathrm{p, tot}}$, bottom panel of Fig.\ref{fig:timeseries_SpotCCF}). The error bars correspond to the 1-$\sigma$ uncertainties derived from the posterior distributions.

To quantify this variability, we calculated the root mean square (RMS) for each season. In Season 1, the star exhibits its highest stability, with an RV RMS of $\sim 0.47$\,km s$^{-1}$ and a $v \sin i$ RMS of $\sim 0.71$\,km s$^{-1}$. During this period, the retrieved $ff_{\mathrm{p, tot}}$ remains consistently low, indicating a phase of minimal star spot coverage. This relative quiescence is directly reflected on the morphology of the LSD profiles, which appear significantly less deformed and more symmetric compared to subsequent epochs (Fig. \ref{fig:total_LSD_grid}). In contrast, other seasons show a marked increase in dispersion. For instance, in Season 2 and 3, fitted with 2- and 3- spots, the RV RMS rises to $\sim 2.15$\,km s$^{-1}$ and the \vsini\ RMS to $\sim 1.35$\,km s$^{-1}$, corresponding to more distorted LSD profile shapes, and higher values of $ff_\mathrm{p, tot}$ compared to Season 1 (Fig.\ref{fig:total_LSD_grid}). Interestingly, The LSDs of Season 4, marked by less distorted profiles, are best fitted with 3-spots. We performed GLS periodogram on both the RV and $v \sin i$ time series to search for potential periodic signals. However, the periodograms did not reveal any significant peaks above the 10\% FAP threshold, indicating that the observed variability lacks a clear, coherent periodicity within our sampling. In Table.\ref{spot_param}, we report the stellar spots parameters.

To investigate whether the lack of periodic signals was due to phase coherence degradation over the long observational baseline or model complexity variations across epochs, we focused our analysis on the high-density months of 2023 March and April. We performed GLS analyses using two different approaches: First, we analysed the parameters derived from the optimal spot configurations selected via Bayesian evidence (which varies between 1, 2, and 3 spots depending on the epoch); second, as a robustness test to eliminate any potential noise or ambiguity from model-complexity variations, we performed the GLS analysis forcing a 2-spot model configuration across these subsets, simplifying the setup to track only a larger and a smaller spot. In both cases, no statistically significant periodic signals were detected above the 10\% FAP threshold. Crucially, the analysis reveals that the total projected filling factor ($ff_{\mathrm{p, tot}}$) and individual spot geometries experience large, intrinsic fluctuations within the same observing season (e.g., Season 3). This rapid physical evolution implies that a static spot configuration modulated purely by stellar rotation fails to track the subsequent line profiles. Instead, the observed short-term variability appears to be dominated by a complex interplay of rapid spot emergence and decay occurring on timescales shorter than or comparable to the rotation period, which remains challenging to disentangle given the current daily-to-weekly sampling cadence.

To examine the reliability of our multi-spot solutions, we examined the physical consistency of the retrieved spot parameters across consecutive observations. As \texttt{SpotCCF} treats each epoch independently, we can not inherently maintain the identity of a specific surface feature; for instance, the parameters labelled as 'Spot 1' in one observation might correspond to 'Spot 3' in the next. To overcome this stochastic labelling and track the evolution of individual active regions, we implemented a re-labelling procedure based on the spot radius. We assumed that, over short timescales, the physical dimensions of the spots remain more stable than their perceived positions due to rotation. Consequently, we categorised the features into dominant (largest), intermediate, and minor (smallest) spots.

Using this systematic classification, we analysed the temporal evolution of both the longitudinal distribution and the spot radii (Fig.\ref{fig:spot_radii_long}). Based on the hypothesis that the stellar surface is stable or evolving slowly, the values of these parameters should be consistent in observations taken on the same night or in close succession. However, the GLS periodogram analysis performed on the time series of the individual spot radii and longitudes did not reveal any significant periodic signals above the 10\% FAP threshold. This suggests that the stochastic nature of the spot emergence, combined with our sampling frequency, likely masks any coherent rotational modulation that might be present in the individual spot parameters. In Fig.\ref{sys_rv_res}, we show the residuals of the fitted $RV_{1}$ and $RV_{2}$ model to the RV measurements in Sect.\ref{rv_analysis} as a function of the systemic velocity. We do not observe any strong linear correlation.

\begin{figure}
    \centering
\includegraphics[viewport= 0 33 590 510, clip,width=0.57\textwidth]{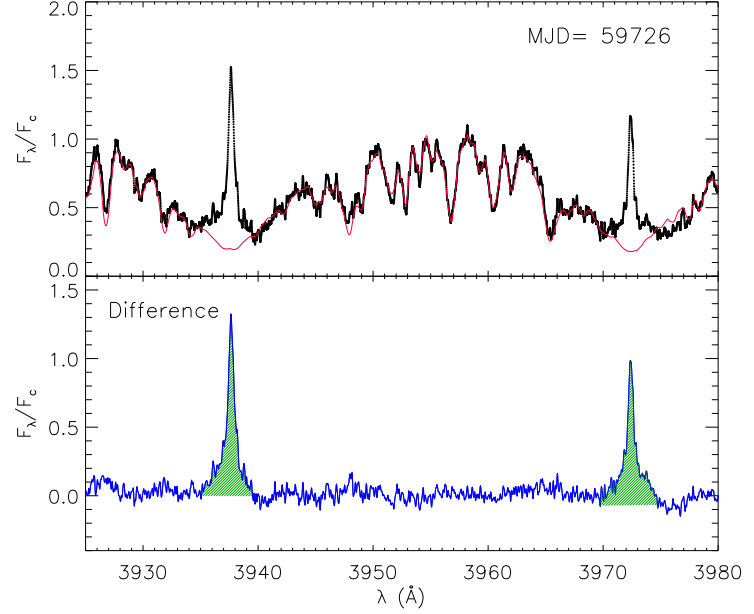}
\vspace{-18pt}
      \caption{Spectral subtraction in the \ion{Ca}{ii} H\&K region. Upper panel: RY Lup spectrum taken on 2022 May 27 (black dots) and the template spectrum (red line). Lower panel: The difference between the observed spectrum and template (blue line). The green shaded regions represent the integrated flux of \ion{Ca}{ii} H\&K lines.} 
\label{fig:emission_cahk}
\end{figure}

As a proxy for stellar magnetic activity and magnetospheric accretion, we analysed the \ion{Ca}{ii} H\,$\&$\,K lines $\lambda\lambda\sim$ 3968.47, 3933.66 {\text{\AA}}. We adopted as diagnostics the $R'_{\rm{HK}}$ index, which represents the ratio of the chromospheric emission luminosity to the total bolometric luminosity of the star, corrected for the photospheric contribution. To account for the photospheric contribution, we subtracted {the best non-active template, which is represented by the HARPS spectrum of HD\,40105 (K1\,V-IV)}
rotationally broadened at the \vsini\ measured by \rotfit\ (Sect.\ref{stellar_properties}), and artificially veiled adopting the value we measured at $\lambda=4000$\,\AA\ ($r_{400}$). 
This allowed us to calculate the fluxes of \ion{Ca}{ii} K and \ion{Ca}{ii} H + H$\epsilon$ by integrating the residual emission obtained subtracting the template from the observed spectrum. The H$\epsilon$  $\lambda$\,=\,3970.08\,\AA\ line is fully blended with the broad \ion{Ca}{ii} H line and its contribution cannot be decoupled from it. In Fig.\ref{fig:emission_cahk}, the shaded green areas represent the integrated flux region $f_{\text{line}}$ used to derive the equivalent widths (EW) of both lines:

\begin{equation}
EW_i = \int_{line} \left(\frac{F_{\lambda,i}^T}{F_c} - \frac{F_{\lambda,i}^O}{F_c}\right ) d\lambda ~,
\end{equation}

where $F_{\lambda,i}^T/F_c$ that of the photospheric template and $F_{\lambda,i}^O/F_c$ is the continuum-normalised flux of the observed spectrum.

We evaluated the surface flux in the \ion{Ca}{ii} H\&K lines and the luminosity ratio, $R'_{\text{HK}}$, as:
\begin{equation}
F_{\rm CaII-K}  =  F_{3933}EW_{\rm CaII-K}\,(1+r_{400})\, , 
\end{equation}
\begin{equation}
F_{\rm CaII-H}  =  F_{3968}EW_{\rm CaII-H}\,(1+r_{400}) \,, 
\end{equation}
\begin{equation}
R'_{HK}  =  (L_{\rm CaII-K}+L_{\rm CaII-H})/L_{\rm bol} = (F_{\rm CaII-K}+F_{\rm CaII-H})/(\sigma T_{\rm eff}^4)\, ,
\end{equation}

\noindent where the factor $(1+r_{400})$ corrects the EWs for the veiling, while $F_{3933}$ and $F_{3968}$ are the continuum fluxes per unit stellar surface area at the centre of the \ion{Ca}{ii} K and H lines, respectively. The latter were evaluated from BT-Settl spectra \citep{2012RSPTA.370.2765A} at the specific \teff\ and \logg\ of the target. The uncertainties in the resulting surface fluxes were estimated by propagating the error in $EW_{\rm CaII-K}$ (or $EW_{\rm CaII-H}$)  and the uncertainty in the continuum flux ($F_{3933}$ or $F_{3968}$); the latter was determined by considering the errors of \teff\  and \logg. Finally, $\sigma \approx 5.67 \times 10^{-8}\,\text{W\,m}^{-2}\,\text{K}^{-4}$ is the Stefan-Boltzmann constant.

\begin{figure}
    \centering
\includegraphics[trim=0 0 0 0 , clip,width=\linewidth]{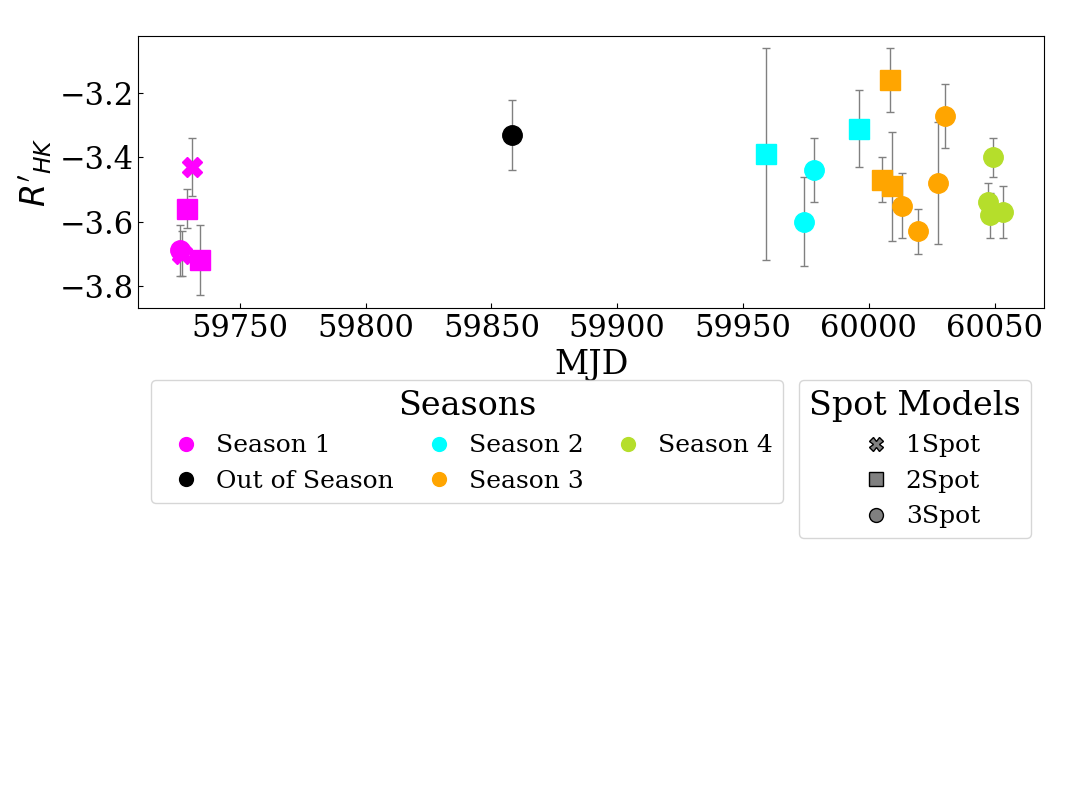}
\vspace{-80pt}
      \caption{$R'_{\mathrm{HK}}$ index values across the observing seasons. Marker shapes denote the the spot models fitted to the LSD profiles. Crosses, squares, and circles represent 1-, 2-, and 3-spot configurations, respectively. The colours denote the different observing seasons.}
\label{log_RHK}
\end{figure}

In Fig.\ref{log_RHK}, we show the measured values of log $R'_{\text{HK}}$ across RY Lup observing seasons. The highest measured values correspond to Season 3, followed by Season 2 with mean value $\sim -3.44 \pm  0.15$ and  $-3.44 \pm  0.1$, respectively. This corresponds to the complex LSD profiles and the higher spots coverage of these seasons. Season 1, followed by Season 4 mark the lowest values of log $R'_{\rm{HK}}$ (Fig.\ref{log_RHK}) with mean value $-3.6\pm 0.1$, and $-3.5\pm 0.07$ , linked to the minimal starspot coverage, and less deformed LSD profiles (Fig.\ref{fig:timeseries_SpotCCF}).

While the majority of main sequence stars show values in the range $-$5.2 < log $R'_{\rm{HK}}$ < $-$4.4 \citep{2018A&A...616A.108B}, younger stars typically have log $R'_{\rm{HK}}$ $\sim -4.2$ \citep{2019ApJ...887...84Z}. Our measurements for RY Lup, ranging from $-$3.8 < log $R'_{\rm{HK}}$ < $-$3.2 are consistent with high chromospheric activity levels of young stars. The GLS periodogram analysis shows no significant period above the $1\%$ FAP. We only detect a period at $P \sim 1.16$ days at $\sim 10\%$ FAP level. In Table.\ref{log_rhk}, we report 
log $R'_{\text{HK}}$ values.

\section{Discussion}
\label{discussion}

The analysis described in the previous section showed that the data can be described by either the presence of spots on the surface of a single star, or by the presence of a binary system. Here we discuss the implications of each of these scenarios, starting from the binary option.

\subsection{Binary orbital solution}

\begin{figure}
    \centering
    \includegraphics[width=0.5\textwidth]{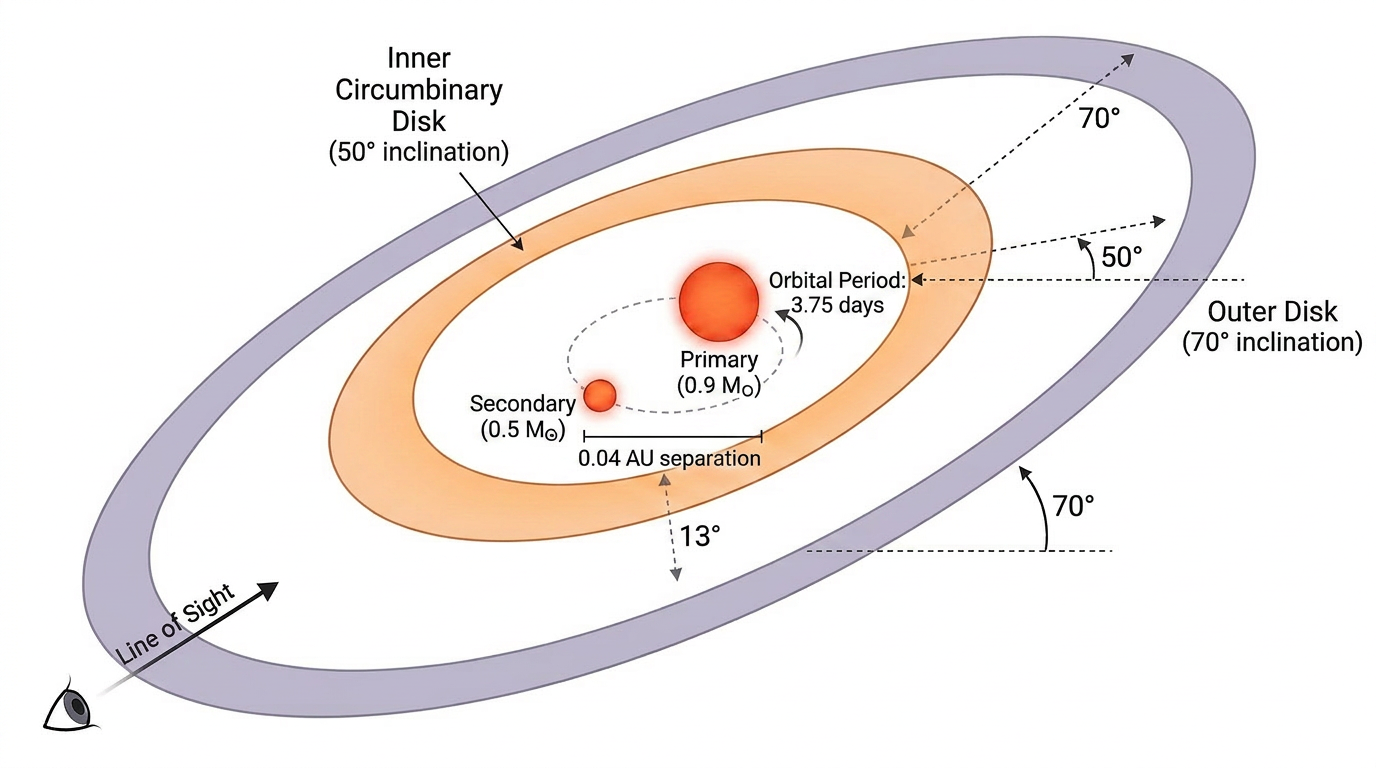}
    \caption{An illustration of the derived binary configuration, showing the potential binary surrounded by a misaligned inner and outer discs.}
    \label{fig:binary_config}
\end{figure}

The derived orbital solution of RY Lup points to a potentially eccentric configuration ($e\sim 0.23$). Given the derived mass ratio from the RV solution $q=0.65$ and the total ALMA dynamical mass $\sim1.3-1.5\,M_{\odot}$ \citep{2018A&A...616A.100Y, 2021ApJ...908...46B}, the mass of the primary  is estimated to be $0.79-0.91\, M_{\odot}$. While the secondary star is $0.51-0.59\, M_{\odot}$. Such solution yields a binary inclination of $i=13.4 ^{\circ}-12.7 ^{\circ}$, revealing a major misalignment to both the inner and outer discs of RY Lup as shown using VLT/SPHERE and ALMA observations \citep[]{2018A&A...614A..88L, 2020ApJ...892..111F} (Fig.\ref{fig:binary_config}). The orbital plane of the binary appears nearly face-on, while the disc is more inclined. At this face-on inclination, the resulting orbital separation is $a=0.045\, \text{au} \approx 9.67\,R_{\odot}$. Using the luminosity ($L_{\star}=1.65\,L_{\odot}$) from \citet{2017A&A...600A..20A}, and  ($L_{\star}=2.5\,L_{\odot}$) from \citet{2020A&A...642A.162G}, we derive $R_{\star}=1.60$--$1.97\,R_{\odot}$, this separation scales to roughly $4.91$--$6.04\,R_{\star}$.

This close configuration places the system well within the 0.12 au inner dust rim \citep{2020A&A...642A.162G}. While the derived configuration of RY Lup can not explain ALMA millimetre dust grain cavity of $\sim 60$\,au, the cavity size derived from GRAVITY, which traces smaller dust grains and gas, is possibly in line with a compact companion. Dynamically, a binary can clear a cavity measuring roughly 2-4 its own semi-major axis \citep{1994ApJ...421..651A}. However, recent 3D hydrodynamical simulations by \citet{2020MNRAS.499.3362R} demonstrate that this clearing can be substantially enhanced by the orbital configuration of the system. Massive companions and higher eccentricities, or orbital misalignments excite eccentricity within the circumbinary disc, expanding the dust cavity boundary even further. This provides a route to eccentric binaries to mimic the large cavities attributed to much wider systems. In the case of RY Lup, a compact binary alone cannot explain the 60 au dust cavity , potentially requiring additional companions or alternative mechanisms within the disc \citep{2025A&A...698A.102R}. 

The derived binary solution introduces geometric and physical tension when compared to the derived stellar parameters in Sect.\ref{stellar_properties}. Specifically, the low orbital inclination of the binary $i\sim13.4 ^{\circ}-12.7 ^{\circ}$ stands in contrast to the derived \vsini $\sim 32 $\,\unit{km.s^{-1}}. Given a stellar radius $R_{\star}=1.60-1.97\,R_{\odot}$ and assuming spin-orbit synchronisation ($P_{\mathrm{rot}}=P_{\mathrm{orb}}=3.75$\,days), the equatorial velocity would be $V_{\mathrm{eq}}\approx 21- 26.6$\,\unit{km.s^{-1}}. This synchronised rotation would yield a \vsini $\approx4.8 - 5.9$ \,\unit{km.s^{-1}}. To match the observed \vsini $\sim 32$\,\unit{km.s^{-1}} at such a face-on orientation, the system would require an unphysical $V_{\mathrm{eq}} \sim 140$\,\unit{km.s^{-1}}, strongly disfavouring a low-inclination binary scenario.

Even if we abandon the single-star framework and assume that the observed \vsini $\sim 32 $\,\unit{km.s^{-1}} is a blending artifact of two distinct stellar components, a nearly-face-on binary remains incompatible. The individual $v \sin i$ of each component cannot be significantly lower than $20\rm\, km\ s^{-1}$. Otherwise, the LSD profiles would exhibit clearly resolved peaks at quadrature, alongside pronounced periodic variations in the line width-neither of which is observed in our data. Adopting the proposed orbital inclination of $12.7^\circ$-$13.4^\circ$ for the components, a minimum individual $v\sin i$ of $20\rm\, km\ s^{-1}$ would translate to $V_{\rm eq}$ of roughly $90\rm\, km\ s^{-1}$. Such a high value could only be reconciled by either assuming stellar radii of about $7\, R_\odot$ or, alternatively, maintaining a radius of $1.60-1.97\, R_\odot$ while invoking a rotation period three times faster than the orbital one, which implies a highly asynchronous system.

Moreover, the derived binary orbit is eccentric with $e\sim 0.23$, which is anomalous for PMS systems. Theoretical models of tidal dissipation in fully convective PMS stars indicate that orbits with $P_{\mathrm{orb}}< 7.5-8$\,days should circularise on timescales much shorter than a few Myr \citep{1989A&A...223..112Z}.
In particular, by adopting the component's masses and binary separation derived under the binary hypothesis, the timescales for circularisation and synchronisation calculated according to \citet{1989A&A...223..112Z} are $\tau_{\rm circ}\sim 1-5$\,Myr and $\tau_{\rm sync}\sim 0.05-0.10$\,Myr, respectively. Therefore, the system is theoretically expected to be fully synchronised and circularised, or at the very least, to display a negligible eccentricity. \\
In our analysis, we show that the RV variations can be modelled by a close-in companion of a separation $a\sim 0.04$\,au. Our derived period of $P=3.75$\,days is inline with what was reported previously in the literature using photometry \citep{1989A&A...211..115G, 2009A&A...499..137M}. This constant periodicity that is observed across various data sets supports the binary nature of RY Lup, rather than a timely variable effect as stellar spots. The inner disc geometry of RY Lup also supports the existence of such a companion with the inner rim at $a=0.12$\,au. However, the inherent complexity of such young stars, which is endorsed by stellar magnetic activity, accretion, and the high $v \sin i$, makes a two-component Gaussian model insufficient for capturing the line profile morphologies at certain epochs (Fig.\ref{fig:total_LSD_grid}). This has led to the exclusion of such profiles from the orbital fitting analysis. This suggests that despite the presence of two RV signatures in the ESPRESSO spectra of RY Lup, the current fitting approach can not fully explain the present variations in the data, which hints at other effects such as variable configurations of stellar spots, that cause additional variations.   

 \subsection{Insights from the multi-spot modelling of RY Lup }
 \label{spot configuration}

The analysis of the LSD profiles through the \texttt{SpotCCF} tool provides a detailed characterisation of the line-profile variations in RY Lup. While the iterative modelling of surface spots achieves a formal improvement in fit quality compared to the dual-Gaussian fit (Fig.\ref{fig:LSDs_SB2}), the resulting physical parameters, specifically $v \sin i$ and the RV, exhibit behaviours that challenge the standard single-star activity scenario.

The simultaneous minimisation of dispersion in RV, $v \sin i$, and filling factor during Season 1 confirms that the \texttt{SpotCCF} tool effectively tracks the baseline level of surface activity during periods of relative quiescence. However, the persistence of a $\sim$15\% dispersion in the $v \sin i$, despite the flexibility of the multi-spot model and the absence of a clear periodic signal, suggests that the observed profile deformations are not exclusively localised. Instead, they appear to affect the global morphology of the LSD profiles. This could be interpreted as the result of extreme surface activity involving very large-scale inhomogeneities that the discrete spot model can only partially approximate. Alternatively, such a systematic change in the apparent broadening could indicate unresolved binarity, where the tool compensates for the dynamic blending of two stellar profiles by adjusting the $v \sin i$ value.

As shown in Fig. \ref{fig:timeseries_SpotCCF}, the $ff_{\mathrm{p, tot}}$ retrieved by \texttt{SpotCCF} ranges from approximately 2\% to 30\%. While the lower values correspond to epochs with less pronounced profile asymmetries (e.g., Season 1), the peak values observed during Season 3 ($\sim$26-29\%) are remarkably consistent with the typical spot coverages reported for T Tauri stars (15--45\%; \citealt{2008ApJ...678..472H, 2009A&A...508.1313F,2017ApJ...836..200G, 2022A&A...667A.124G, 2025ApJ...992L..33P}). However, a key limitation to consider is that \texttt{SpotCCF} operates under the assumption of completely 'black' spots ($T = 0\,\text{ K}$), providing the maximum possible flux contrast. In reality, star spots possess finite temperatures, typically $\sim$1000\,K cooler than the photosphere. To reproduce the observed deep profile distortions with a more realistic temperature contrast, a significantly larger spotted area would be required, potentially pushing the filling factor toward even more extreme values and further challenging the physical plausibility of the single-star scenario. 

As shown in Fig. \ref{fig:Example_SpotCCF}, the \texttt{SpotCCF} model provides an excellent fit to the observed LSD profile, effectively minimising the residuals and accurately reproducing the morphology of the 'bump'. In comparison, a simpler two-component Gaussian model (Fig. \ref{fig:LSDs_SB2}), used here as a proxy for a binary configuration, yields a formally less precise fit with larger residuals. However, this superior statistical performance of \texttt{SpotCCF} does not necessarily imply that stellar activity is the correct physical interpretation.

Since the width of the observed 'bump' strictly constrains the longitudinal extent of the spot, any increase in area must be achieved by extending the spot distribution in latitude. While stellar activity often manifests within preferred latitude bands, frequently appearing symmetrically in both hemispheres \citep[e.g., ][]{Strassmeier2009A&ARv..17..251S, Hathaway2015LRSP...12....4H}, as observed in the Sun, reproducing these profiles would necessitate an extensive and highly concentrated distribution of spots along specific longitudes. While such a complex surface morphology cannot be entirely ruled out for an active, accreting star like RY Lup, this requirement for massive, 'ribbon-like' spot structures makes the single-star activity model increasingly complex. Conversely, while the Gaussian decomposition in Fig. \ref{fig:LSDs_SB2} is a first-order approximation, the presence of an unresolved stellar companion provides a simpler explanation, producing the asymmetric peaks with variable intensity in the LSD profile. 

Finally, it is important to address the lack of significant periodic signals in our analysis. The temporal cadence and the total number of observations in our dataset are likely insufficient to perform a robust periodogram analysis of spot migration or orbital motion. The sparse sampling, characterised by long gaps between seasons and irregular intervals within them, significantly limits our sensitivity to coherent periodicities. Consequently, the absence of a clear signal in the spot properties time series should not be interpreted as evidence against binarity or rotational modulation, but rather as a limitation imposed by the current observational window. In light of these considerations, our analysis does not allow us to definitively confirm or exclude either the single-star activity model or the binary system hypothesis.

Ultimately, it is highly probable that these two scenarios are not mutually exclusive. Even if RY Lup is indeed an unresolved spectroscopic binary, it could be intrinsically characterised by high levels of surface magnetic activity as shown by the $R'_{\mathrm{HK}}$ estimates (Fig.\ref{log_RHK}).  In such a scenario, the orbital motion and the spot-induced deformations would be deeply intertwined, making it impossible to treat the dynamical and atmospheric effects as decoupled. Our results suggest that the 'apparent' variations in $v \sin i$ and RV are likely a convolution of both phenomena. To break the current degeneracy between temperature-induced spot signals and purely geometric Doppler shifts from a companion, a more extensive observational strategy is required. Only a significantly higher-cadence monitoring, combined with multi-wavelength observations to leverage the wavelength-dependence of spot contrasts, would provide the necessary constraints to disentangle the orbital signature from the stellar activity.

\section{Conclusions}
In this work, we study the RV variations of RY Lup, emission lines, activity indices, and stellar spots configurations on the stellar surface of RY Lup, using the high-resolution spectra of the VLT/ESPRESSO over $\sim 327$ days. We tested both the binary and stellar spots hypotheses. We discuss both approaches and we conclude the following:

\begin{itemize}

\item The observed RV variations can be modelled by a close-in companion at a separation of $\sim0.04$\,\text{au} with an orbital period of $3.75$\,days. While this close-in orbit is spatially compatible with the inner disc geometry-as its semi-major axis places it well within the $a \sim 0.12$\,\text{au} inner disc rim derived from VLTI/GRAVITY-its nearly face-on orientation ($i \sim 13^\circ$) and the eccentricity of $ e \sim 0.23$ introduces severe geometric tension with the large observed $v \sin i \approx 32\,\text{km s}^{-1}$ and the strong tidal dissipation expected for such a short-period, close-in orbit. 
    
     \item Although the modelling of the LSD profile with the spot models by \texttt{SpotCCF} provides a superior statistical fit to the LSD profile asymmetries, the resulting extreme and potentially unphysical spot geometries-alongside the high $15\%$ dispersion in \vsini\- suggest that it is compensating for a more complex underlying reality. Instead, this constant change across the observing seasons supports the binary hypothesis, as the orbital motion of an unresolved companion would introduce this line variation.

       \item  The total spot filling factor peaks during subsequent seasons as in 2023 January, February, and March with $ff_\mathrm{p,tot}$ reaching $\sim 26-29\%$, consistent with typical coverages of T Tauri stars. 

    \item We do not detect a periodic signal from the stellar spots parameters such as the longitudinal distribution and the spot radii due to the stochastic nature of the spot emergence and the sampling frequency.

 \item Modelling the LSD profiles across different epochs reveals that the initial observing season in  2022 May represents the period of lowest surface magnetic activity, with stellar activity intensifying in consecutive seasons. 

         \item Elevated magnetic stellar activity is confirmed by $R'_{\mathrm{HK}}$ index estimates, which peak in early 2023, and coincide with complex LSD morphologies and higher number of fitted spots.

     \item The binary and magnetic activity scenarios are likely not mutually exclusive. If RY Lup is an unresolved spectroscopic binary, its orbital motion would be deeply intertwined with the high surface magnetic activity indicated by the $R'_{\mathrm{HK}}$ estimates, requiring a comprehensive future follow-up to jointly model both dynamical and atmospheric effects.

\end{itemize}

This multi-approach analysis carried on the active and fast-rotating star RY Lup shows the complexity of such systems and the difficulty of tracing and understanding the RV variations in the presence of accretion and stellar spots. With this analysis, we demonstrate the necessity to observe with long and high-cadence observing runs such systems to better characterise the physics and the overall evolution, including the surrounding disc. Future RV campaigns should aim at observing multiple periods of such stars in a well-sampled regime with a wide wavelength coverage. This would allow to better constrain the stellar spots effects and clarify the nature of such systems. 

\section{Data availability}

Additional data for this article are available at \url{https://doi.org/10.5281/zenodo.20327276}

\begin{acknowledgements}
We thank the anonymous referee for the constructive reports which improved the quality of the paper. Based on observations collected at the European Organization for Astronomical Research in the Southern Hemisphere under ESO programs 106.20Z8.007, 110.2483.001, and 111.250K.001.
Funded by the European Union (ERC, WANDA, 101039452). Views and opinions expressed are however those of the author(s) only and do not necessarily reflect those of the European Union or the European Research Council Executive Agency. Neither the European Union nor the granting authority can be held responsible for them. AF acknowledges funding from the Large Grant INAF-2024 ``Spectral Key features of Young stellar objects: Wind-Accretion LinKs Explored in the infraRed (SKYWALKER)''. ER acknowledges financial support from the European Union’s Horizon Europe research and innovation programme
under the Marie Skłodowska-Curie grant agreement No. 101102964 (ORBITD),
including a secondment carried out at the ESO Headquarters in Garching,
Germany. ER also acknowledges support from the European Union (ERC Starting Grant
DiscEvol, project number 101039651). Figure \ref{fig:binary_config} was generated via the \tt{FigureLabs} platform \url{https://www.figurelabs.ai}.
\end{acknowledgements}

\bibliographystyle{aa}
\bibliography{bibliography_RYLup.bib}

@ARTICLE{2021ApJ...908...46B,
       author = {{Braun}, Teresa A.~M. and {Yen}, Hsi-Wei and {Koch}, Patrick M. and {Manara}, Carlo F. and {Miotello}, Anna and {Testi}, Leonardo},
        title = "{Dynamical Stellar Masses of Pre-main-sequence Stars in Lupus and Taurus Obtained with ALMA Surveys in Comparison with Stellar Evolutionary Models}",
      journal = {\apj},
         year = 2021,
        month = feb,
       volume = {908},
       number = {1},
          eid = {46},
        pages = {46},
          doi = {10.3847/1538-4357/abd24f},
archivePrefix = {arXiv},
       eprint = {2012.07441},
 primaryClass = {astro-ph.SR},
       adsurl = {https://ui.adsabs.harvard.edu/abs/2021ApJ...908...46B}
}

@ARTICLE{Freudling13,
   author = {{Freudling}, W. and {Romaniello}, M. and {Bramich}, D.~M. and
{Ballester}, P. and {Forchi}, V. and {Garc{\'{\i}}a-Dabl{\'o}}, C.~E. and
{Moehler}, S. and {Neeser}, M.~J.},
    title = "{Automated data reduction workflows for astronomy. The ESO Reflex environment}",
  journal = {\aap},
archivePrefix = "arXiv",
   eprint = {1311.5411},
 primaryClass = "astro-ph.IM",
     year = 2013,
    month = nov,
   volume = 559,
      eid = {A96},
    pages = {A96},
      doi = {10.1051/0004-6361/201322494},
   adsurl = {http://adsabs.harvard.edu/abs/2013A%26A...559A..96F}
}

@ARTICLE{2023MNRAS.523.3318P,
       author = {{Picogna}, Giovanni and {Sch{\"a}fer}, Carolina and {Ercolano}, Barbara and {Rab}, Christian and {Franz}, Rafael and {G{\'a}rate}, Mat{\'\i}as},
        title = "{Observability of photoevaporation signatures in the dust continuum emission of transition discs}",
      journal = {\mnras},
         year = 2023,
        month = aug,
       volume = {523},
       number = {3},
        pages = {3318-3327},
          doi = {10.1093/mnras/stad1504},
archivePrefix = {arXiv},
       eprint = {2305.06014},
 primaryClass = {astro-ph.EP},
       adsurl = {https://ui.adsabs.harvard.edu/abs/2023MNRAS.523.3318P}
}

@ARTICLE{2015A&A...575A...4F,
       author = {{Frasca}, A. and {Biazzo}, K. and {Lanzafame}, A.~C. and {Alcal{\'a}}, J.~M. and {Brugaletta}, E. and {Klutsch}, A. and {Stelzer}, B. and {Sacco}, G.~G. and {Spina}, L. and {Jeffries}, R.~D. and {Montes}, D. and {Alfaro}, E.~J. and {Barentsen}, G. and {Bonito}, R. and {Gameiro}, J.~F. and {L{\'o}pez-Santiago}, J. and {Pace}, G. and {Pasquini}, L. and {Prisinzano}, L. and {Sousa}, S.~G. and {Gilmore}, G. and {Randich}, S. and {Micela}, G. and {Bragaglia}, A. and {Flaccomio}, E. and {Bayo}, A. and {Costado}, M.~T. and {Franciosini}, E. and {Hill}, V. and {Hourihane}, A. and {Jofr{\'e}}, P. and {Lardo}, C. and {Maiorca}, E. and {Masseron}, T. and {Morbidelli}, L. and {Worley}, C.~C.},
        title = "{The Gaia-ESO Survey: Chromospheric emission, accretion properties, and rotation in {\ensuremath{\gamma}} Velorum and Chamaeleon I{\ensuremath{\star}}{\ensuremath{\star}}{\ensuremath{\star}}}",
      journal = {\aap},
         year = 2015,
        month = mar,
       volume = {575},
          eid = {A4},
        pages = {A4},
          doi = {10.1051/0004-6361/201424409},
archivePrefix = {arXiv},
       eprint = {1412.4153},
 primaryClass = {astro-ph.SR},
       adsurl = {https://ui.adsabs.harvard.edu/abs/2015A&A...575A...4F}
}

@ARTICLE{2021A&A...650A.196M,
       author = {{Manara}, C.~F. and {Frasca}, A. and {Venuti}, L. and {Siwak}, M. and {Herczeg}, G.~J. and {Calvet}, N. and {Hernandez}, J. and {Tychoniec}, {\L}. and {Gangi}, M. and {Alcal{\'a}}, J.~M. and {Boffin}, H.~M.~J. and {Nisini}, B. and {Robberto}, M. and {Briceno}, C. and {Campbell-White}, J. and {Sicilia-Aguilar}, A. and {McGinnis}, P. and {Fedele}, D. and {K{\'o}sp{\'a}l}, {\'A}. and {{\'A}brah{\'a}m}, P. and {Alonso-Santiago}, J. and {Antoniucci}, S. and {Arulanantham}, N. and {Bacciotti}, F. and {Banzatti}, A. and {Beccari}, G. and {Benisty}, M. and {Biazzo}, K. and {Bouvier}, J. and {Cabrit}, S. and {Caratti o Garatti}, A. and {Coffey}, D. and {Covino}, E. and {Dougados}, C. and {Eisl{\"o}ffel}, J. and {Ercolano}, B. and {Espaillat}, C.~C. and {Erkal}, J. and {Facchini}, S. and {Fang}, M. and {Fiorellino}, E. and {Fischer}, W.~J. and {France}, K. and {Gameiro}, J.~F. and {Garcia Lopez}, R. and {Giannini}, T. and {Ginski}, C. and {Grankin}, K. and {G{\"u}nther}, H.~M. and {Hartmann}, L. and {Hillenbrand}, L.~A. and {Hussain}, G.~A.~J. and {James}, M.~M. and {Koutoulaki}, M. and {Lodato}, G. and {Mauc{\'o}}, K. and {Mendigut{\'\i}a}, I. and {Mentel}, R. and {Miotello}, A. and {Oudmaijer}, R.~D. and {Rigliaco}, E. and {Rosotti}, G.~P. and {Sanchis}, E. and {Schneider}, P.~C. and {Spina}, L. and {Stelzer}, B. and {Testi}, L. and {Thanathibodee}, T. and {Vink}, J.~S. and {Walter}, F.~M. and {Williams}, J.~P. and {Zsidi}, G.},
        title = "{PENELLOPE: The ESO data legacy program to complement the Hubble UV Legacy Library of Young Stars (ULLYSES). I. Survey presentation and accretion properties of Orion OB1 and {\ensuremath{\sigma}}-Orionis}",
      journal = {\aap},
         year = 2021,
        month = jun,
       volume = {650},
          eid = {A196},
        pages = {A196},
          doi = {10.1051/0004-6361/202140639},
archivePrefix = {arXiv},
       eprint = {2103.12446},
 primaryClass = {astro-ph.SR},
       adsurl = {https://ui.adsabs.harvard.edu/abs/2021A&A...650A.196M}
}

@ARTICLE{2026arXiv260207731B,
       author = {{Blakely}, Dori and {Thompson}, William and {Johnstone}, Doug and {Speedie}, Jessica and {Xuan}, Jerry W. and {Blouin}, Simon and {Zhang}, Jingwen and {Ruffio}, Jean-Baptiste and {Nielsen}, Eric and {Bowler}, Brendan P. and {Franson}, Kyle and {Roberson}, William and {Cloutier}, Ryan and {Fogal}, Andre and {Hessel}, Kaitlyn and {Marois}, Christian and {Rochon}, Alexandra},
        title = "{Dynamical Mass Constraints on Transition Disk Perturbers with the G23H Catalog}",
      journal = {arXiv e-prints},
         year = 2026,
        month = feb,
          eid = {arXiv:2602.07731},
        pages = {arXiv:2602.07731},
          doi = {10.48550/arXiv.2602.07731},
archivePrefix = {arXiv},
       eprint = {2602.07731},
 primaryClass = {astro-ph.EP},
       adsurl = {https://ui.adsabs.harvard.edu/abs/2026arXiv260207731B}
}

@ARTICLE{2009A&A...499..137M,
       author = {{Manset}, N. and {Bastien}, P. and {M{\'e}nard}, F. and {Bertout}, C. and {Le van Suu}, A. and {Boivin}, L.},
        title = "{Photometric and polarimetric clues to the circumstellar environment of RY Lupi}",
      journal = {\aap},
         year = 2009,
        month = may,
       volume = {499},
       number = {1},
        pages = {137-148},
          doi = {10.1051/0004-6361/200810945},
       adsurl = {https://ui.adsabs.harvard.edu/abs/2009A&A...499..137M}
}

@ARTICLE{1989A&A...211..115G,
       author = {{Gahm}, G.~F. and {Fischerstrom}, C. and {Liseau}, R. and {Lindroos}, K.~P.},
        title = "{Long- and short-term variability of the T Tauri star RY Lupi.}",
      journal = {\aap},
         year = 1989,
        month = feb,
       volume = {211},
        pages = {115-130},
       adsurl = {https://ui.adsabs.harvard.edu/abs/1989A&A...211..115G}
}

@ARTICLE{2017A&A...600A..20A,
       author = {{Alcal{\'a}}, J.~M. and {Manara}, C.~F. and {Natta}, A. and {Frasca}, A. and {Testi}, L. and {Nisini}, B. and {Stelzer}, B. and {Williams}, J.~P. and {Antoniucci}, S. and {Biazzo}, K. and {Covino}, E. and {Esposito}, M. and {Getman}, F. and {Rigliaco}, E.},
        title = "{X-shooter spectroscopy of young stellar objects in Lupus. Accretion properties of class II and transitional objects}",
      journal = {\aap},
         year = 2017,
        month = apr,
       volume = {600},
          eid = {A20},
        pages = {A20},
          doi = {10.1051/0004-6361/201629929},
archivePrefix = {arXiv},
       eprint = {1612.07054},
 primaryClass = {astro-ph.SR},
       adsurl = {https://ui.adsabs.harvard.edu/abs/2017A&A...600A..20A}
}

@ARTICLE{2018A&A...616A...1G,
       author = {{Gaia Collaboration} and {Brown}, A.~G.~A. and {Vallenari}, A. and {Prusti}, T. and {de Bruijne}, J.~H.~J. and {Babusiaux}, C. and {Bailer-Jones}, C.~A.~L. and {Biermann}, M. and {Evans}, D.~W. and {Eyer}, L. and {Jansen}, F. and {Jordi}, C. and {Klioner}, S.~A. and {Lammers}, U. and {Lindegren}, L. and {Luri}, X. and {Mignard}, F. and {Panem}, C. and {Pourbaix}, D. and {Randich}, S. and {Sartoretti}, P. and {Siddiqui}, H.~I. and {Soubiran}, C. and {van Leeuwen}, F. and {Walton}, N.~A. and {Arenou}, F. and {Bastian}, U. and {Cropper}, M. and {Drimmel}, R. and {Katz}, D. and {Lattanzi}, M.~G. and {Bakker}, J. and {Cacciari}, C. and {Casta{\~n}eda}, J. and {Chaoul}, L. and {Cheek}, N. and {De Angeli}, F. and {Fabricius}, C. and {Guerra}, R. and {Holl}, B. and {Masana}, E. and {Messineo}, R. and {Mowlavi}, N. and {Nienartowicz}, K. and {Panuzzo}, P. and {Portell}, J. and {Riello}, M. and {Seabroke}, G.~M. and {Tanga}, P. and {Th{\'e}venin}, F. and {Gracia-Abril}, G. and {Comoretto}, G. and {Garcia-Reinaldos}, M. and {Teyssier}, D. and {Altmann}, M. and {Andrae}, R. and {Audard}, M. and {Bellas-Velidis}, I. and {Benson}, K. and {Berthier}, J. and {Blomme}, R. and {Burgess}, P. and {Busso}, G. and {Carry}, B. and {Cellino}, A. and {Clementini}, G. and {Clotet}, M. and {Creevey}, O. and {Davidson}, M. and {De Ridder}, J. and {Delchambre}, L. and {Dell'Oro}, A. and {Ducourant}, C. and {Fern{\'a}ndez-Hern{\'a}ndez}, J. and {Fouesneau}, M. and {Fr{\'e}mat}, Y. and {Galluccio}, L. and {Garc{\'\i}a-Torres}, M. and {Gonz{\'a}lez-N{\'u}{\~n}ez}, J. and {Gonz{\'a}lez-Vidal}, J.~J. and {Gosset}, E. and {Guy}, L.~P. and {Halbwachs}, J.-L. and {Hambly}, N.~C. and {Harrison}, D.~L. and {Hern{\'a}ndez}, J. and {Hestroffer}, D. and {Hodgkin}, S.~T. and {Hutton}, A. and {Jasniewicz}, G. and {Jean-Antoine-Piccolo}, A. and {Jordan}, S. and {Korn}, A.~J. and {Krone-Martins}, A. and {Lanzafame}, A.~C. and {Lebzelter}, T. and {L{\"o}ffler}, W. and {Manteiga}, M. and {Marrese}, P.~M. and {Mart{\'\i}n-Fleitas}, J.~M. and {Moitinho}, A. and {Mora}, A. and {Muinonen}, K. and {Osinde}, J. and {Pancino}, E. and {Pauwels}, T. and {Petit}, J.-M. and {Recio-Blanco}, A. and {Richards}, P.~J. and {Rimoldini}, L. and {Robin}, A.~C. and {Sarro}, L.~M. and {Siopis}, C. and {Smith}, M. and {Sozzetti}, A. and {S{\"u}veges}, M. and {Torra}, J. and {van Reeven}, W. and {Abbas}, U. and {Abreu Aramburu}, A. and {Accart}, S. and {Aerts}, C. and {Altavilla}, G. and {{\'A}lvarez}, M.~A. and {Alvarez}, R. and {Alves}, J. and {Anderson}, R.~I. and {Andrei}, A.~H. and {Anglada Varela}, E. and {Antiche}, E. and {Antoja}, T. and {Arcay}, B. and {Astraatmadja}, T.~L. and {Bach}, N. and {Baker}, S.~G. and {Balaguer-N{\'u}{\~n}ez}, L. and {Balm}, P. and {Barache}, C. and {Barata}, C. and {Barbato}, D. and {Barblan}, F. and {Barklem}, P.~S. and {Barrado}, D. and {Barros}, M. and {Barstow}, M.~A. and {Bartholom{\'e} Mu{\~n}oz}, S. and {Bassilana}, J.-L. and {Becciani}, U. and {Bellazzini}, M. and {Berihuete}, A. and {Bertone}, S. and {Bianchi}, L. and {Bienaym{\'e}}, O. and {Blanco-Cuaresma}, S. and {Boch}, T. and {Boeche}, C. and {Bombrun}, A. and {Borrachero}, R. and {Bossini}, D. and {Bouquillon}, S. and {Bourda}, G. and {Bragaglia}, A. and {Bramante}, L. and {Breddels}, M.~A. and {Bressan}, A. and {Brouillet}, N. and {Br{\"u}semeister}, T. and {Brugaletta}, E. and {Bucciarelli}, B. and {Burlacu}, A. and {Busonero}, D. and {Butkevich}, A.~G. and {Buzzi}, R. and {Caffau}, E. and {Cancelliere}, R. and {Cannizzaro}, G. and {Cantat-Gaudin}, T. and {Carballo}, R. and {Carlucci}, T. and {Carrasco}, J.~M. and {Casamiquela}, L. and {Castellani}, M. and {Castro-Ginard}, A. and {Charlot}, P. and {Chemin}, L. and {Chiavassa}, A. and {Cocozza}, G. and {Costigan}, G. and {Cowell}, S. and {Crifo}, F. and {Crosta}, M. and {Crowley}, C. and {Cuypers}, J. and {Dafonte}, C. and {Damerdji}, Y. and {Dapergolas}, A. and {David}, P. and {David}, M. and {de Laverny}, P. and {De Luise}, F.},
        title = "{Gaia Data Release 2. Summary of the contents and survey properties}",
      journal = {\aap},
         year = 2018,
        month = aug,
       volume = {616},
          eid = {A1},
        pages = {A1},
          doi = {10.1051/0004-6361/201833051},
archivePrefix = {arXiv},
       eprint = {1804.09365},
 primaryClass = {astro-ph.GA},
       adsurl = {https://ui.adsabs.harvard.edu/abs/2018A&A...616A...1G}
}

@ARTICLE{2018A&A...616A.100Y,
       author = {{Yen}, Hsi-Wei and {Koch}, Patrick M. and {Manara}, Carlo F. and {Miotello}, Anna and {Testi}, Leonardo},
        title = "{Stellar masses and disk properties of Lupus young stellar objects traced by velocity-aligned stacked ALMA $^{13}$CO and C$^{18}$O spectra}",
      journal = {\aap},
         year = 2018,
        month = aug,
       volume = {616},
          eid = {A100},
        pages = {A100},
          doi = {10.1051/0004-6361/201732196},
archivePrefix = {arXiv},
       eprint = {1804.06272},
 primaryClass = {astro-ph.GA},
       adsurl = {https://ui.adsabs.harvard.edu/abs/2018A&A...616A.100Y}
}

@ARTICLE{2020A&A...642A.162G,
       author = {{GRAVITY Collaboration} and {Bouarour}, Y.-I. and {Perraut}, K. and {M{\'e}nard}, F. and {Brandner}, W. and {Caratti O Garatti}, A. and {Caselli}, P. and {van Dishoeck}, E. and {Dougados}, C. and {Garcia-Lopez}, R. and {Grellmann}, R. and {Henning}, T. and {Klarmann}, L. and {Labadie}, L. and {Natta}, A. and {Sanchez-Bermudez}, J. and {Thi}, W.-F. and {de Zeeuw}, P.~T. and {Amorim}, A. and {Baub{\"o}ck}, M. and {Benisty}, M. and {Berger}, J.-P. and {Clenet}, Y. and {Coud{\'e} Du Foresto}, V. and {Duvert}, G. and {Eckart}, A. and {Eisenhauer}, F. and {Eupen}, F. and {Filho}, M. and {Gao}, F. and {Garcia}, P. and {Gendron}, E. and {Genzel}, R. and {Gillessen}, S. and {Jim{\'e}nez-Rosales}, A. and {Jocou}, L. and {Hippler}, S. and {Horrobin}, M. and {Hubert}, Z. and {Kervella}, P. and {Lacour}, S. and {Le Bouquin}, J.-B. and {L{\'e}na}, P. and {Ott}, T. and {Paumard}, T. and {Perrin}, G. and {Pfuhl}, O. and {Rousset}, G. and {Scheithauer}, S. and {Shangguan}, J. and {Stadler}, J. and {Straub}, O. and {Straubmeier}, C. and {Sturm}, E. and {Vincent}, F.~H. and {von Fellenberg}, S.~D. and {Widmann}, F. and {Wiest}, M.},
        title = "{The GRAVITY young stellar object survey. III. The dusty disk of RY Lup}",
      journal = {\aap},
         year = 2020,
        month = oct,
       volume = {642},
          eid = {A162},
        pages = {A162},
          doi = {10.1051/0004-6361/202038249},
archivePrefix = {arXiv},
       eprint = {2008.08527},
 primaryClass = {astro-ph.SR},
       adsurl = {https://ui.adsabs.harvard.edu/abs/2020A&A...642A.162G}
}

@ARTICLE{2018ApJ...855...98A,
       author = {{Arulanantham}, N. and {France}, K. and {Hoadley}, K. and {Manara}, C.~F. and {Schneider}, P.~C. and {Alcal{\'a}}, J.~M. and {Banzatti}, A. and {G{\"u}nther}, H.~M. and {Miotello}, A. and {van der Marel}, N. and {van Dishoeck}, E.~F. and {Walsh}, C. and {Williams}, J.~P.},
        title = "{A UV-to-NIR Study of Molecular Gas in the Dust Cavity around RY Lupi}",
      journal = {\apj},
         year = 2018,
        month = mar,
       volume = {855},
       number = {2},
          eid = {98},
        pages = {98},
          doi = {10.3847/1538-4357/aaaf65},
archivePrefix = {arXiv},
       eprint = {1802.05275},
 primaryClass = {astro-ph.SR},
       adsurl = {https://ui.adsabs.harvard.edu/abs/2018ApJ...855...98A}
}

@ARTICLE{2020ApJ...892..111F,
       author = {{Francis}, Logan and {van der Marel}, Nienke},
        title = "{Dust-depleted Inner Disks in a Large Sample of Transition Disks through Long-baseline ALMA Observations}",
      journal = {\apj},
         year = 2020,
        month = apr,
       volume = {892},
       number = {2},
          eid = {111},
        pages = {111},
          doi = {10.3847/1538-4357/ab7b63},
archivePrefix = {arXiv},
       eprint = {2003.00079},
 primaryClass = {astro-ph.EP},
       adsurl = {https://ui.adsabs.harvard.edu/abs/2020ApJ...892..111F}
}

@ARTICLE{2018A&A...614A..88L,
       author = {{Langlois}, M. and {Pohl}, A. and {Lagrange}, A.-M. and {Maire}, A.-L. and {Mesa}, D. and {Boccaletti}, A. and {Gratton}, R. and {Denneulin}, L. and {Klahr}, H. and {Vigan}, A. and {Benisty}, M. and {Dominik}, C. and {Bonnefoy}, M. and {Menard}, F. and {Avenhaus}, H. and {Cheetham}, A. and {Van Boekel}, R. and {de Boer}, J. and {Chauvin}, G. and {Desidera}, S. and {Feldt}, M. and {Galicher}, R. and {Ginski}, C. and {Girard}, J.~H. and {Henning}, T. and {Janson}, M. and {Kopytova}, T. and {Kral}, Q. and {Ligi}, R. and {Messina}, S. and {Peretti}, S. and {Pinte}, C. and {Sissa}, E. and {Stolker}, T. and {Zurlo}, A. and {Magnard}, Y. and {Blanchard}, P. and {Buey}, T. and {Suarez}, M. and {Cascone}, E. and {Moller-Nilsson}, O. and {Weber}, L. and {Petit}, C. and {Pragt}, J.},
        title = "{First scattered light detection of a nearly edge-on transition disk around the T Tauri star RY Lupi}",
      journal = {\aap},
         year = 2018,
        month = jun,
       volume = {614},
          eid = {A88},
        pages = {A88},
          doi = {10.1051/0004-6361/201731624},
archivePrefix = {arXiv},
       eprint = {1802.03995},
 primaryClass = {astro-ph.EP},
       adsurl = {https://ui.adsabs.harvard.edu/abs/2018A&A...614A..88L}
}

@ARTICLE{2016ApJ...828...46A,
       author = {{Ansdell}, M. and {Williams}, J.~P. and {van der Marel}, N. and {Carpenter}, J.~M. and {Guidi}, G. and {Hogerheijde}, M. and {Mathews}, G.~S. and {Manara}, C.~F. and {Miotello}, A. and {Natta}, A. and {Oliveira}, I. and {Tazzari}, M. and {Testi}, L. and {van Dishoeck}, E.~F. and {van Terwisga}, S.~E.},
        title = "{ALMA Survey of Lupus Protoplanetary Disks. I. Dust and Gas Masses}",
      journal = {\apj},
         year = 2016,
        month = sep,
       volume = {828},
       number = {1},
          eid = {46},
        pages = {46},
          doi = {10.3847/0004-637X/828/1/46},
archivePrefix = {arXiv},
       eprint = {1604.05719},
 primaryClass = {astro-ph.EP},
       adsurl = {https://ui.adsabs.harvard.edu/abs/2016ApJ...828...46A}
}

@ARTICLE{2026A&A...706A.228A,
       author = {{Alqubelat}, Hala and {Manara}, Carlo F. and {Campbell-White}, Justyn and {Petr-Gotzens}, Monika G. and {Tofflemire}, Benjamin M. and {Banzatti}, Andrea and {Ragusa}, Enrico and {Whelan}, Emma T. and {Bourdarot}, Guillaume and {Dougados}, Catherine and {Fiorellino}, Eleonora and {Mills}, Sean I.},
        title = "{Coordinated space-and ground-based monitoring of accretion bursts in a protoplanetary disc: The orbital and accretion properties of DQ Tau}",
      journal = {\aap},
         year = 2026,
        month = feb,
       volume = {706},
          eid = {A228},
        pages = {A228},
          doi = {10.1051/0004-6361/202557425},
archivePrefix = {arXiv},
       eprint = {2511.08311},
 primaryClass = {astro-ph.SR},
       adsurl = {https://ui.adsabs.harvard.edu/abs/2026A&A...706A.228A}
}

@ARTICLE{2021A&A...645A..96P,
       author = {{Pepe}, F. and {Cristiani}, S. and {Rebolo}, R. and {Santos}, N.~C. and {Dekker}, H. and {Cabral}, A. and {Di Marcantonio}, P. and {Figueira}, P. and {Lo Curto}, G. and {Lovis}, C. and {Mayor}, M. and {M{\'e}gevand}, D. and {Molaro}, P. and {Riva}, M. and {Zapatero Osorio}, M.~R. and {Amate}, M. and {Manescau}, A. and {Pasquini}, L. and {Zerbi}, F.~M. and {Adibekyan}, V. and {Abreu}, M. and {Affolter}, M. and {Alibert}, Y. and {Aliverti}, M. and {Allart}, R. and {Allende Prieto}, C. and {{\'A}lvarez}, D. and {Alves}, D. and {Avila}, G. and {Baldini}, V. and {Bandy}, T. and {Barros}, S.~C.~C. and {Benz}, W. and {Bianco}, A. and {Borsa}, F. and {Bourrier}, V. and {Bouchy}, F. and {Broeg}, C. and {Calderone}, G. and {Cirami}, R. and {Coelho}, J. and {Conconi}, P. and {Coretti}, I. and {Cumani}, C. and {Cupani}, G. and {D'Odorico}, V. and {Damasso}, M. and {Deiries}, S. and {Delabre}, B. and {Demangeon}, O.~D.~S. and {Dumusque}, X. and {Ehrenreich}, D. and {Faria}, J.~P. and {Fragoso}, A. and {Genolet}, L. and {Genoni}, M. and {G{\'e}nova Santos}, R. and {Gonz{\'a}lez Hern{\'a}ndez}, J.~I. and {Hughes}, I. and {Iwert}, O. and {Kerber}, F. and {Knudstrup}, J. and {Landoni}, M. and {Lavie}, B. and {Lillo-Box}, J. and {Lizon}, J.-L. and {Maire}, C. and {Martins}, C.~J.~A.~P. and {Mehner}, A. and {Micela}, G. and {Modigliani}, A. and {Monteiro}, M.~A. and {Monteiro}, M.~J.~P.~F.~G. and {Moschetti}, M. and {Murphy}, M.~T. and {Nunes}, N. and {Oggioni}, L. and {Oliveira}, A. and {Oshagh}, M. and {Pall{\'e}}, E. and {Pariani}, G. and {Poretti}, E. and {Rasilla}, J.~L. and {Rebord{\~a}o}, J. and {Redaelli}, E.~M. and {Santana Tschudi}, S. and {Santin}, P. and {Santos}, P. and {S{\'e}gransan}, D. and {Schmidt}, T.~M. and {Segovia}, A. and {Sosnowska}, D. and {Sozzetti}, A. and {Sousa}, S.~G. and {Span{\`o}}, P. and {Su{\'a}rez Mascare{\~n}o}, A. and {Tabernero}, H. and {Tenegi}, F. and {Udry}, S. and {Zanutta}, A.},
        title = "{ESPRESSO at VLT. On-sky performance and first results}",
      journal = {\aap},
         year = 2021,
        month = jan,
       volume = {645},
          eid = {A96},
        pages = {A96},
          doi = {10.1051/0004-6361/202038306},
archivePrefix = {arXiv},
       eprint = {2010.00316},
 primaryClass = {astro-ph.IM},
       adsurl = {https://ui.adsabs.harvard.edu/abs/2021A&A...645A..96P}
}

@INPROCEEDINGS{2020ASPC..527..667M,
       author = {{Modigliani}, A. and {Freudling}, W. and {Anderson}, R.~I. and {Lovis}, C. and {Sosnowska}, D. and {Segovia}, A.},
        title = "{The EsoReflex Workflow to Reduce ESPRESSO Data}",
    booktitle = {Astronomical Data Analysis Software and Systems XXIX},
         year = 2020,
       editor = {{Pizzo}, R. and {Deul}, E.~R. and {Mol}, J.~D. and {de Plaa}, J. and {Verkouter}, H.},
       series = {Astronomical Society of the Pacific Conference Series},
       volume = {527},
        month = jan,
        pages = {667},
       adsurl = {https://ui.adsabs.harvard.edu/abs/2020ASPC..527..667M}
}

@ARTICLE{2026A&A...705A.238V,
       author = {{Vioque}, Miguel and {Booth}, Richard A. and {Ragusa}, Enrico and {Ribas}, {\'A}lvaro and {Kurtovic}, Nicol{\'a}s T. and {Rosotti}, Giovanni P. and {Penoyre}, Zephyr and {Facchini}, Stefano and {Garufi}, Antonio and {Manara}, Carlo F. and {Hu{\'e}lamo}, Nuria and {Winter}, Andrew and {P{\'e}rez}, Sebasti{\'a}n and {Benisty}, Myriam and {Mendigut{\'\i}a}, Ignacio and {Cuello}, Nicol{\'a}s and {Penzlin}, Anna B.~T. and {Castro-Ginard}, Alfred and {Teague}, Richard},
        title = "{Astrometric view of companions in the inner dust cavities of protoplanetary discs}",
      journal = {\aap},
         year = 2026,
        month = jan,
       volume = {705},
          eid = {A238},
        pages = {A238},
          doi = {10.1051/0004-6361/202557086},
archivePrefix = {arXiv},
       eprint = {2512.00157},
 primaryClass = {astro-ph.EP},
       adsurl = {https://ui.adsabs.harvard.edu/abs/2026A&A...705A.238V}
}

@INPROCEEDINGS{2023ASPC..534..605B,
       author = {{Benisty}, M. and {Dominik}, C. and {Follette}, K. and {Garufi}, A. and {Ginski}, C. and {Hashimoto}, J. and {Keppler}, M. and {Kley}, W. and {Monnier}, J.},
        title = "{Optical and Near-infrared View of Planet-forming Disks and Protoplanets}",
    booktitle = {Protostars and Planets VII},
         year = 2023,
       editor = {{Inutsuka}, S. and {Aikawa}, Y. and {Muto}, T. and {Tomida}, K. and {Tamura}, M.},
       series = {Astronomical Society of the Pacific Conference Series},
       volume = {534},
        month = jul,
        pages = {605},
          doi = {10.48550/arXiv.2203.09991},
archivePrefix = {arXiv},
       eprint = {2203.09991},
 primaryClass = {astro-ph.EP},
       adsurl = {https://ui.adsabs.harvard.edu/abs/2023ASPC..534..605B}}

@ARTICLE{2018A&A...617A..44K,
       author = {{Keppler}, M. and {Benisty}, M. and {M{\"u}ller}, A. and {Henning}, Th. and {van Boekel}, R. and {Cantalloube}, F. and {Ginski}, C. and {van Holstein}, R.~G. and {Maire}, A.-L. and {Pohl}, A. and {Samland}, M. and {Avenhaus}, H. and {Baudino}, J.-L. and {Boccaletti}, A. and {de Boer}, J. and {Bonnefoy}, M. and {Chauvin}, G. and {Desidera}, S. and {Langlois}, M. and {Lazzoni}, C. and {Marleau}, G.-D. and {Mordasini}, C. and {Pawellek}, N. and {Stolker}, T. and {Vigan}, A. and {Zurlo}, A. and {Birnstiel}, T. and {Brandner}, W. and {Feldt}, M. and {Flock}, M. and {Girard}, J. and {Gratton}, R. and {Hagelberg}, J. and {Isella}, A. and {Janson}, M. and {Juhasz}, A. and {Kemmer}, J. and {Kral}, Q. and {Lagrange}, A.-M. and {Launhardt}, R. and {Matter}, A. and {M{\'e}nard}, F. and {Milli}, J. and {Molli{\`e}re}, P. and {Olofsson}, J. and {P{\'e}rez}, L. and {Pinilla}, P. and {Pinte}, C. and {Quanz}, S.~P. and {Schmidt}, T. and {Udry}, S. and {Wahhaj}, Z. and {Williams}, J.~P. and {Buenzli}, E. and {Cudel}, M. and {Dominik}, C. and {Galicher}, R. and {Kasper}, M. and {Lannier}, J. and {Mesa}, D. and {Mouillet}, D. and {Peretti}, S. and {Perrot}, C. and {Salter}, G. and {Sissa}, E. and {Wildi}, F. and {Abe}, L. and {Antichi}, J. and {Augereau}, J.-C. and {Baruffolo}, A. and {Baudoz}, P. and {Bazzon}, A. and {Beuzit}, J.-L. and {Blanchard}, P. and {Brems}, S.~S. and {Buey}, T. and {De Caprio}, V. and {Carbillet}, M. and {Carle}, M. and {Cascone}, E. and {Cheetham}, A. and {Claudi}, R. and {Costille}, A. and {Delboulb{\'e}}, A. and {Dohlen}, K. and {Fantinel}, D. and {Feautrier}, P. and {Fusco}, T. and {Giro}, E. and {Gluck}, L. and {Gry}, C. and {Hubin}, N. and {Hugot}, E. and {Jaquet}, M. and {Le Mignant}, D. and {Llored}, M. and {Madec}, F. and {Magnard}, Y. and {Martinez}, P. and {Maurel}, D. and {Meyer}, M. and {M{\"o}ller-Nilsson}, O. and {Moulin}, T. and {Mugnier}, L. and {Orign{\'e}}, A. and {Pavlov}, A. and {Perret}, D. and {Petit}, C. and {Pragt}, J. and {Puget}, P. and {Rabou}, P. and {Ramos}, J. and {Rigal}, F. and {Rochat}, S. and {Roelfsema}, R. and {Rousset}, G. and {Roux}, A. and {Salasnich}, B. and {Sauvage}, J.-F. and {Sevin}, A. and {Soenke}, C. and {Stadler}, E. and {Suarez}, M. and {Turatto}, M. and {Weber}, L.},
        title = "{Discovery of a planetary-mass companion within the gap of the transition disk around PDS 70}",
      journal = {\aap},
         year = 2018,
        month = sep,
       volume = {617},
          eid = {A44},
        pages = {A44},
          doi = {10.1051/0004-6361/201832957},
archivePrefix = {arXiv},
       eprint = {1806.11568},
 primaryClass = {astro-ph.EP},
       adsurl = {https://ui.adsabs.harvard.edu/abs/2018A&A...617A..44K}
}

@ARTICLE{2024A&A...685L...1C,
       author = {{Christiaens}, V. and {Samland}, M. and {Henning}, Th. and {Portilla-Revelo}, B. and {Perotti}, G. and {Matthews}, E. and {Absil}, O. and {Decin}, L. and {Kamp}, I. and {Boccaletti}, A. and {Tabone}, B. and {Marleau}, G.-D. and {van Dishoeck}, E.~F. and {G{\"u}del}, M. and {Lagage}, P.-O. and {Barrado}, D. and {Caratti o Garatti}, A. and {Glauser}, A.~M. and {Olofsson}, G. and {Ray}, T.~P. and {Scheithauer}, S. and {Vandenbussche}, B. and {Waters}, L.~B.~F.~M. and {Arabhavi}, A.~M. and {Grant}, S.~L. and {Jang}, H. and {Kanwar}, J. and {Schreiber}, J. and {Schwarz}, K. and {Temmink}, M. and {{\"O}stlin}, G.},
        title = "{MINDS: JWST/NIRCam imaging of the protoplanetary disk PDS 70. A spiral accretion stream and a potential third protoplanet}",
      journal = {\aap},
         year = 2024,
        month = may,
       volume = {685},
          eid = {L1},
        pages = {L1},
          doi = {10.1051/0004-6361/202349089},
archivePrefix = {arXiv},
       eprint = {2403.04855},
 primaryClass = {astro-ph.EP},
       adsurl = {https://ui.adsabs.harvard.edu/abs/2024A&A...685L...1C}
}

@ARTICLE{2008ApJ...678L..59I,
       author = {{Ireland}, M.~J. and {Kraus}, A.~L.},
        title = "{The Disk Around CoKu Tauri/4: Circumbinary, Not Transitional}",
      journal = {\apjl},
         year = 2008,
        month = may,
       volume = {678},
       number = {1},
        pages = {L59},
          doi = {10.1086/588216},
archivePrefix = {arXiv},
       eprint = {0803.2044},
 primaryClass = {astro-ph},
       adsurl = {https://ui.adsabs.harvard.edu/abs/2008ApJ...678L..59I}
}

@ARTICLE{Espaillat2014,
    author = {{Espaillat}, C. and {Muzerolle}, J. and {Najita}, J. and {Andrews}, S. and {Zweibel}, E. and {Calvet}, N. and {Carr}, J. and {Hartmann}, L. and {Herczeg}, G. and {Ingleby}, L. and {Alexander}, R. and {Gorti}, U. and {Hollenbach}, D. and {Pontoppidan}, K. and {Wood}, K. and {Adams}, F. and {Blake}, G. and {Bouvier}, J. and {Close}, L. and {Cieza}, L. and {D'Alessio}, P. and {Fischer}, W. and {Isella}, A. and {Jang-Condell}, H. and {Kastner}, J. and {Lada}, C. and {Manara}, C. and {McClure}, M. and {Mendigut{\'\i}a}, I. and {Meyer}, M. and {Montesinos}, E. and {Muzerolle}, J. and {Pascucci}, I. and {Perez}, L. and {Pontoppidan}, K. and {Rigliaco}, E. and {Salyk}, C. and {Schneider}, A. and {Sicilia-Aguilar}, A. and {Testi}, L. and {Thi}, W. and {Williams}, J. and {Wright}, C.},
    title = "{On the Nature of Transitional Disks}",
    journal = {Protostars and Planets VI},
    pages = {497},
    year = 2014,
    adsurl = {https://ui.adsabs.harvard.edu/abs/2014prpl.conf..497E}
}

@ARTICLE{2024A&A...685A..52G,
       author = {{Ginski}, C. and {Garufi}, A. and {Benisty}, M. and {Tazaki}, R. and {Dominik}, C. and {Ribas}, {\'A}. and {Engler}, N. and {Birnstiel}, T. and {Chauvin}, G. and {Columba}, G. and {Facchini}, S. and {Goncharov}, A. and {Hagelberg}, J. and {Henning}, T. and {Hogerheijde}, M. and {van Holstein}, R.~G. and {Huang}, J. and {Muto}, T. and {Pinilla}, P. and {Kanagawa}, K. and {Kim}, S. and {Kurtovic}, N. and {Langlois}, M. and {Manara}, C. and {Milli}, J. and {Momose}, M. and {Orihara}, R. and {Pawellek}, N. and {Pinte}, C. and {Rab}, C. and {Schmidt}, T.~O.~B. and {Snik}, F. and {Wahhaj}, Z. and {Williams}, J. and {Zurlo}, A.},
        title = "{The SPHERE view of the Chamaeleon I star-forming region. The full census of planet-forming disks with GTO and DESTINYS programs}",
      journal = {\aap},
         year = 2024,
        month = may,
       volume = {685},
          eid = {A52},
        pages = {A52},
          doi = {10.1051/0004-6361/202244005},
archivePrefix = {arXiv},
       eprint = {2403.02149},
 primaryClass = {astro-ph.GA},
       adsurl = {https://ui.adsabs.harvard.edu/abs/2024A&A...685A..52G}
}

@ARTICLE{2025A&A...704A.221V,
       author = {{van Capelleveen}, Richelle F. and {Kenworthy}, Matthew A. and {Ginski}, Christian and {Mamajek}, Eric E. and {Bohn}, Alexander J. and {Landman}, Rico and {Stolker}, Tomas and {Zhang}, Yapeng and {van der Marel}, Nienke and {Snellen}, Ignas},
        title = "{WIde Separation Planets In Time (WISPIT): Two directly imaged exoplanets around the Sun-like stellar binary WISPIT 1}",
      journal = {\aap},
         year = 2025,
        month = dec,
       volume = {704},
          eid = {A221},
        pages = {A221},
          doi = {10.1051/0004-6361/202556584},
archivePrefix = {arXiv},
       eprint = {2508.18456},
 primaryClass = {astro-ph.EP},
       adsurl = {https://ui.adsabs.harvard.edu/abs/2025A&A...704A.221V}
}

@ARTICLE{2017A&A...602A..33F,
       author = {{Frasca}, A. and {Biazzo}, K. and {Alcal{\'a}}, J.~M. and {Manara}, C.~F. and {Stelzer}, B. and {Covino}, E. and {Antoniucci}, S.},
        title = "{X-shooter spectroscopy of young stellar objects in Lupus. Atmospheric parameters, membership, and activity diagnostics}",
      journal = {\aap},
         year = 2017,
        month = jun,
       volume = {602},
          eid = {A33},
        pages = {A33},
          doi = {10.1051/0004-6361/201630108},
archivePrefix = {arXiv},
       eprint = {1703.01251},
 primaryClass = {astro-ph.SR},
       adsurl = {https://ui.adsabs.harvard.edu/abs/2017A&A...602A..33F}
}

@ARTICLE{2018A&A...616A..79G,
       author = {{Ginski}, C. and {Benisty}, M. and {van Holstein}, R.~G. and {Juh{\'a}sz}, A. and {Schmidt}, T.~O.~B. and {Chauvin}, G. and {de Boer}, J. and {Wilby}, M. and {Manara}, C.~F. and {Delorme}, P. and {M{\'e}nard}, F. and {Pinilla}, P. and {Birnstiel}, T. and {Flock}, M. and {Keller}, C. and {Kenworthy}, M. and {Milli}, J. and {Olofsson}, J. and {P{\'e}rez}, L. and {Snik}, F. and {Vogt}, N.},
        title = "{First direct detection of a polarized companion outside a resolved circumbinary disk around CS Chamaeleonis}",
      journal = {\aap},
         year = 2018,
        month = aug,
       volume = {616},
          eid = {A79},
        pages = {A79},
          doi = {10.1051/0004-6361/201732417},
archivePrefix = {arXiv},
       eprint = {1805.02261},
 primaryClass = {astro-ph.EP},
       adsurl = {https://ui.adsabs.harvard.edu/abs/2018A&A...616A..79G}
}

@ARTICLE{2020MNRAS.496.3257B,
       author = {{Bredall}, J.~W. and {Shappee}, B.~J. and {Gaidos}, E. and {Jayasinghe}, T. and {Vallely}, P. and {Stanek}, K.~Z. and {Kochanek}, C.~S. and {Gagn{\'e}}, J. and {Hart}, K. and {Holoien}, T.~W.-S. and {Prieto}, J.~L. and {Van Saders}, J.},
        title = "{The ASAS-SN catalogue of variable stars - VIII. 'Dipper' stars in the Lupus star-forming region}",
      journal = {\mnras},
         year = 2020,
        month = aug,
       volume = {496},
       number = {3},
        pages = {3257-3269},
          doi = {10.1093/mnras/staa1588},
archivePrefix = {arXiv},
       eprint = {2005.14201},
 primaryClass = {astro-ph.SR},
       adsurl = {https://ui.adsabs.harvard.edu/abs/2020MNRAS.496.3257B}
}

@ARTICLE{2024A&A...683A.239D,
       author = {{Di Maio}, C. and {Petralia}, A. and {Micela}, G. and {Lanza}, A.~F. and {Rainer}, M. and {Malavolta}, L. and {Benatti}, S. and {Affer}, L. and {Maldonado}, J. and {Colombo}, S. and {Damasso}, M. and {Maggio}, A. and {Biazzo}, K. and {Bignamini}, A. and {Borsa}, F. and {Boschin}, W. and {Cabona}, L. and {Cecconi}, M. and {Claudi}, R. and {Covino}, E. and {Di Fabrizio}, L. and {Gratton}, R. and {Lorenzi}, V. and {Mancini}, L. and {Messina}, S. and {Molinari}, E. and {Molinaro}, M. and {Nardiello}, D. and {Poretti}, E. and {Sozzetti}, A.},
        title = "{The GAPS programme at TNG. LII. Spot modelling of V1298 Tau using the SpotCCF tool}",
      journal = {\aap},
         year = 2024,
        month = mar,
       volume = {683},
          eid = {A239},
        pages = {A239},
          doi = {10.1051/0004-6361/202348223},
archivePrefix = {arXiv},
       eprint = {2312.14269},
 primaryClass = {astro-ph.SR},
       adsurl = {https://ui.adsabs.harvard.edu/abs/2024A&A...683A.239D}
}

@ARTICLE{2020ARA&A..58..483A,
       author = {{Andrews}, Sean M.},
        title = "{Observations of Protoplanetary Disk Structures}",
      journal = {\araa},
         year = 2020,
        month = aug,
       volume = {58},
        pages = {483-528},
          doi = {10.1146/annurev-astro-031220-010302},
archivePrefix = {arXiv},
       eprint = {2001.05007},
 primaryClass = {astro-ph.EP},
       adsurl = {https://ui.adsabs.harvard.edu/abs/2020ARA&A..58..483A}
}

@ARTICLE{2022A&A...658A.183B,
       author = {{Bohn}, A.~J. and {Benisty}, M. and {Perraut}, K. and {van der Marel}, N. and {W{\"o}lfer}, L. and {van Dishoeck}, E.~F. and {Facchini}, S. and {Manara}, C.~F. and {Teague}, R. and {Francis}, L. and {Berger}, J.-P. and {Garcia-Lopez}, R. and {Ginski}, C. and {Henning}, T. and {Kenworthy}, M. and {Kraus}, S. and {M{\'e}nard}, F. and {M{\'e}rand}, A. and {P{\'e}rez}, L.~M.},
        title = "{Probing inner and outer disk misalignments in transition disks. Constraints from VLTI/GRAVITY and ALMA observations}",
      journal = {\aap},
         year = 2022,
        month = feb,
       volume = {658},
          eid = {A183},
        pages = {A183},
          doi = {10.1051/0004-6361/202142070},
archivePrefix = {arXiv},
       eprint = {2112.00123},
 primaryClass = {astro-ph.EP},
       adsurl = {https://ui.adsabs.harvard.edu/abs/2022A&A...658A.183B}
}

@ARTICLE{2023EPJP..138..225V,
       author = {{van der Marel}, Nienke},
        title = "{Transition disks: the observational revolution from SEDs to imaging}",
      journal = {European Physical Journal Plus},
         year = 2023,
        month = mar,
       volume = {138},
       number = {3},
          eid = {225},
        pages = {225},
          doi = {10.1140/epjp/s13360-022-03628-0},
archivePrefix = {arXiv},
       eprint = {2210.05539},
 primaryClass = {astro-ph.EP},
       adsurl = {https://ui.adsabs.harvard.edu/abs/2023EPJP..138..225V}
}

@ARTICLE{2008ApJ...678..472H,
       author = {{Huerta}, Marcos and {Johns-Krull}, Christopher M. and {Prato}, L. and {Hartigan}, Patrick and {Jaffe}, D.~T.},
        title = "{Starspot-Induced Radial Velocity Variability in LkCa 19}",
      journal = {\apj},
         year = 2008,
        month = may,
       volume = {678},
       number = {1},
        pages = {472-482},
          doi = {10.1086/526415},
archivePrefix = {arXiv},
       eprint = {0711.2505},
 primaryClass = {astro-ph},
       adsurl = {https://ui.adsabs.harvard.edu/abs/2008ApJ...678..472H}
}

@INPROCEEDINGS{2024AAS...24322503E,
       author = {{Erba}, Christiana},
        title = "{SpecpolFlow: a new, open-source, pythonic workflow for optical spectropolarimetry}",
    booktitle = {American Astronomical Society Meeting Abstracts \#243},
         year = 2024,
       series = {American Astronomical Society Meeting Abstracts},
       volume = {243},
        month = feb,
          eid = {225.03},
        pages = {225.03},
       adsurl = {https://ui.adsabs.harvard.edu/abs/2024AAS...24322503E}
}

@ARTICLE{2025A&A...698A.102R,
       author = {{Ragusa}, Enrico and {Lodato}, Giuseppe and {Cuello}, Nicol{\'a}s and {Vioque}, Miguel and {Manara}, Carlo F. and {Toci}, Claudia},
        title = "{The likelihood of not detecting cavity-carving companions in transition discs {\textendash} A statistical approach}",
      journal = {\aap},
         year = 2025,
        month = jun,
       volume = {698},
          eid = {A102},
        pages = {A102},
          doi = {10.1051/0004-6361/202554462},
archivePrefix = {arXiv},
       eprint = {2504.06337},
 primaryClass = {astro-ph.EP},
       adsurl = {https://ui.adsabs.harvard.edu/abs/2025A&A...698A.102R}
}

@ARTICLE{2026A&A...708A.230D,
       author = {{Donati}, J.-F. and {Cristofari}, P.~I. and {Carmona}, A. and {Lavail}, A. and {Moutou}, C. and {Bouvier}, J. and {Perraut}, K. and {Alencar}, S.~H.~P. and {M{\'e}nard}, F. and {Audard}, M. and {Petit}, P. and {Alecian}, E. and {Ray}, T.},
        title = "{Monitoring the magnetospheric accretion of the classical T Tauri star DO Tau with SPIRou}",
      journal = {\aap},
         year = 2026,
        month = apr,
       volume = {708},
          eid = {A230},
        pages = {A230},
          doi = {10.1051/0004-6361/202558694},
archivePrefix = {arXiv},
       eprint = {2602.24078},
 primaryClass = {astro-ph.SR},
       adsurl = {https://ui.adsabs.harvard.edu/abs/2026A&A...708A.230D}
}

@ARTICLE{1997MNRAS.291..658D,
       author = {{Donati}, J.-F. and {Semel}, M. and {Carter}, B.~D. and {Rees}, D.~E. and {Collier Cameron}, A.},
        title = "{Spectropolarimetric observations of active stars}",
      journal = {\mnras},
         year = 1997,
        month = nov,
       volume = {291},
       number = {4},
        pages = {658-682},
          doi = {10.1093/mnras/291.4.658},
       adsurl = {https://ui.adsabs.harvard.edu/abs/1997MNRAS.291..658D}
}

@ARTICLE{2025ApJ...992L..33P,
       author = {{P{\'e}rez Paolino}, Facundo and {Hillenbrand}, Lynne A. and {Bary}, Jeffrey S.},
        title = "{Separating Photospheric and Starspot Magnetic Fields in Pre-main-sequence Stars Using IGRINS Spectroscopy}",
      journal = {\apjl},
         year = 2025,
        month = oct,
       volume = {992},
       number = {2},
          eid = {L33},
        pages = {L33},
          doi = {10.3847/2041-8213/ae0cb1},
archivePrefix = {arXiv},
       eprint = {2509.20475},
 primaryClass = {astro-ph.SR},
       adsurl = {https://ui.adsabs.harvard.edu/abs/2025ApJ...992L..33P}
}

@ARTICLE{2017ApJ...836..200G,
       author = {{Gully-Santiago}, Michael A. and {Herczeg}, Gregory J. and {Czekala}, Ian and {Somers}, Garrett and {Grankin}, Konstantin and {Covey}, Kevin R. and {Donati}, J.~F. and {Alencar}, Silvia H.~P. and {Hussain}, Gaitee A.~J. and {Shappee}, Benjamin J. and {Mace}, Gregory N. and {Lee}, Jae-Joon and {Holoien}, T.~W.-S. and {Jose}, Jessy and {Liu}, Chun-Fan},
        title = "{Placing the Spotted T Tauri Star LkCa 4 on an HR Diagram}",
      journal = {\apj},
         year = 2017,
        month = feb,
       volume = {836},
       number = {2},
          eid = {200},
        pages = {200},
          doi = {10.3847/1538-4357/836/2/200},
archivePrefix = {arXiv},
       eprint = {1701.06703},
 primaryClass = {astro-ph.SR},
       adsurl = {https://ui.adsabs.harvard.edu/abs/2017ApJ...836..200G}
}

@ARTICLE{2022A&A...667A.124G,
       author = {{Gangi}, M. and {Antoniucci}, S. and {Biazzo}, K. and {Frasca}, A. and {Nisini}, B. and {Alcal{\'a}}, J.~M. and {Giannini}, T. and {Manara}, C.~F. and {Giunta}, A. and {Harutyunyan}, A. and {Munari}, U. and {Vitali}, F.},
        title = "{GIARPS High-resolution Observations of T Tauri stars (GHOsT). IV. Accretion properties of the Taurus-Auriga young association}",
      journal = {\aap},
         year = 2022,
        month = nov,
       volume = {667},
          eid = {A124},
        pages = {A124},
          doi = {10.1051/0004-6361/202244042},
archivePrefix = {arXiv},
       eprint = {2208.14895},
 primaryClass = {astro-ph.SR},
       adsurl = {https://ui.adsabs.harvard.edu/abs/2022A&A...667A.124G}
}

@ARTICLE{Feroz2019OJAp....2E..10F,
       author = {{Feroz}, Farhan and {Hobson}, Michael P. and {Cameron}, Ewan and {Pettitt}, Anthony N.},
        title = "{Importance Nested Sampling and the MultiNest Algorithm}",
      journal = {The Open Journal of Astrophysics},
         year = 2019,
        month = nov,
       volume = {2},
       number = {1},
          eid = {10},
        pages = {10},
          doi = {10.21105/astro.1306.2144},
archivePrefix = {arXiv},
       eprint = {1306.2144},
 primaryClass = {astro-ph.IM},
       adsurl = {https://ui.adsabs.harvard.edu/abs/2019OJAp....2E..10F}
}

@article{Kass10.2307/2291091,
 ISSN = {01621459},
 author = {Robert E. Kass and Adrian E. Raftery},
 journal = {Journal of the American Statistical Association},
 number = {430},
 pages = {773--795},
 publisher = {[American Statistical Association, Taylor & Francis, Ltd.]},
 title = {Bayes Factors},
 volume = {90},
 year = {1995}
}

@ARTICLE{Zechmeister2009A&A...496..577Z,
	author = {{Zechmeister}, M. and {K{\"u}rster}, M.},
	title = "{The generalised Lomb-Scargle periodogram. A new formalism for the floating-mean and Keplerian periodograms}",
	journal = {\aap},
	year = 2009,
	month = mar,
	volume = {496},
	number = {2},
	pages = {577-584},
	doi = {10.1051/0004-6361:200811296},
	archivePrefix = {arXiv},
	eprint = {0901.2573},
	primaryClass = {astro-ph.IM},
	adsurl = {https://ui.adsabs.harvard.edu/abs/2009A&A...496..577Z}
}

@ARTICLE{Hathaway2015LRSP...12....4H,
       author = {{Hathaway}, David H.},
        title = "{The Solar Cycle}",
      journal = {Living Reviews in Solar Physics},
         year = 2015,
        month = dec,
       volume = {12},
       number = {1},
          eid = {4},
        pages = {4},
          doi = {10.1007/lrsp-2015-4},
archivePrefix = {arXiv},
       eprint = {1502.07020},
 primaryClass = {astro-ph.SR},
       adsurl = {https://ui.adsabs.harvard.edu/abs/2015LRSP...12....4H}
}

@ARTICLE{Strassmeier2009A&ARv..17..251S,
       author = {{Strassmeier}, Klaus G.},
        title = "{Starspots}",
      journal = {\aapr},
         year = 2009,
        month = sep,
       volume = {17},
       number = {3},
        pages = {251-308},
          doi = {10.1007/s00159-009-0020-6},
       adsurl = {https://ui.adsabs.harvard.edu/abs/2009A&ARv..17..251S}
}

@ARTICLE{2015MNRAS.452.2396M,
       author = {{Miranda}, Ryan and {Lai}, Dong},
        title = "{Tidal truncation of inclined circumstellar and circumbinary discs in young stellar binaries}",
      journal = {\mnras},
         year = 2015,
        month = sep,
       volume = {452},
       number = {3},
        pages = {2396-2409},
          doi = {10.1093/mnras/stv1450},
archivePrefix = {arXiv},
       eprint = {1504.02917},
 primaryClass = {astro-ph.EP},
       adsurl = {https://ui.adsabs.harvard.edu/abs/2015MNRAS.452.2396M}
}

@ARTICLE{1994ApJ...421..651A,
       author = {{Artymowicz}, Pawel and {Lubow}, Stephen H.},
        title = "{Dynamics of Binary-Disk Interaction. I. Resonances and Disk Gap Sizes}",
      journal = {\apj},
         year = 1994,
        month = feb,
       volume = {421},
        pages = {651},
          doi = {10.1086/173679},
       adsurl = {https://ui.adsabs.harvard.edu/abs/1994ApJ...421..651A}
}

@ARTICLE{2018A&A...616A.108B,
       author = {{Boro Saikia}, S. and {Marvin}, C.~J. and {Jeffers}, S.~V. and {Reiners}, A. and {Cameron}, R. and {Marsden}, S.~C. and {Petit}, P. and {Warnecke}, J. and {Yadav}, A.~P.},
        title = "{Chromospheric activity catalogue of 4454 cool stars. Questioning the active branch of stellar activity cycles}",
      journal = {\aap},
         year = 2018,
        month = aug,
       volume = {616},
          eid = {A108},
        pages = {A108},
          doi = {10.1051/0004-6361/201629518},
archivePrefix = {arXiv},
       eprint = {1803.11123},
 primaryClass = {astro-ph.SR},
       adsurl = {https://ui.adsabs.harvard.edu/abs/2018A&A...616A.108B}
}

@ARTICLE{2019ApJ...887...84Z,
       author = {{Zhang}, Jiajun and {Zhao}, Jingkun and {Oswalt}, Terry D. and {Fang}, Xiangsong and {Zhao}, Gang and {Liang}, Xilong and {Ye}, Xianhao and {Zhong}, Jing},
        title = "{Stellar Chromospheric Activity and Age Relation from Open Clusters in the LAMOST Survey}",
      journal = {\apj},
         year = 2019,
        month = dec,
       volume = {887},
       number = {1},
          eid = {84},
        pages = {84},
          doi = {10.3847/1538-4357/ab4efe},
archivePrefix = {arXiv},
       eprint = {1909.13520},
 primaryClass = {astro-ph.SR},
       adsurl = {https://ui.adsabs.harvard.edu/abs/2019ApJ...887...84Z}
}

@ARTICLE{1986A&A...165..110B,
       author = {{Bouvier}, J. and {Bertout}, C. and {Benz}, W. and {Mayor}, M.},
        title = "{Rotation in T Tauri stars. I. Obervations and immediate analysis.}",
      journal = {\aap},
         year = 1986,
        month = sep,
       volume = {165},
        pages = {110-119},
       adsurl = {https://ui.adsabs.harvard.edu/abs/1986A&A...165..110B}
}

@ARTICLE{2016A&A...587A..28H,
       author = {{Hackman}, T. and {Lehtinen}, J. and {Ros{\'e}n}, L. and {Kochukhov}, O. and {K{\"a}pyl{\"a}}, M.~J.},
        title = "{Zeeman-Doppler imaging of active young solar-type stars}",
      journal = {\aap},
         year = 2016,
        month = mar,
       volume = {587},
          eid = {A28},
        pages = {A28},
          doi = {10.1051/0004-6361/201527320},
archivePrefix = {arXiv},
       eprint = {1509.02285},
 primaryClass = {astro-ph.SR},
       adsurl = {https://ui.adsabs.harvard.edu/abs/2016A&A...587A..28H}
}

@ARTICLE{2019A&A...625A..79H,
       author = {{Hackman}, T. and {Ilyin}, I. and {Lehtinen}, J.~J. and {Kochukhov}, O. and {K{\"a}pyl{\"a}}, M.~J. and {Piskunov}, N. and {Willamo}, T.},
        title = "{Starspot activity of HD 199178. Doppler images from 1994-2017}",
      journal = {\aap},
         year = 2019,
        month = may,
       volume = {625},
          eid = {A79},
        pages = {A79},
          doi = {10.1051/0004-6361/201834763},
archivePrefix = {arXiv},
       eprint = {1812.02013},
 primaryClass = {astro-ph.SR},
       adsurl = {https://ui.adsabs.harvard.edu/abs/2019A&A...625A..79H}
}

@ARTICLE{2003A&A...405..149F,
       author = {{Frasca}, A. and {Alcal{\'a}}, J.~M. and {Covino}, E. and {Catalano}, S. and {Marilli}, E. and {Paladino}, R.},
        title = "{Further identification of ROSAT all-sky survey sources in Orion}",
      journal = {\aap},
         year = 2003,
        month = jul,
       volume = {405},
        pages = {149-163},
          doi = {10.1051/0004-6361:20030644},
       adsurl = {https://ui.adsabs.harvard.edu/abs/2003A&A...405..149F}
}

@ARTICLE{2015PhyS...90e4005R,
       author = {{Ryabchikova}, T. and {Piskunov}, N. and {Kurucz}, R.~L. and {Stempels}, H.~C. and {Heiter}, U. and {Pakhomov}, Yu and {Barklem}, P.~S.},
        title = "{A major upgrade of the VALD database}",
      journal = {\physscr},
         year = 2015,
        month = may,
       volume = {90},
       number = {5},
          eid = {054005},
        pages = {054005},
          doi = {10.1088/0031-8949/90/5/054005},
       adsurl = {https://ui.adsabs.harvard.edu/abs/2015PhyS...90e4005R}
}

@ARTICLE{2009A&A...508.1313F,
       author = {{Frasca}, A. and {Covino}, E. and {Spezzi}, L. and {Alcal{\'a}}, J.~M. and {Marilli}, E. and {F{\.z}r{\'e}sz}, G. and {Gandolfi}, D.},
        title = "{REM near-IR and optical photometric monitoring of pre-main sequence stars in Orion. Rotation periods and starspot parameters}",
      journal = {\aap},
         year = 2009,
        month = dec,
       volume = {508},
       number = {3},
        pages = {1313-1330},
          doi = {10.1051/0004-6361/200913327},
       adsurl = {https://ui.adsabs.harvard.edu/abs/2009A&A...508.1313F}
}

@ARTICLE{Buchner2014A&A...564A.125B,
       author = {{Buchner}, J. and {Georgakakis}, A. and {Nandra}, K. and {Hsu}, L. and {Rangel}, C. and {Brightman}, M. and {Merloni}, A. and {Salvato}, M. and {Donley}, J. and {Kocevski}, D.},
        title = "{X-ray spectral modelling of the AGN obscuring region in the CDFS: Bayesian model selection and catalogue}",
      journal = {\aap},
         year = 2014,
        month = apr,
       volume = {564},
          eid = {A125},
        pages = {A125},
          doi = {10.1051/0004-6361/201322971},
archivePrefix = {arXiv},
       eprint = {1402.0004},
 primaryClass = {astro-ph.HE},
       adsurl = {https://ui.adsabs.harvard.edu/abs/2014A&A...564A.125B}
}

@ARTICLE{1989A&A...223..112Z,
       author = {{Zahn}, J.-P. and {Bouchet}, L.},
        title = "{Tidal evolution of close binary stars. II. Orbital circularization oflate-type binaries.}",
      journal = {\aap},
         year = 1989,
        month = oct,
       volume = {223},
        pages = {112-118},
       adsurl = {https://ui.adsabs.harvard.edu/abs/1989A&A...223..112Z}
}

@ARTICLE{2012RSPTA.370.2765A,
       author = {{Allard}, F. and {Homeier}, D. and {Freytag}, B.},
        title = "{Models of very-low-mass stars, brown dwarfs and exoplanets}",
      journal = {Philosophical Transactions of the Royal Society of London Series A},
         year = 2012,
        month = jun,
       volume = {370},
       number = {1968},
        pages = {2765-2777},
          doi = {10.1098/rsta.2011.0269},
archivePrefix = {arXiv},
       eprint = {1112.3591},
 primaryClass = {astro-ph.SR},
       adsurl = {https://ui.adsabs.harvard.edu/abs/2012RSPTA.370.2765A}
}

@ARTICLE{2020MNRAS.499.3362R,
       author = {{Ragusa}, Enrico and {Alexander}, Richard and {Calcino}, Josh and {Hirsh}, Kieran and {Price}, Daniel J.},
        title = "{The evolution of large cavities and disc eccentricity in circumbinary discs}",
      journal = {\mnras},
         year = 2020,
        month = dec,
       volume = {499},
       number = {3},
        pages = {3362-3380},
          doi = {10.1093/mnras/staa2954},
archivePrefix = {arXiv},
       eprint = {2009.10738},
 primaryClass = {astro-ph.EP},
       adsurl = {https://ui.adsabs.harvard.edu/abs/2020MNRAS.499.3362R}
}

\begin{appendix}

\section{ROTFIT analysis log}

In this section, we present the parameters values\\ derived using {\rm ROTFIT} in Sect.\ref{stellar_properties}.

\begin{sidewaystable}[p]
    \centering
    \caption{Photospheric properties from the ESPRESSO spectra fit with {\rm ROTFIT}, including the effective temperature ($T_\mathrm{eff}$), surface gravity log g, iron abundance [Fe/H],\\ spectral type, projected rotational velocity ($v\, \sin i$), and the veiling ($r$).}
    \label{ROTFIT_table}

        \setlength{\tabcolsep}{1.7pt}

    \tiny                  
    
    \resizebox{0.8\textwidth}{!}{%
        \begin{tabular}{lcccccccccccccccccc}
\hline  
\hline
DATE-OBS & MJD & $T_{\text{eff}}$ & err & $\log g$ & err & [Fe/H] & err  & $v \sin i$ & err & $r_{400}$ & $r_{420}$ & $r_{450}$ & $r_{500}$ & $r_{550}$ & $r_{600}$ & $r_{650}$ \\
         &     & (K)              & (K) & (dex)    & (dex)& (dex)  & (dex)&        & ($\text{km s}^{-1}$) & ($\text{km s}^{-1}$) & & & & & & & \\
\hline

2022-05-27T01-09-01-521 & 59726.047930  &    5168 &  60 &  4.16 &  0.24 &  0.05 &  0.08 &   32.3 &  0.8 & 0.30 & 0.04 & 0.06 & 0.00 & 0.12 & 0.09 & 0.04 \\
2022-05-28T01-07-56-958 &  59727.047190 &    5138 &  66 &  4.17 &  0.23 &  0.04 &  0.08 &   32.1 &  0.8 & 0.20 & 0.01 & 0.00 & 0.00 & 0.05 & 0.06 & 0.05 \\
2022-05-30T01-26-20-958 &  59729.059960 &    5168 &  75 &  4.04 &  0.19 &  0.02 &  0.07 &    33.2 &  1.3 & 0.00 & 0.00 & 0.00 & 0.00 & 0.00 & 0.00 & 0.00 \\
2022-05-31T23-43-12-883 &  59730.988340 &    5176 &  65 &  4.05 &  0.22 &  0.05 &  0.08 &   31.8 &  0.7 & 0.40 & 0.08 & 0.08 & 0.01 & 0.12 & 0.09 & 0.07 \\
2022-06-04T00-42-02-815 &  59734.029200 &    5200 &  73 &  4.10 &  0.20 &  0.03 &  0.07 &   33.7 &  1.6 & 0.00 & 0.00 & 0.00 & 0.00 & 0.00 & 0.00 & 0.00 \\
2022-10-06T00-02-16-084 &  59858.001580 &    5148 &  65 &  4.34 &  0.20 &  0.08 &  0.08 &   34.2 &  1.2 & 0.50 & 0.17 & 0.06 & 0.00 & 0.02 & 0.08 & 0.03 \\
2023-01-15T07-59-51-548 &  59959.333240 &    5200 &  64 &  4.11 &  0.20 &  0.05 &  0.09 &   33.9 &  2.2 & 0.80 & 0.72 & 0.38 & 0.08 & 0.11 & 0.11 & 0.11 \\
2023-01-30T08-37-27-330 &  59974.359340 &    5172 &  67 &  4.47 &  0.14 &  0.05 &  0.08 &    31.4 &  1.2 & 0.40 & 0.08 & 0.07 & 0.00 & 0.09 & 0.10 & 0.07 \\
2023-02-03T08-24-15-020 &  59978.350170 &    5187 &  67 &  4.15 &  0.23 &  0.08 &  0.09 &    30.4 &  1.3 & 0.50 & 0.19 & 0.13 & 0.01 & 0.10 & 0.08 & 0.05 \\
2023-02-21T06-01-38-983 &  59996.251150 &    5211 &  68 &  4.21 &  0.18 &  0.04 &  0.08 &    34.2 &  1.3 & 0.70 & 0.20 & 0.08 & 0.01 & 0.10 & 0.08 & 0.09 \\
2023-03-02T05-27-04-925 &  60005.227140 &    5148 &  70 &  3.97 &  0.27 &  0.02 &  0.07 &    31.6 &  1.0 & 0.40 & 0.08 & 0.07 & 0.05 & 0.10 & 0.07 & 0.13 \\
2023-03-05T07-22-07-170 &  60008.307030 &    5152 &  63 &  4.45 &  0.16 &  0.08 &  0.09 &    31.3 &  1.4 & 0.70 & 0.29 & 0.16 & 0.08 & 0.17 & 0.07 & 0.10 \\
2023-03-06T05-21-45-169 &  60009.223440 &    5198 &  65 &  4.23 &  0.20 &  0.06 &  0.09 &    32.5 &  1.3 & 0.70 & 0.22 & 0.08 & 0.05 & 0.08 & 0.10 & 0.10 \\
2023-03-10T07-24-02-870 &  60013.308370 &    5161 &  68 &  4.07 &  0.22 &  0.05 &  0.08 &    31.9 &  0.8 & 0.30 & 0.07 & 0.05 & 0.01 & 0.12 & 0.06 & 0.06 \\
2023-03-16T09-13-13-194 &  60019.384180 &    5137 &  62 &  3.94 &  0.22 &  0.04 &  0.08 &    32.4 &  1.0 & 0.20 & 0.05 & 0.00 & 0.00 & 0.08 & 0.07 & 0.01 \\
2023-03-24T08-36-32-601 &  60027.358710 &    5198 &  69 &  4.34 &  0.14 &  0.03 &  0.08 &    34.8 &  1.3 & 0.70 & 0.26 & 0.08 & 0.04 & 0.04 & 0.07 & 0.04 \\
2023-03-27T03-58-53-646 &  60030.165900 &    5145 &  62 &  4.24 &  0.17 &  0.06 &  0.08 &    32.8 &  1.1 & 0.40 & 0.16 & 0.03 & 0.01 & 0.05 & 0.07 & 0.06 \\
2023-04-13T07-24-13-013 &  60047.308480 &    5128 &  61 &  3.96 &  0.23 &  0.04 &  0.08 &   33.8 &  1.0 & 0.20 & 0.04 & 0.05 & 0.00 & 0.04 & 0.03 & 0.06 \\
2023-04-14T04-11-56-103 &  60048.174950 &    5125 &  60 &  4.04 &  0.22 &  0.04 &  0.08 &    31.5 &  0.9 & 0.20 & 0.04 & 0.04 & 0.00 & 0.05 & 0.06 & 0.05 \\
2023-04-15T09-23-50-205 &  60049.391550 &    5139 &  62 &  4.35 &  0.22 &  0.06 &  0.08 &   32.2 &  0.9 & 0.20 & 0.05 & 0.04 & 0.00 & 0.09 & 0.05 & 0.00 \\
2023-04-19T03-46-18-024 &  60053.157150 &    5156 &  65 &  3.99 &  0.22 &  0.02 &  0.08 &   31.2 &  1.1 & 0.30 & 0.04 & 0.06 & 0.00 & 0.08 & 0.06 & 0.06 \\
\hline
\end{tabular}
    }

    \footnotesize
    \tablefoot{The measured veiling at different wavelengths is reported in column labelled 'r' followed by the wavelength in nm.}
\end{sidewaystable}

\clearpage

\section{LSD profiles}

The LSD profiles of RY Lup across the observing campaign.

\begin{figure}[H] 
    \raggedright %

    \begin{subfigure}[b]{0.35\textwidth}
        \includegraphics[width=\textwidth]{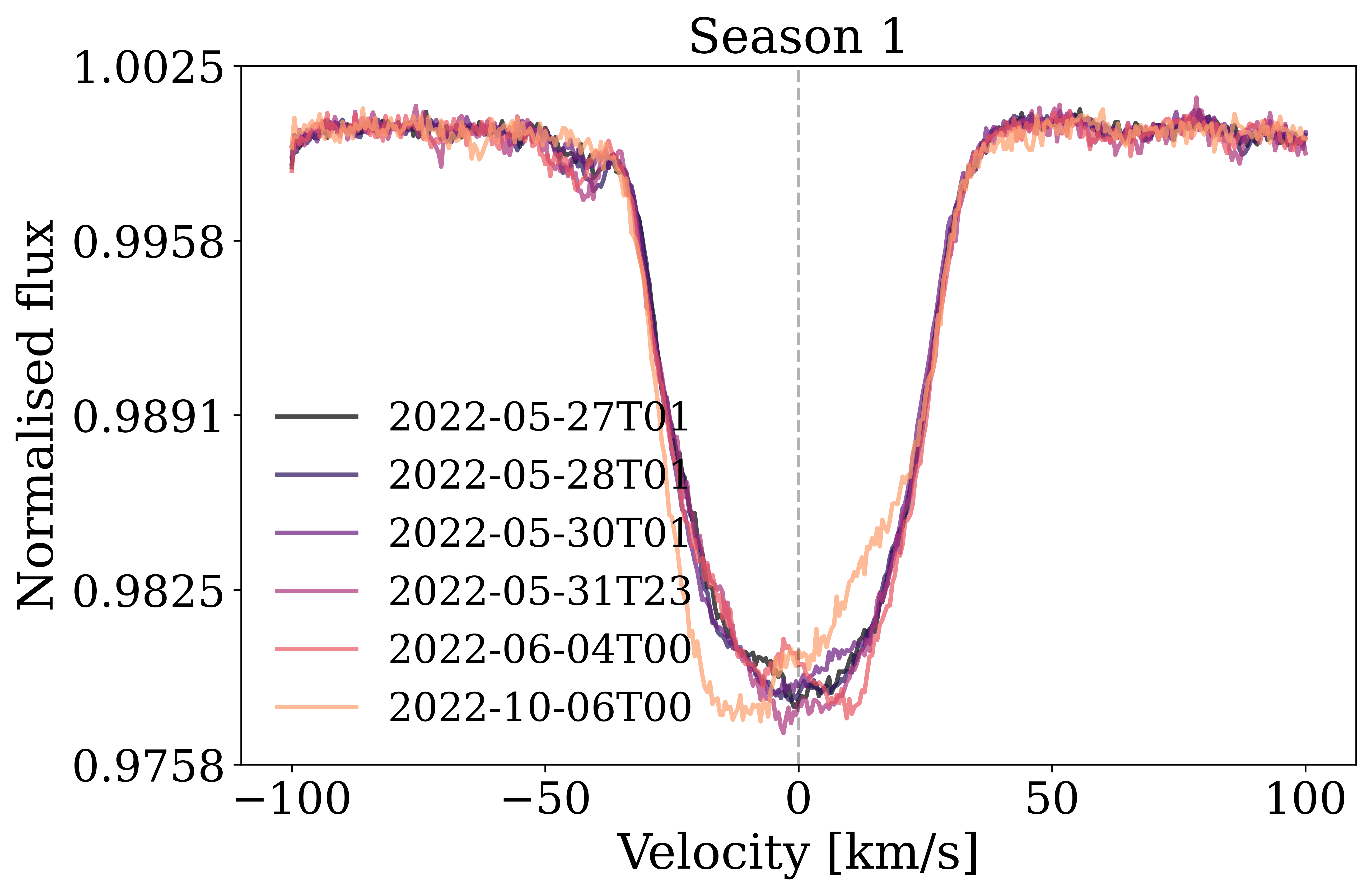}
        \label{subfig:LSD_may_june}
    \end{subfigure}%
    \hspace{10pt}%
    \begin{subfigure}[b]{0.35\textwidth}
        \includegraphics[width=\textwidth]{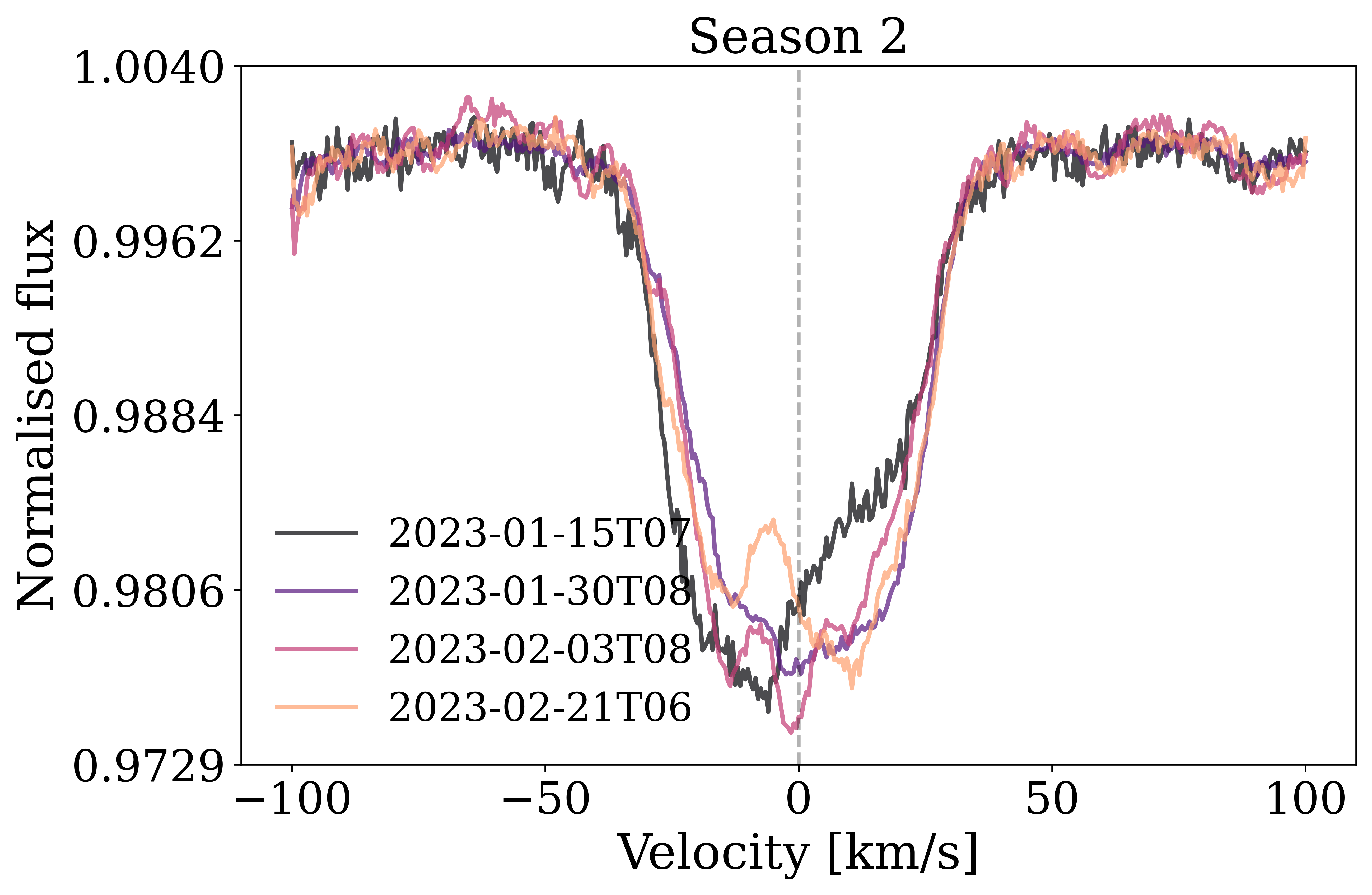}
        \label{subfig:LSD_jan_feb}
    \end{subfigure}

    \vspace{10pt} 

    \begin{subfigure}[b]{0.35\textwidth}
        \includegraphics[width=\textwidth]{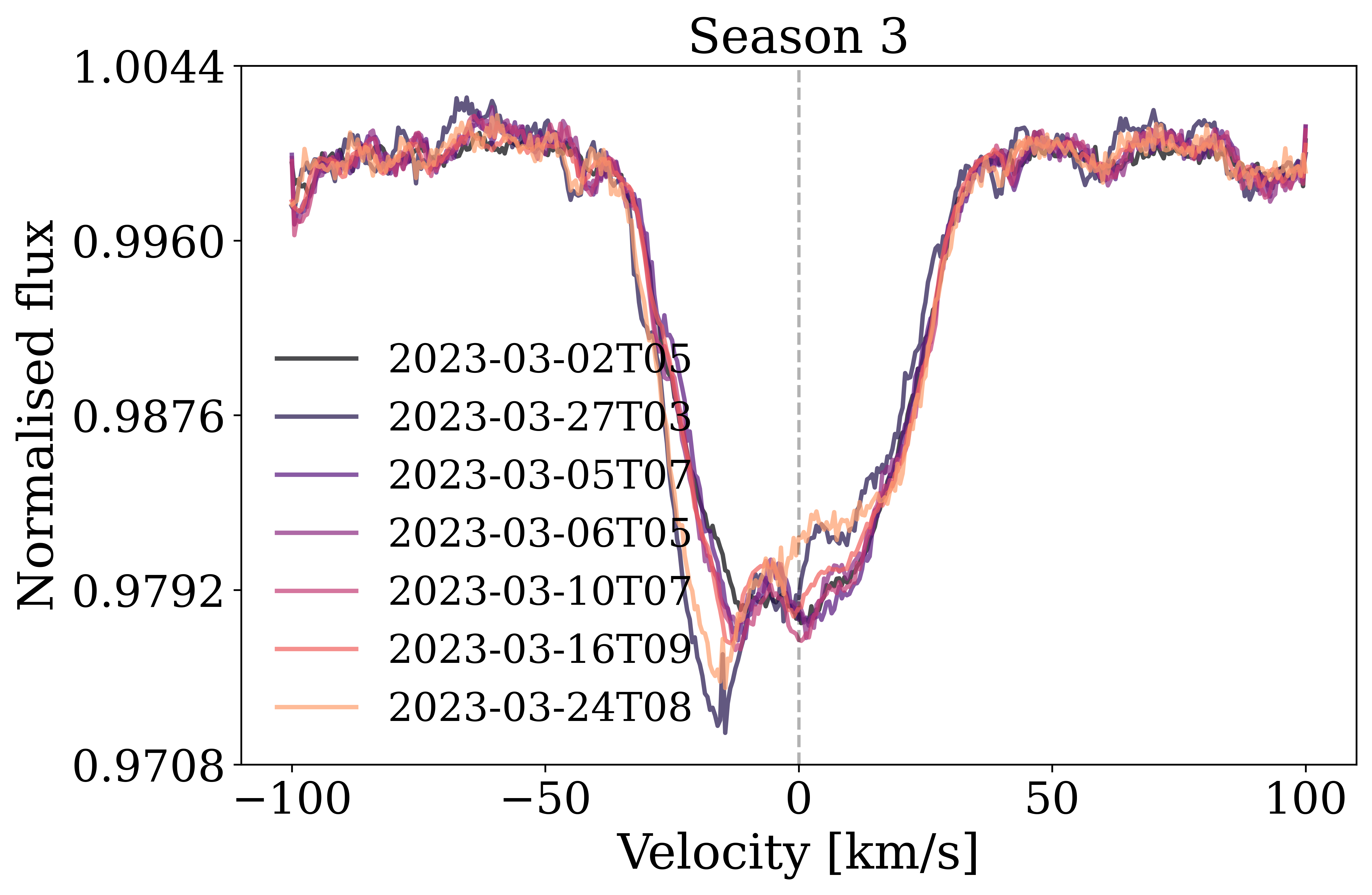}
        \label{subfig:LSD_march}
    \end{subfigure}%
    \hspace{10pt}%
    \begin{subfigure}[b]{0.35\textwidth}
        \includegraphics[width=\textwidth]{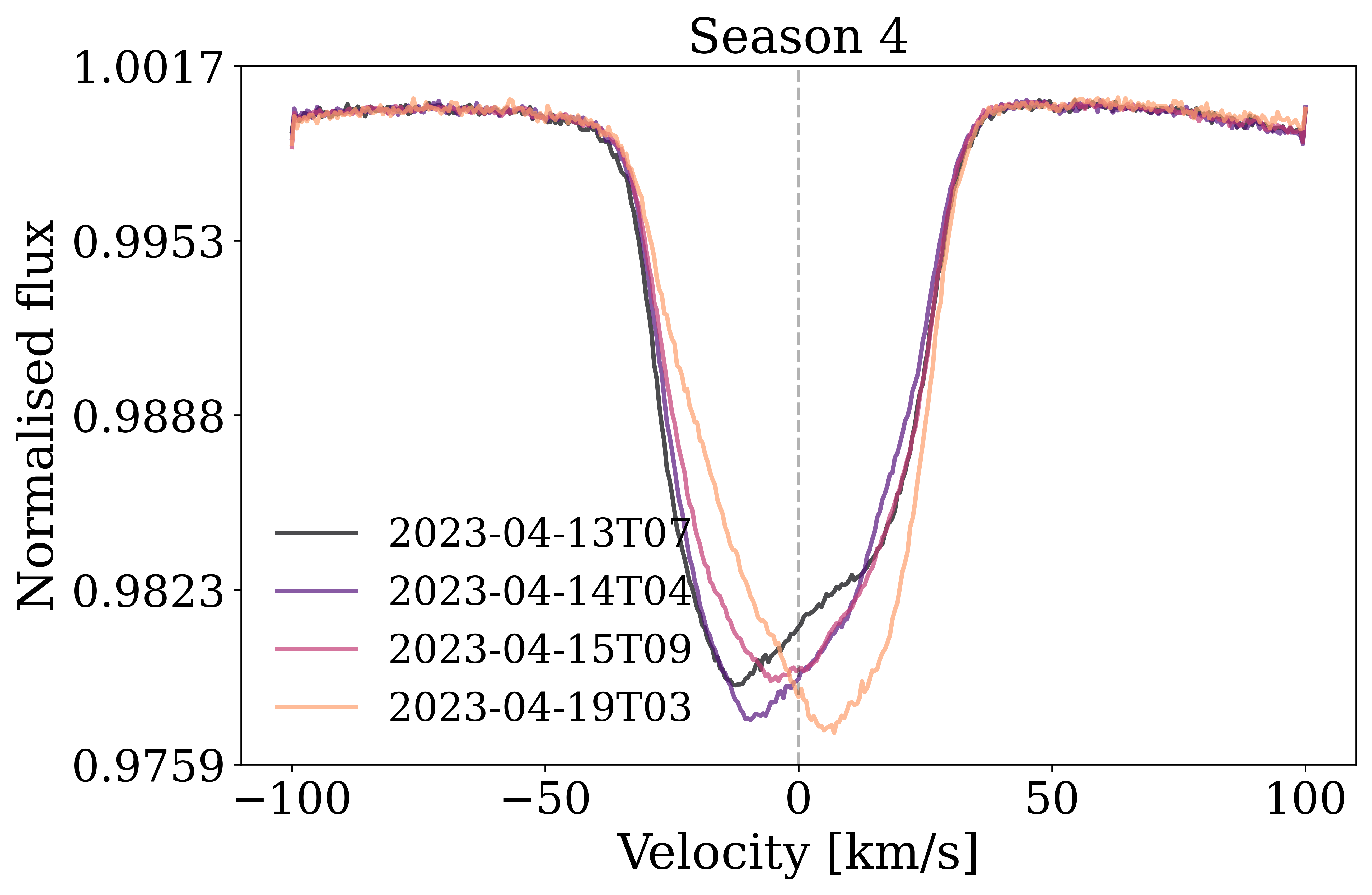}
        \label{subfig:LSD_april}
    \end{subfigure}

    \caption{LSD profiles of RY Lup across different observing seasons. Season 1 corresponds to 2022 May, June, and 'Out of season' epoch from 2022 October. Season 2 profiles are taken over 37 days in 2023 January and February. Season 3 are taken over 25 days in 2023 March. Season 4 corresponds to six consecutive profiles taken in 2023 April.}
    \label{fig:total_LSD_grid}
\end{figure}

\begin{figure}[htbp]
    \centering
  \begin{subfigure}[b]{0.35\textwidth}
        \centering
        \includegraphics[width=\textwidth]{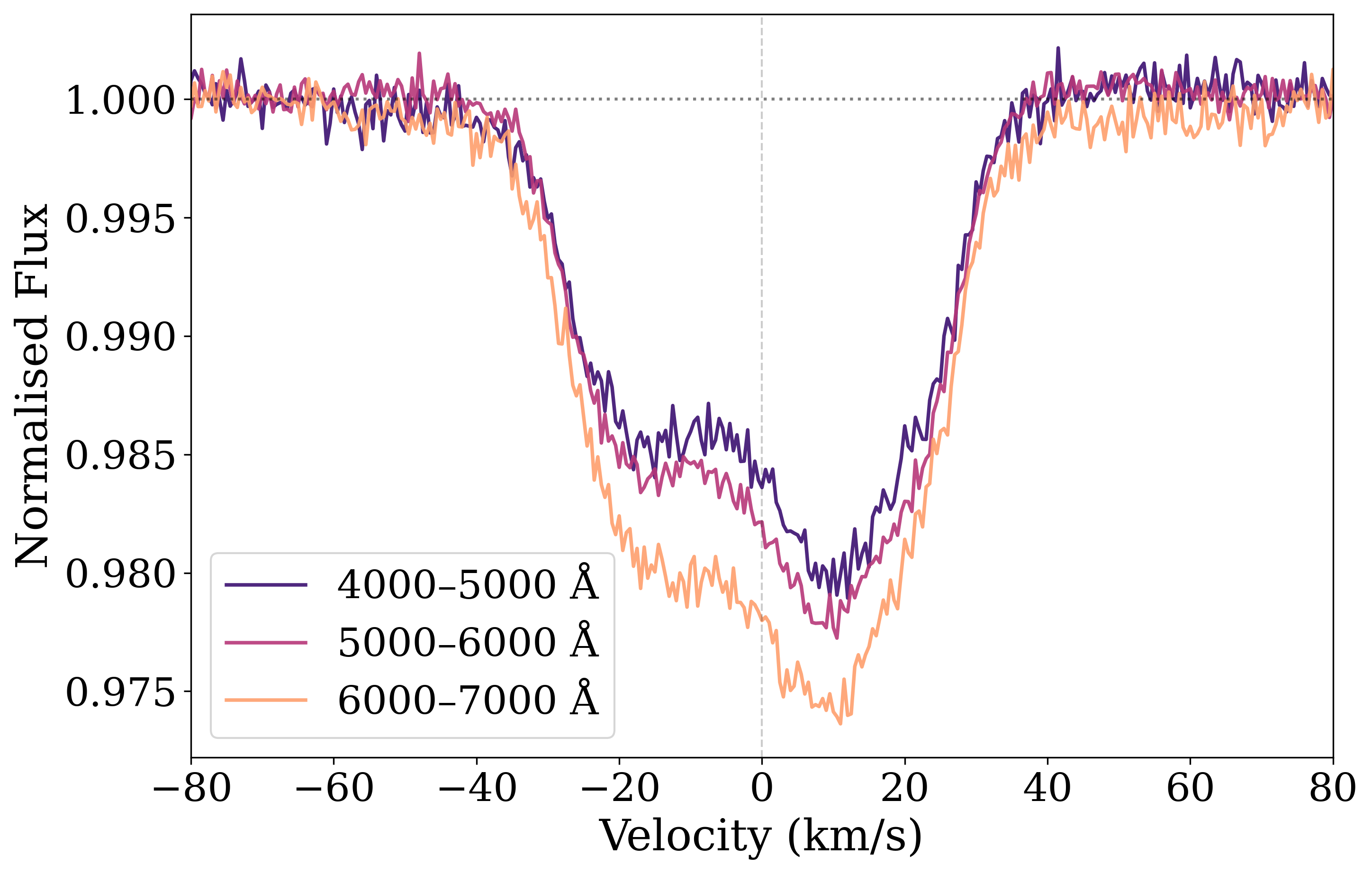}
    \end{subfigure}
    
    \begin{subfigure}[b]{0.35\textwidth}
        \centering
        \includegraphics[width=\textwidth]{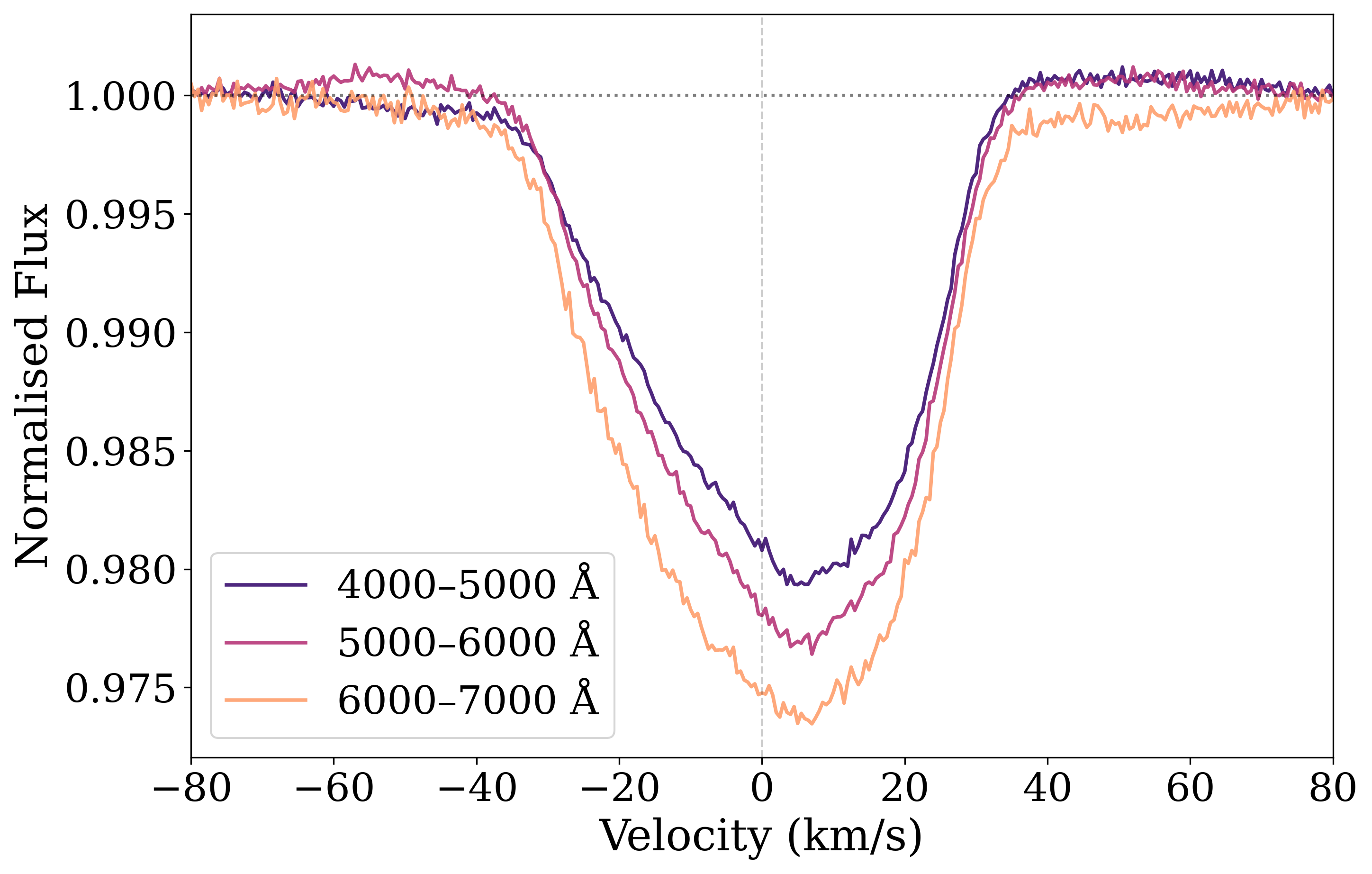}
    \end{subfigure}
      \caption{LSD profiles of RY Lup calculated across different wavelength ranges of ESPRESSO for the epochs 2023 February 21 and 2023 April 19.}
    \label{fig:total_comparison}
\end{figure}

\section{RV fitting}
The fit of the RV curve described in Sect.~\ref{RV calculation and fitting}
 is performed with an MCMC tool, which allows one to explore the correlation between the posterior distribution of the fitted values. This is shown in Fig.~\ref{fig:corner_plot_orig}. 

\begin{figure}
    \includegraphics[width=0.55\textwidth]{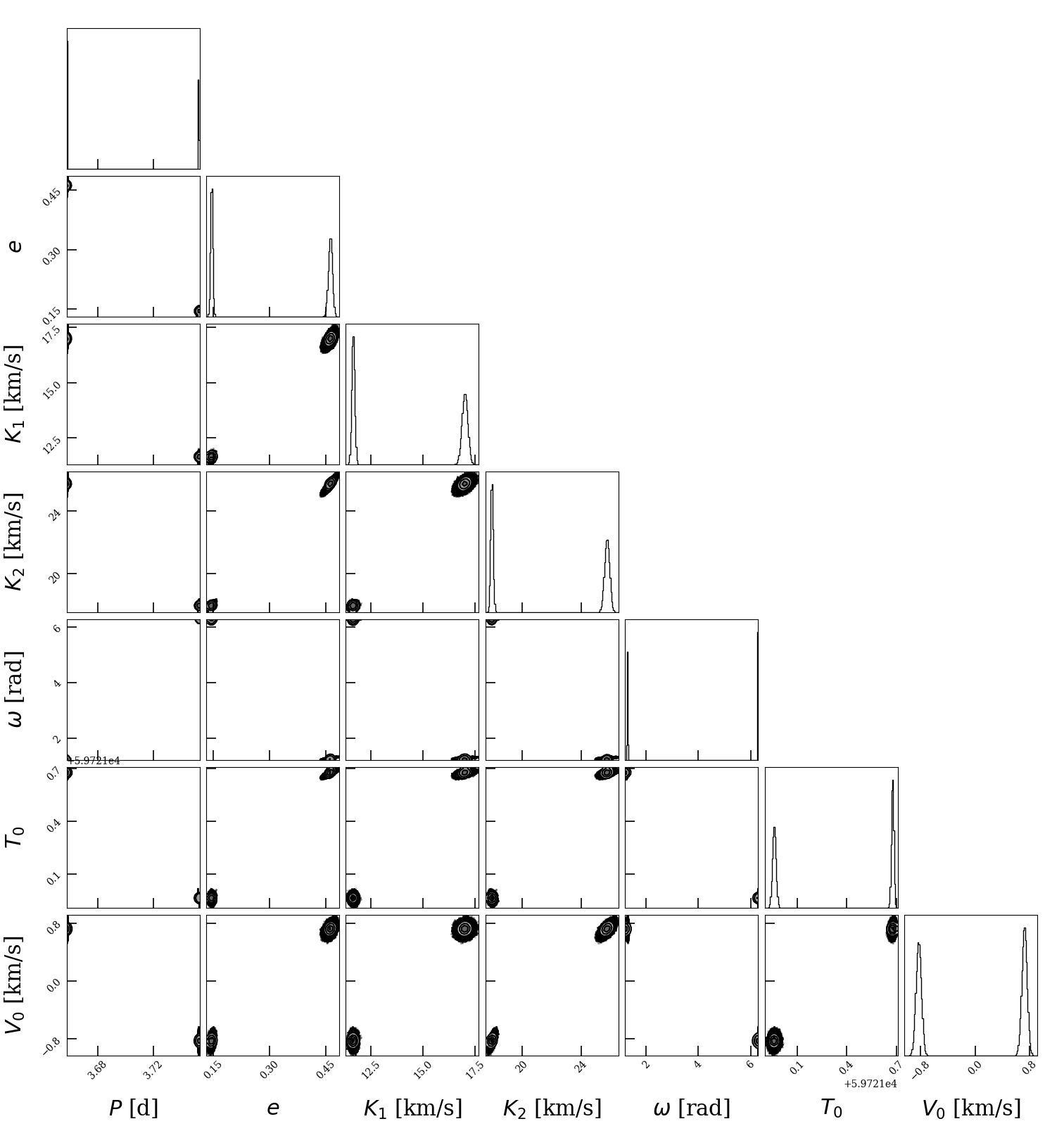} 
    \caption{Corner plot of the orbital solution for RY Lup shows bimodal posterior, which indicates two distinct set of solutions.}
    \label{fig:corner_plot_orig}
\end{figure}

 The fit of the RV curve described in Sect.~\ref{RV calculation and fitting}, which corresponds to the lower eccentricity solution (Fig.~\ref{fig:cleaned_corner}). 
The distributions are single-peaked, with Gaussian profiles. The orange lines in. It  highlights the location of the best likelihood estimates. The off-diagonal 2D plots represent the bivariate distributions between each pair of parameters which provide an immediate estimate of their correlation. The argument of periastron ($\omega$) distribution is not well constrained.

\begin{figure}[h!]
    \centering
    \includegraphics[width=0.57\textwidth]{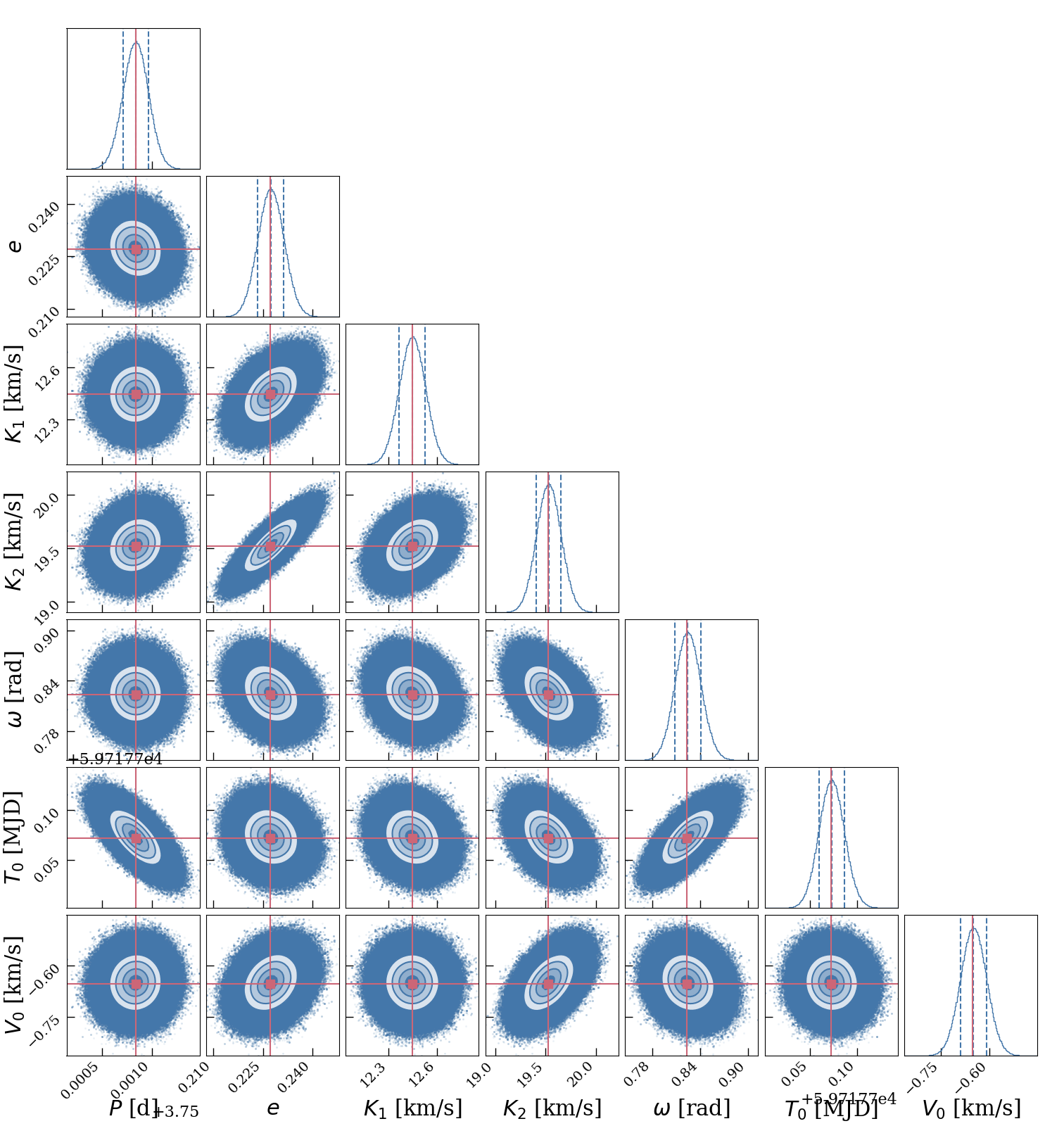} 
    \caption{ Corner plot of the orbital solution for RY Lup adapted in this work. The vertical dashed lines indicating the 16th, 50th and 84th percentile. The vertical red lines correspond to the best likelihood estimates of each distribution.}
    \label{fig:cleaned_corner}
\end{figure}

\clearpage

\section{log of $R'_{\text{HK}}$ index measurements}

In this section, we provide the values of  equivalent\\ width of Ca H \& K lines, integrated fluxes, and $R'_{\text{HK}}$\\ index measurements.
\label{log_rhk}

\begin{sidewaystable}[p]
    \centering
    \caption{Log of the equivalent width (EW) measurements of both the Ca H\&K line, the derived fluxes a the stellar surface ($F_{k}$,$F_{H}$), $R'_{\text{HK}}$ index, the error $e_{R'_{\text{HK}}}$,\\ and the veiling (${r}$). }
    \label{logrhk_table}
    
    \setlength{\tabcolsep}{1.9pt} 

    \renewcommand{\arraystretch}{0.75}

    \tiny
    
    \resizebox{0.8\textwidth}{!}{%
        \begin{tabular}{lccccccc}
\hline 
\hline
DATE-OBS & MJD & $EW_{K}$ & $EW_{H}$ & $F_{K} / 10^6$ & $F_{H} / 10^6$ & $R'_{\mathrm{HK}}$ & $r$ \\
 & & (\AA) & (\AA) & ($\text{erg cm}^{-2}\text{ s}^{-1}$) & ($\text{erg cm}^{-2}\text{ s}^{-1}$) & & \\
\hline
2022 May 27     & 59726.04793 & $-1.88 \pm 0.26$ & $-1.20 \pm 0.24$ & $5.06 \pm 0.84$  & $3.20 \pm 0.72$  & $-3.69 \pm 0.08$ & 0.3 \\
2022 May 28     & 59727.04719 & $-1.67 \pm 0.22$ & $-1.30 \pm 0.23$ & $4.50 \pm 0.72$  & $3.45 \pm 0.70$  & $-3.70 \pm 0.07$ & 0.2 \\
2022 May 30     & 59729.05996 & $-2.26 \pm 0.26$ & $-1.85 \pm 0.26$ & $6.08 \pm 0.91$  & $4.97 \pm 0.85$  & $-3.56 \pm 0.06$ & 0.0 \\
2022 May 31     & 59730.98834 & $-3.01 \pm 0.64$ & $-2.55 \pm 0.58$ & $8.1 \pm 1.9$    & $6.8 \pm 1.7$    & $-3.43 \pm 0.09$ & 0.4 \\
2022 June 4     & 59734.02920 & $-1.44 \pm 0.40$ & $-1.40 \pm 0.42$ & $3.9 \pm 1.1$    & $3.7 \pm 1.2$    & $-3.72 \pm 0.11$ & 0.0 \\
2022 October 6  & 59858.00158 & $-3.80 \pm 0.90$ & $-3.19 \pm 0.96$ & $10.2 \pm 2.6$   & $8.6 \pm 2.7$    & $-3.33 \pm 0.11$ & 0.5 \\
2023 January 15 & 59959.33324 & $-3.25 \pm 3.66$ & $-2.81 \pm 3.20$ & $8.8 \pm 9.9$    & $7.6 \pm 8.7$    & $-3.39 \pm 0.33$ & 0.8 \\
2023 January 30 & 59974.35934 & $-2.19 \pm 0.68$ & $-1.60 \pm 0.65$ & $5.8 \pm 1.9$    & $4.2 \pm 1.8$    & $-3.60 \pm 0.14$ & 0.4 \\
2023 February 3 & 59978.35017 & $-2.96 \pm 0.61$ & $-2.50 \pm 0.65$ & $8.0 \pm 1.8$    & $6.6 \pm 1.9$    & $-3.44 \pm 0.10$ & 0.5 \\
2023 February 21& 59996.25115 & $-4.45 \pm 1.20$ & $-3.0 \pm 1.08$  & $11.9 \pm 3.4$   & $8.0 \pm 3.0$    & $-3.31 \pm 0.12$ & 0.7 \\
2023 March 2    & 60005.22714 & $-2.83 \pm 0.43$ & $-2.26 \pm 0.40$ & $7.6 \pm 1.4$    & $6.1 \pm 1.2$    & $-3.47 \pm 0.07$ & 0.4 \\
2023 March 5    & 60008.30703 & $-5.63 \pm 1.36$ & $-4.80 \pm 1.24$ & $15.1 \pm 3.9$   & $12.8 \pm 3.5$   & $-3.16 \pm 0.10$ & 0.7 \\
2023 March 6    & 60009.22344 & $-2.47 \pm 1.0$  & $-2.40 \pm 1.32$ & $6.6 \pm 2.7$    & $6.3 \pm 3.6$    & $-3.49 \pm 0.17$ & 0.7 \\
2023 March 10   & 60013.30837 & $-2.27 \pm 0.43$ & $-2.17 \pm 0.5403$& $6.1 \pm 1.3$    & $5.8 \pm 1.6$    & $-3.53 \pm 0.09$ & 0.3 \\
2023 March 16   & 60019.38418 & $-1.93 \pm 0.24$ & $-1.55 \pm 0.2328$& $5.20 \pm 0.81$  & $4.17 \pm 0.74$  & $-3.63 \pm 0.07$ & 0.2 \\
2023 March 24   & 60027.35871 & $-2.91 \pm 1.20$ & $-2.0 \pm 1.4745$ & $7.8 \pm 3.2$    & $5.4 \pm 4.0$    & $-3.48 \pm 0.19$ & 0.7 \\
2023 March 27   & 60030.16590 & $-4.20 \pm 0.92$ & $-3.74 \pm 1.0$  & $11.4 \pm 2.7$   & $10.1 \pm 2.8$   & $-3.27 \pm 0.10$ & 0.4 \\
2023 April 13   & 60047.30848 & $-2.44 \pm 0.21$ & $-1.90 \pm 0.21$ & $6.58 \pm 0.84$  & $5.10 \pm 0.75$  & $-3.54 \pm 0.06$ & 0.2 \\
2023 April 14   & 60048.17495 & $-2.08 \pm 0.25$ & $-1.80 \pm 0.30$ & $5.63 \pm 0.87$  & $4.84 \pm 0.92$  & $-3.58 \pm 0.07$ & 0.2 \\
2023 April 15   & 60049.39155 & $-3.245 \pm 0.31$& $-2.70 \pm 0.27$ & $8.7 \pm 1.2$    & $7.21 \pm 0.99$  & $-3.40 \pm 0.06$ & 0.2 \\
2023 April 19   & 60053.15715 & $-2.21 \pm 0.40$ & $-1.90 \pm 0.35$ & $5.9 \pm 1.2$    & $5.0 \pm 1.0$    & $-3.57 \pm 0.08$ & 0.3 \\
\hline
\end{tabular}

    }

    \footnotesize

    \tablefoot{$EW_K$ and $EW_H$ are  in \AA; $F_{K}$, $F_{H}$ are in units of $\text{erg cm}^{-2}\text{ s}^{-1}$.}
\end{sidewaystable}

\clearpage

\section{Stellar spots}\label{appendix:spots}

Examples of different stellar spots models over different times of RY Lup observational campaign.

 \begin{figure}[htbp]
     \centering
     \begin{subfigure}[b]{0.48\textwidth}
         \centering
         \includegraphics[width=\textwidth]{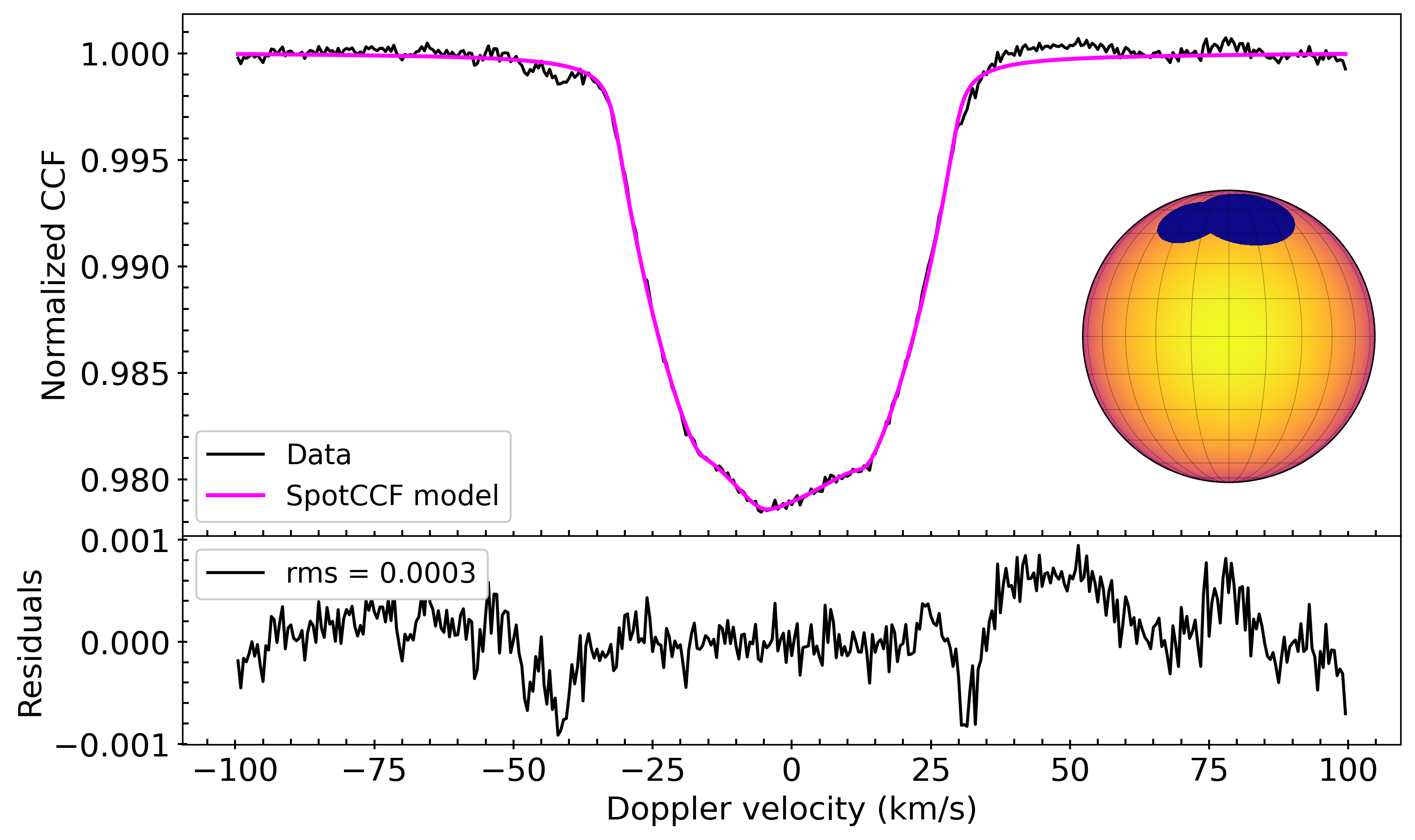}
         \label{fig:example_season1}
     \end{subfigure}
     \hfill 
     \begin{subfigure}[b]{0.48\textwidth}
         \centering
         \includegraphics[width=\textwidth]{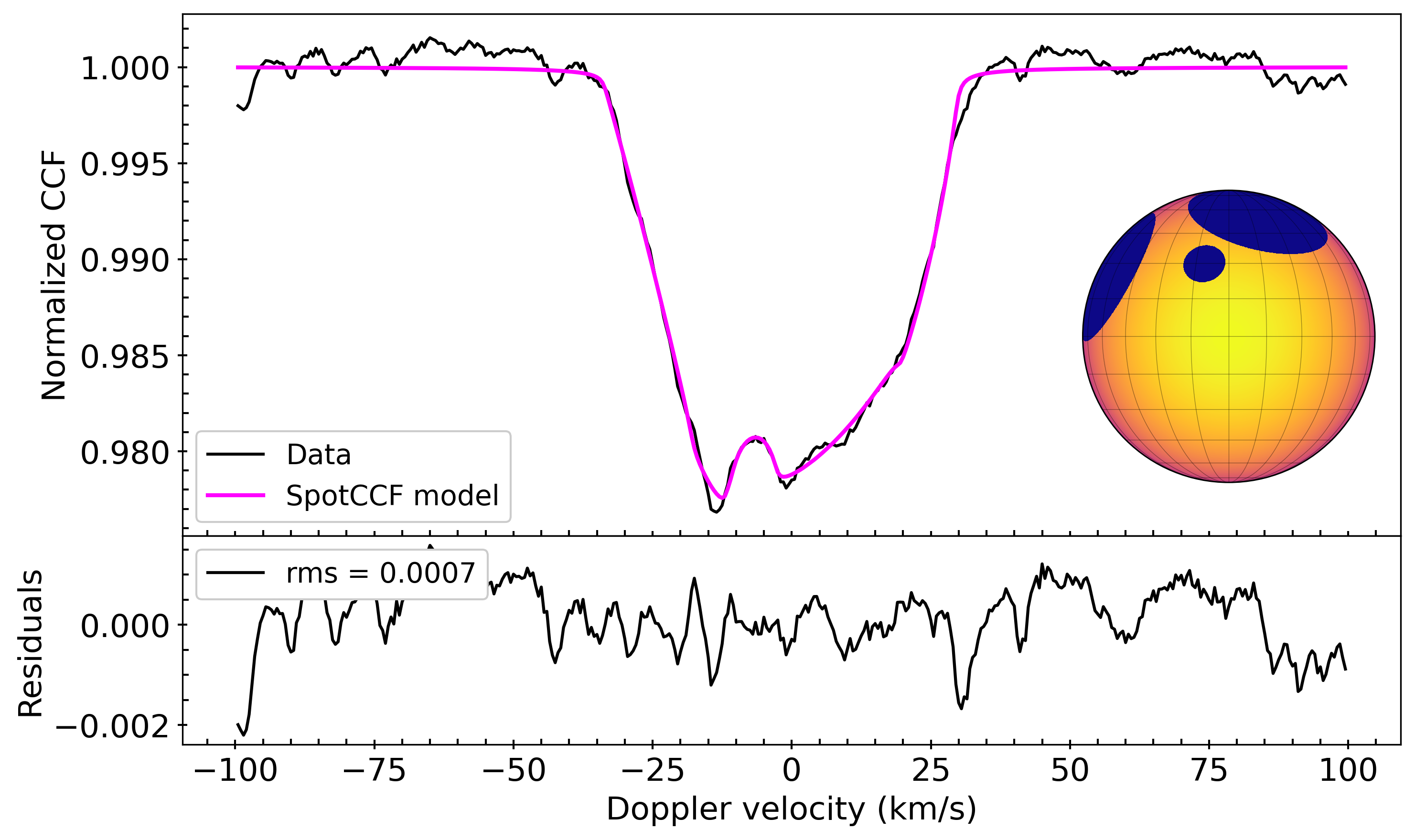}
         \label{fig:season3}
     \end{subfigure}
     
     \vspace{0.5cm} 
     
     \begin{subfigure}[b]{0.48\textwidth}
         \centering
         \includegraphics[width=\textwidth]{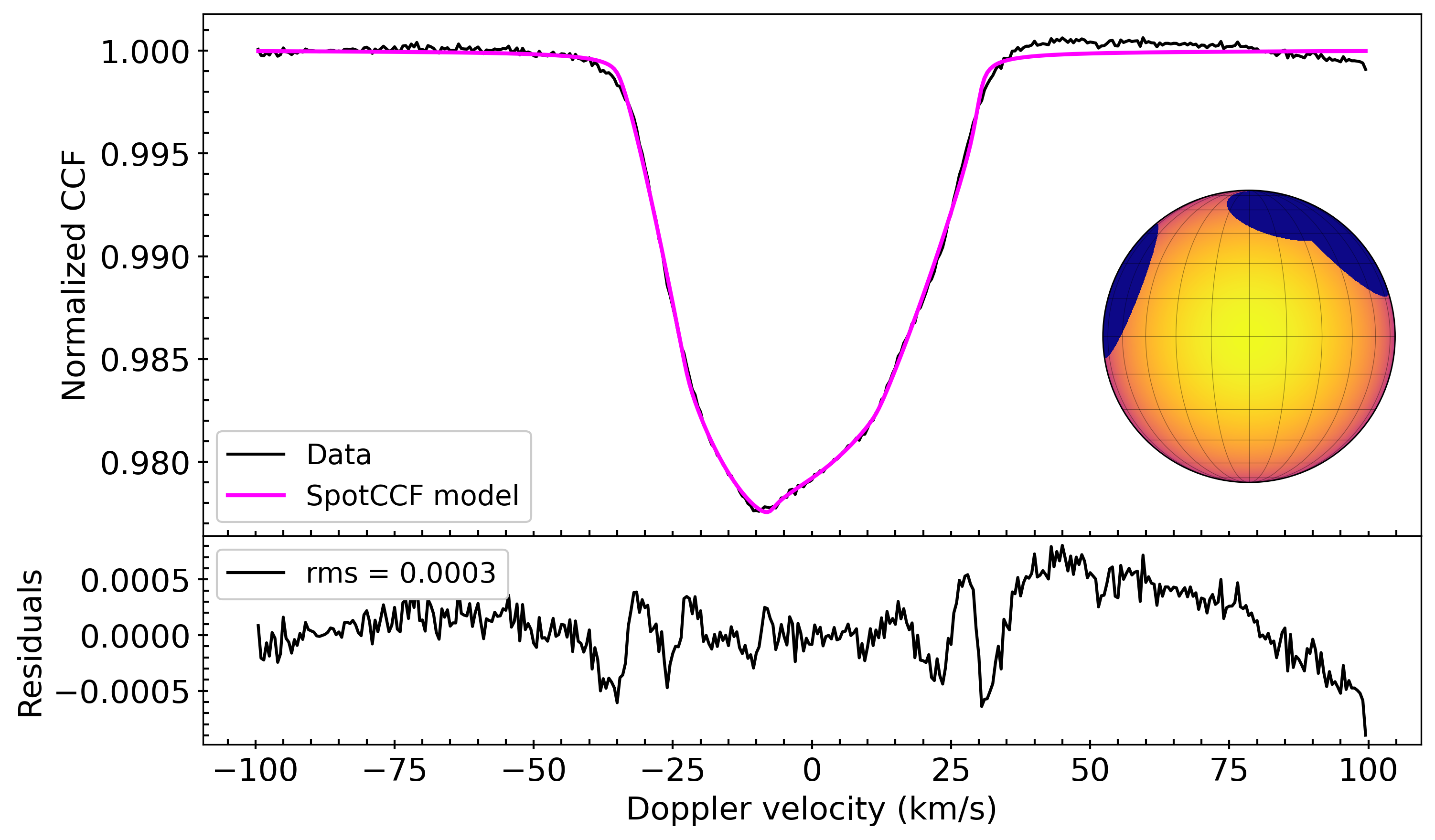}
         \label{fig:season4}
     \end{subfigure}
     
     \caption{LSD profiles of RY Lup modelled with \texttt{SpotCCF}. (Top panel) LSD profile of 2022 May 30, (middle panel) LSD profile of 2023 March 16, and (bottom panel) the LSD profile of 2023 April 14. The displayed spot configurations represent the best-fit models according to the Bayesian evidence, selected from a comparison of models ranging from 0 to 3 spots. In each subplot, the top panel shows the observed LSD profile (blue) and the \texttt{SpotCCF} model (orange), while the bottom panel displays the residuals.}
     \label{fig:Example_SpotCCF_extra}
\end{figure}

In Fig.\ref{sys_rv_res}, we do not observe a correlation between the RV residuals from the potential binary model fitted to the RV data in Sect.\ref{rv_analysis} and the systemic RV measured in Sect.\ref{Spot modelling}. 
 \begin{figure}
    \centering
\includegraphics[trim=0 0 0 0 , clip,width=\linewidth]{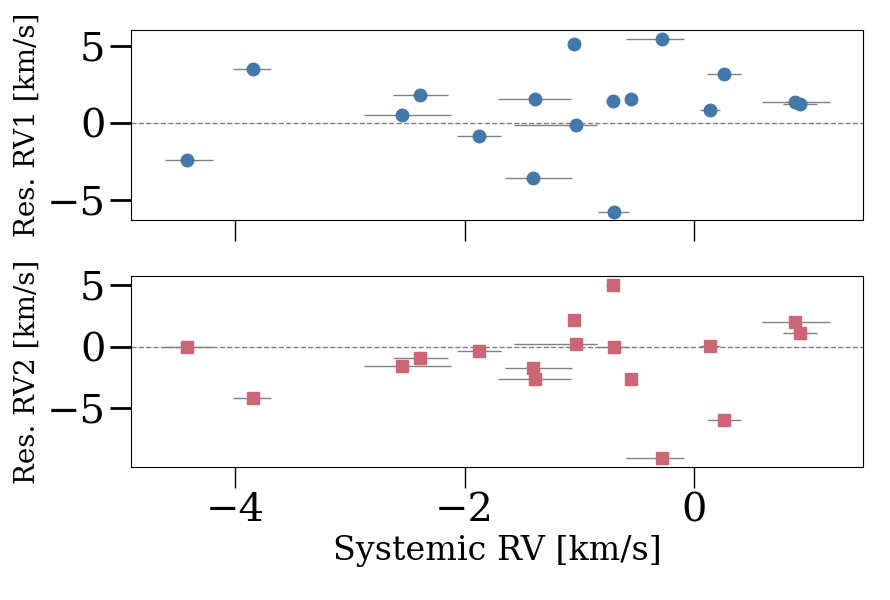}
      \caption{Residuals of the LSD RV measurements as a function of systemic velocity.}
\label{sys_rv_res}
\end{figure}

In Fig.\ref{fig:spot_radii_long}, we show the distribution of the stellar spots configurations using the longitude and stellar radius across the observing seasons. Season 3 shows the highest longitudinal scatter,marking a period of increased activity. The spot radii remain largely confined to the $0.3\text{--}0.5\,R_*$ range across the observing seasons.

\begin{figure}
     \centering
 \includegraphics[width=\linewidth]{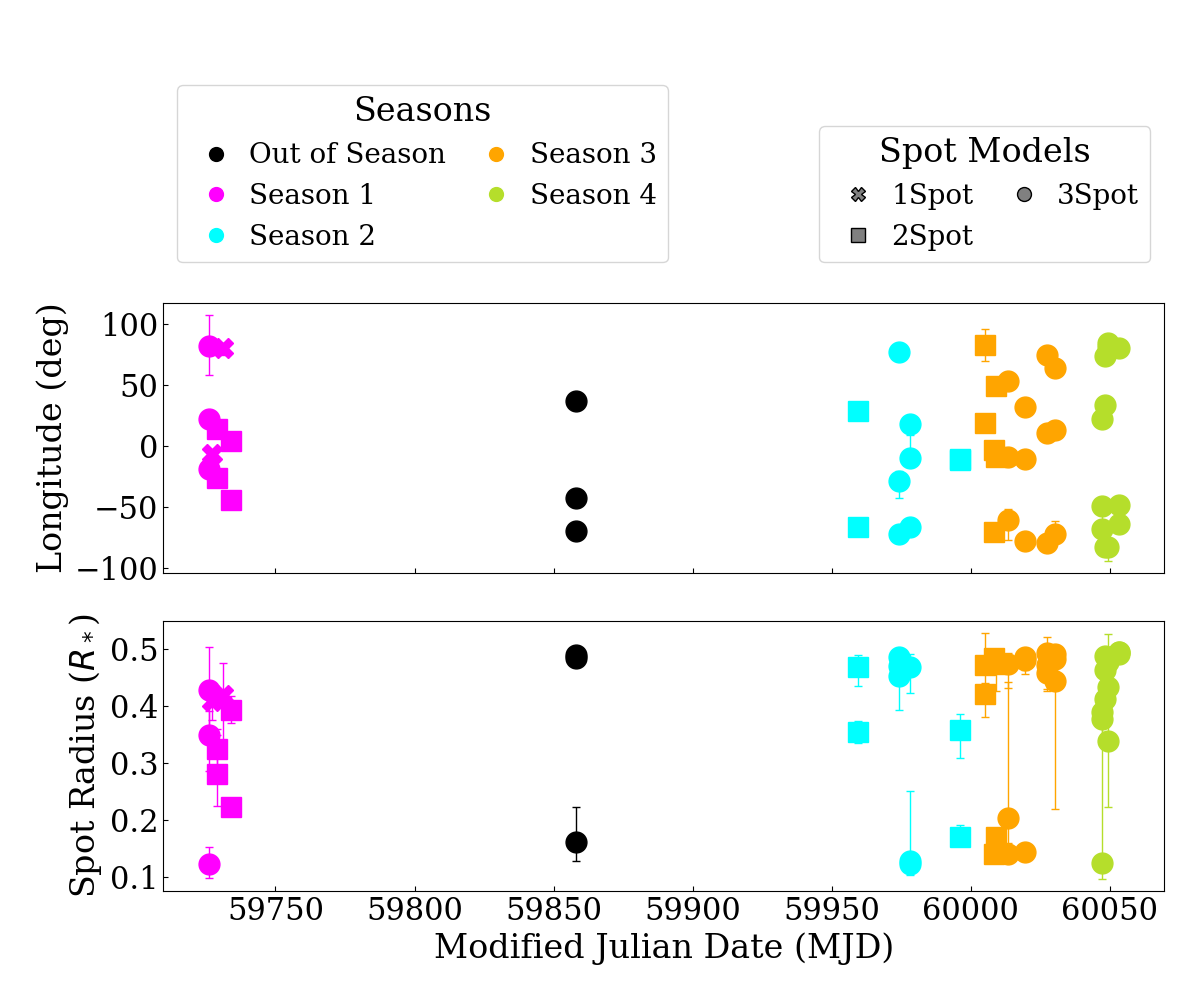}
 \vspace{-10pt}
       \caption{Temporal evolution of the spot longitudes (top panel) across the time, and the spot radius (bottom panel) derived from fitting the LSD profiles of RY Lup with \texttt{SpotCCF}. Marker colours denote the different observing seasons. The 'Out of Season' data point (black circle) represents the isolated observation taken on October 6, 2022. Marker shapes indicate the optimal model configuration selected by Bayesian evidence: crosses, squares, and circles represent 1-, 2-, and 3-spot configurations, respectively.
       }
 \label{fig:spot_radii_long}
 \end{figure}

\clearpage

\section{Log of spot models parameters}

In this section, we provide the systemic RV, the $v \sin i$, their associated errors, and the total spot filling factor values, derived in Sect.\ref{Spot modelling}.

\begin{sidewaystable}[p]
    \centering
    \setlength{\tabcolsep}{0.5pt} 
    \renewcommand{\arraystretch}{0.3}
    
    \tiny 

     \caption{Log of the parameters of the spot models fitted to the LSD profiles in Sect.\ref{Spot modelling}.}
    
    \resizebox{0.9\textwidth}{!}{%
        \begin{tabular}{lcccccccccc}
\hline
\hline
Observation date & MJD & Season & Model & $V_{0}$ & $RV_\mathrm{err,low}$ & $RV_\mathrm{err,high}$ & $v \sin i$ & $v \sin i_\mathrm{err,low}$ & $v \sin i_\mathrm{err,high}$ & $ff_\mathrm{p,tot}$ \\
& & & selected & (km s$^{-1}$) & (km s$^{-1}$) & (km s$^{-1}$) & (km s$^{-1}$) & (km s$^{-1}$) & (km s$^{-1}$) & \\
 \hline  
2022 May 27 & 59726.047930 &  Season 1 &	3Spot & -0.0497 &	0.0983 & 0.1020 & 32.0352 &	0.1625 & 0.1520 & 0.0363\\
2022 May 28 & 59727.047190 &	Season 1 &	1Spot &	-0.5438 &	0.0528 &	0.0523 & 31.1849 &	0.1082 & 0.1072 & 0.0432\\
2022 May 30 & 59729.059960 &	Season 1 &	2Spot &	-1.0467 &	0.0361 & 0.0362 &	30.8315 & 0.0759 &	0.0774 & 0.0705\\
2022 May 31 & 59730.988340 &	Season 1 &	1Spot &	0.2622 &	0.1496 &	0.1488 &	32.8467 &	0.2164 &	0.2192 &	0.0246\\
2022 June 4 & 59734.029200 &	Season 1 &	2Spot &	-0.7063 &	0.0579 &	0.0580 &	31.4492 &	0.0939 &	0.0957 &	0.0848\\
2022 October 6 & 59858.001580 &	Out of Season &	3Spot &	-4.4222 &	0.1960 & 0.2263 &	35.6134 &	0.2845 &	0.2500 &	0.1404\\
2023 January 15 & 59959.333240 &	Season 2 &	2Spot &	-4.4973 &	0.3771 &	0.3843 &	34.9998 &	0.4988 &	0.4831 &	0.1526\\
2023 January 30 & 59974.359340 &	Season 2 &	3Spot &	0.8863 &	0.2896 &	0.3029 &	34.5710 &	0.4164 &	0.3972 &	0.0970\\
2023 February 3 & 59978.350170 &	Season 2 &	3Spot &	-2.5501 &	0.3316 &	0.4317 &	32.4699 &	0.5567 &	0.4447 &	0.0907\\
2023 February 21 & 59996.251150 &	Season 2 &	2Spot &	0.1450 &	0.0866 &	0.0871 &	31.8030 &	0.1535 &	0.1455 &	0.0919\\
2023 March 2 & 60005.227140 &	Season 3 &	2Spot &	-0.2783 &	0.3130 &	0.1900 &	31.4112 &	0.2375 &	0.4024 &	0.0507\\
2023 March 5 & 60008.307030 &	Season 3 &	2Spot &	-2.3861 &	0.2384 &	0.2397 &	32.2968 &	0.3330 &	0.3278 &	0.0791\\
2023 March 6 & 60009.223440 &	Season 3 &	2Spot &	-0.6995 &	0.1323 &	0.1321 &	30.5976 &	0.1886 &	0.2017 &	0.0876\\
2023 March 10 & 60013.308370 &	Season 3 &	3Spot &	-1.0305 &	0.5423 &	0.1878 &	31.0799 &	0.2468 &	0.7128 &	0.0641\\
2023 March 16 & 60019.384180 &	Season 3 &	3Spot &	-1.8767 &	0.1914 &	0.1920 &	31.6370 &	0.2594 &	0.3012 &	0.1073\\
2023 March 24 & 60027.358710 &	Season 3 &	3Spot &	-1.3855 &	0.3201 &	 0.3134 &	34.0704 &	0.6035 &	0.5781 &	0.2615\\
2023 March 27 & 60030.165900  &	Season 3 &	3Spot &	-2.7556 &	0.4891 &	0.8168 &	33.4054 &	1.2748 &	0.6650 &	0.2905\\
2023 April 13 & 60047.308480 & 	Season 4 &	3Spot &	-3.8424 &	0.1742 &	0.1544 &	34.4881 &	0.1880 &	0.2177 &	0.1039\\
2023 April 14 & 60048.174950 &	Season 4 &	3Spot &	-2.2082 &	0.1575 &	0.1739 &	32.4494 &	0.2157 &	0.1888 &	0.1064\\
2023 April 15 & 60049.391550 &	Season 4 &	3Spot &	-1.4048 &	0.2439 &	0.3416 &	33.6077 &	0.3280 &	0.4755 &	0.0438\\
2023 April 19 & 60053.157150 &	Season 4 &	3Spot &	0.9279 &	0.1457 &	0.1458 &	34.0722 &	0.2188 &	0.1919 &	0.1113\\

\hline
\end{tabular}

    }

    \label{spot_param}

    \footnotesize

\end{sidewaystable}

\clearpage

\section{Log of the parameters of spots}

In this section, we provide the radius and latitude, and the associated errors of each parameter for each modelled spot per epoch.

\begin{sidewaystable}[p]
    \centering
    \caption{Fitted stellar spots models properties.}
    \label{spot_param}

    \setlength{\tabcolsep}{1.pt} 
    \scriptsize    
    \setlength{\tabcolsep}{0.5pt}

    \resizebox{0.8\textwidth}{!}{%

\begin{tabular}{lccccccccccccccccccccc}
\hline
\hline
\shortstack{Obs \\ Date} & 
\shortstack{Model \\ Sel.} & 
MJD & 
\shortstack{$\text{L}_{1}$ \\ Spot} & 
\shortstack{$\text{L}_{1}$ \\ {\tiny -err}} & 
\shortstack{$\text{L}_{1}$ \\ {\tiny +err}} & 
\shortstack{$R_{1}$ \\ Spot} & 
\shortstack{$R_{1}$ \\ {\tiny -err}} & 
\shortstack{$R_{1}$ \\ {\tiny +err}} & 
\shortstack{$\text{L}_{2}$ \\ Spot} & 
\shortstack{$\text{L}_{2}$ \\ {\tiny -err}} & 
\shortstack{$\text{L}_{2}$ \\ {\tiny +err}} & 
\shortstack{$R_{2}$ \\ Spot} & 
\shortstack{$R_{2}$ \\ {\tiny -err}} & 
\shortstack{$R_{2}$ \\ {\tiny +err}} & 
\shortstack{$\text{L}_{3}$ \\ Spot} & 
\shortstack{$\text{L}_{3}$ \\ {\tiny -err}} & 
\shortstack{$\text{L}_{3}$ \\ {\tiny +err}} & 
\shortstack{$R_{3}$ \\ Spot} & 
\shortstack{$R_{3}$ \\ {\tiny -err}} & 
\shortstack{$R_{\mathrm{s},3}$ \\ {\tiny +err}} \\
\hline

2022 May 27&3Spot&59726.047930&22.21&5.70&5.42&0.43&0.10&0.10&81.40&23.10&25.50&0.34&0.06&0.042&-19.04&2.90&3.11&0.14&0.03&0.03\\
2022 May 28 &1Spot&59727.047190&-6.90&3.43&3.50&0.41&0.032&0.033&-&-&-&-&-&-&-&-&-&-&-&-\\
2022 May 30&2Spot&59729.059960&14.2&1.40&1.90&0.32&0.032&0.03&-26.20&2.10&3.50&0.30&0.10&0.10&-&-&-&-&-&-\\
2022 May 31&1Spot&59730.988340&80.30&6.30&4.23&0.42&0.10&0.06&-&-&-&-&-&-&-&-&-&-&-&-\\
2022 June 4&2Spot&59734.029200&-43.90&4.0&3.31&0.40&0.02&0.024&3.72&0.80&0.71&0.22&0.01&0.013&-&-&-&-&-&-\\
2022 October 4&3Spot&59858.001580&36.83&1.50&1.51&0.50&0.01&0.01&-69.20&1.80&1.41&0.50&0.02&0.01&-42.90&7.32&3.70&0.16&0.03&0.062\\
2023 January 15&2Spot&59959.333240&-66.0&2.50&2.0&0.47&0.03&0.022&28.73&1.77&2.0&0.35&0.02&0.02&-&-&-&-&-&-\\
2023 January 30&3Spot&59978.350170&-28.71&13.54&8.46&0.50&0.02&0.01&-71.82&1.83&2.0&0.50&0.03&0.02&77.03&3.50&2.50&0.45&0.06&0.03\\
2023 February 3 &3Spot&59978.350170&-66.0&2.70&3.41&0.50&0.05&0.02&18.07&1.44&1.0&0.13&0.01&0.013&-9.81&2.81&19.0&0.12&0.02&0.13\\
2023 February 21&2Spot&59996.251150&-11.55&2.82&5.15&0.36&0.05&0.03&-10.64&1.40&0.80&0.20&0.01&0.02&-&-&-&-&-&-\\
2023 March 2&2Spot&60005.227140&83.0&13.04&13.20&0.50&0.04&0.05&19.0&3.50&4.05&0.42&0.04&0.02&-&-&-&-&-&-\\
2023 March 5&2Spot&60008.307030&-70.61&2.33&2.40&0.50&0.02&0.01&-3.40&0.85&0.80&0.14&0.01&0.01&-&-&-&-&-&-\\
2023 March 6&2Spot&60009.223440&49.20&5.0&5.24&0.5&0.05&0.02&-8.80&1.30&1.17&0.20&0.02&0.02&-&-&-&-&-&-\\
2023 March 10 &3Spot&60013.308370&53.0&5.50&7.0&0.50&0.042&0.02&-60.54&16.30&9.0&0.20&0.05&0.24&-8.82&1.30&1.35&0.14&0.01&0.015\\
2023 March 16 &3Spot&60019.384180&31.62&2.03&2.17&0.50&0.03&0.014&-78.0&3.0&3.14&0.50&0.02&0.01&-11.05&1.10&1.0&0.14&0.01&0.01\\
2023 March 24&3Spot&60027.358710&11.0&2.34&3.15&0.50&0.06&0.03&74.30&1.20&1.10&0.50&0.01&0.01&-79.63&2.10&2.11&0.46&0.03&0.02\\
2023 March 27&3Spot&60030.165900&12.84&3.40&2.30&0.50&0.01&0.01&64.0&4.0&4.2&0.5&0.02&0.01&-72.35&3.40&11.06&0.44&0.22&0.04\\
2023 April 13&3Spot&60047.308480&-68.0&0.74&0.77&0.40&0.01&0.01&21.70&5.60&3.30&0.40&0.03&0.06&-48.80&21.80&6.0&0.12&0.03&0.24\\
2023 April 14&3Spot&60048.174950&73.70&1.75&1.52&0.50&0.01&0.01&-82.32&1.70&2.15&0.50&0.04&0.02&33.60&5.30&5.53&0.31&0.03&0.04\\
2023 April 15&3Spot&60049.391550&-82.90&11.12&7.40&0.50&0.10&0.05&84.0&1.25&1.34&0.43&0.02&0.015&80.54&5.42&3.41&0.34&0.12&0.10\\
2023 April 19&3Spot&60053.157150&80.15&0.92&0.90&0.50&0.01&00.003&-64.11&1.55&2.0&0.50&0.01&0.01&-48.3&0.90&0.86&0.50&0.01&0.004\\

\hline
\end{tabular}
    }

    \footnotesize
    \tablefoot{\small The parameters $L$, and $R$ define the latitude, and radius of each stellar spot, where the numeric index 1, 2, and 3 correspond to individual modelled star spots. The notation $+$\text{err} and $-$\text{err} denotes the higher and lower uncertainty limits, respectively. Large error bars for spot radii reflect an unbreakable degeneracy between the $L$, $R$, and temperature of the spot.}

\end{sidewaystable}
 
\end{appendix}

\end{document}